\documentclass[twocolumn]{aastex63}
\usepackage{apjfonts}
\usepackage{amsmath}

\newcommand{\lcut}{$10^{25.25}$\ }
\newcommand{\loglcut}{25.25}
\newcommand{\maxz}{2}
\newcommand{\duty}{10}
\newcommand{\samplesize}{150,000}
\newcommand{\lzrgminz}{0.25}
\newcommand{\lzrgmaxz}{0.5}
\newcommand{\izrgminz}{0.5}
\newcommand{\izrgmaxz}{0.8}
\newcommand{\hzrgminzphot}{1}
\newcommand{\hizeff}{1.5}
\newcommand{\izeff}{0.6}
\newcommand{\lozeff}{0.4}
\newcommand{\hzrgfluxcut}{2.5}
\newcommand{\lookbacktime}{10 }
\newcommand{\wtwofaint}{17.25}
\newcommand{\jetdommass}{13}
\newcommand{\hzrgfsat}{1^{+2}_{-0.1}}
\newcommand{\izrgfsat}{13^{+7}_{-4}}
\newcommand{\lzrgfsat}{23^{+19}_{-11}}

\shorttitle{Halo Environments of $z < 2$ LOFAR Radio Galaxies}
\shortauthors{Petter et al.}
\accepted{2024 July 25 to \apj}
\graphicspath{{./}{figures/}}

\begin{document}

\title{Environments of luminous low-frequency radio galaxies since cosmic noon: jet-mode feedback dominates in groups}

\correspondingauthor{Grayson Petter}
\email{Grayson.C.Petter.GR@dartmouth.edu}

\author[0000-0001-6941-8411]{Grayson C. Petter}
\affil{Department of Physics and Astronomy, Dartmouth College, 6127 Wilder Laboratory, Hanover, NH 03755, USA}

\author[0000-0003-1468-9526]{Ryan C. Hickox}
\affiliation{Department of Physics and Astronomy, Dartmouth College, 6127 Wilder Laboratory, Hanover, NH 03755, USA}

\author[0000-0003-0487-6651]{Leah K. Morabito}
\affiliation{Centre for Extragalactic Astronomy, Department of Physics, Durham University, Durham DH1 3LE, UK}
\affiliation{Institute for Computational Cosmology, Department of Physics, University of Durham, South Road, Durham DH1 3LE, UK}

\author[0000-0002-5896-6313]{David M. Alexander}
\affiliation{Centre for Extragalactic Astronomy, Department of Physics, Durham University, Durham DH1 3LE, UK}



\begin{abstract}

Coupling between relativistic jets launched by accreting supermassive black holes and the surrounding gaseous media is a vital ingredient in galaxy evolution models. To constrain the environments in which this feedback takes place over cosmic time, we study the host halo properties of luminous low-frequency radio galaxies ($L_{150 \ \mathrm{MHz}} \gtrsim$ \lcut W/Hz) selected with the International LOFAR Telescope out to $z \sim \maxz$ through tomographic clustering and cosmic microwave background lensing measurements. We find that these systems occupy halos characteristic of galaxy groups ($M_h = 10^{13} - 10^{14} h^{-1} M_{\odot}$), evolving at a rate consistent with the mean growth rate of halos over the past $\sim$\lookbacktime Gyr. The coevolution of the clustering and the luminosity function reveals that the duty cycle of these systems is of order $\sim 10\%$ but has been mildly increasing since $z\sim 2$, while the duty cycle of quasars has been declining. We estimate the characteristic kinetic heating power injected by powerful jets per halo as a function of mass, and compare to the same quantity injected by quasar winds. We find that powerful jet heating dominates over quasar winds in halos $M_h \gtrsim 10^{\jetdommass} h^{-1} M_{\odot}$ at $z < 2$. These results conform to the paradigm of galaxy evolution in which mechanical jet power feedback is the dominant heating mechanism of the gas content of groups and clusters.


\end{abstract}



\section{Introduction} \label{sec:intro}

Galaxies form in the standard $\Lambda$-cold dark matter cosmology when baryons cool onto collapsed dark matter halos \citep{1978MNRAS.183..341W, 1991ApJ...379...52W}. However, galaxy formation models which incorporate only the first-order baryonic physics such as gravitational shock-heating and radiative cooling fail to match a wide array of observations of the most massive systems, galaxy groups and clusters. In particular, the efficiency of gas cooling and in turn galaxy growth is observed to be poorer than these models predict in more massive systems \citep{2001MNRAS.326.1228B, 2003ApJ...599...38B, 2004MNRAS.348.1078B, 2008MNRAS.391..481S, 2010PhR...495...33B,  2022PhR...973....1D}.

The advent of the ROSAT, \textit{Chandra} and \textit{XMM-Newton} X-ray telescopes revealed an unexpected abundance of hot gas in clusters and groups. The cooling time of X-ray gas cores in clusters, groups, and elliptical galaxies is observed to be much shorter than the ages of the systems \citep{1992MNRAS.258..177E, 2008ApJ...683L.107C, 2008MNRAS.385..757D}, implying that thermal energy must be resupplied by an internal mechanism \citep{1993MNRAS.263..323T, 1995MNRAS.276..663B}. 

It is now recognized that a supermassive black hole (SMBH) lies at the center of every massive galaxy \citep{1998AJ....115.2285M, 2013ARA&A..51..511K}, and that accretion onto this object triggers active galactic nucleus (AGN) activity which plays an integral role in shaping the host galaxy's growth and evolution through feedback processes \citep{2006MNRAS.368L..67B, 2007ARA&A..45..117M, 2009Natur.460..213C, 2012NJPh...14e5023M, 2012ARA&A..50..455F, 2012NewAR..56...93A, 2014ARA&A..52..589H, 2020NewAR..8801539H}. This picture was cemented theoretically when cosmological semi-analytic and hydrodynamic simulations including AGN feedback were able to address the cooling flow problem and reproduce observed local stellar mass functions \citep{2006MNRAS.365...11C, 2006MNRAS.370..645B, 2010MNRAS.406..822M}.

Feedback occurs when the AGN activity deposits kinetic energy into the surrounding medium through either radiative or mechanical processes, and these modes of feedback have canonically been associated with the accretion modes of the AGN itself \citep[e.g.,][]{2005MNRAS.363L..91C, 2007MNRAS.380..877S, 2012MNRAS.422.2816B}. Accretion at high Eddington ratios tends to form radiatively efficient disks, while mechanical energy dominates the output of AGN accreting at lower rates. This picture is however oversimplified, as multiple mechanisms can contribute to the kinetic output of both radiatively efficient and inefficient AGN activity \citep{2022MNRAS.516..245W, 2024Galax..12...17H}. For example,  compared to the classical radio-selected AGN lacking a radiatively efficient disk (low-excitation radio galaxies; LERGs), high-excitation radio galaxies (HERGs) host a radiatively-efficient disk but also drive powerful mechanical output. For the purposes of this work, we will explore the relative feedback roles played by AGN injecting energy dominantly through mechanical (jet/radio-mode) or radiative (quasar/wind-mode) processes, but we acknowledge these definitions are not perfectly exclusive.

The discovery of X-ray cavities coincident with radio jets \citep{1993MNRAS.264L..25B, 1994MNRAS.270..173C, 2000MNRAS.318L..65F, 2006MNRAS.366..417F} provided evidence for jet-mode feedback in the aforementioned groups and clusters. These systems serve as laboratories with which to estimate the jet energy deposition \citep{2000ApJ...534L.135M, 2000A&A...356..788C, 2004ApJ...607..800B, 2008ApJ...686..859B, 2010ApJ...720.1066C, 2022A&A...668A..65T}, showing that the mechanical power can greatly exceed that of the synchrotron luminosity, enough to inflate and heat the atmospheres of groups and clusters. 

Alternatively, radiation pressure from efficient accretion can drive quasar/wind-mode feedback \citep{1998A&A...331L...1S, 2000ApJ...545...63E, 2005ARA&A..43..769V, 2005Natur.433..604D, 2005MNRAS.361..776S, 2006ApJS..163....1H, 2006ApJ...653...86T, 2021NatAs...5...13L}, evidenced by observed gaseous outflows coincident with quasar activity \citep{2010MNRAS.402.2211A, 2011ApJ...729L..27R, 2012MNRAS.426.1073H, 2013MNRAS.436.2576L, 2014MNRAS.441.3306H, 2014ApJ...788...54G, 2015MNRAS.446.2394B, 2016MNRAS.459.3144Z, 2017A&A...601A.143F, 2019MNRAS.488.4126P}. However to date, the complete picture of the relative role that jet-mode and wind-mode feedback play is elusive. Radio galaxies are thought to be preferentially triggered by an advection-dominated accretion flow \citep{1994ApJ...428L..13N}, when the SMBH is fed by condensing hot gas in a massive halo. Meanwhile, quasar activity is generated by a thin accretion disk \citep{1973A&A....24..337S}, expected to be fed by cold gas streams in lower-mass halos \citep{2009MNRAS.395..160K}. This suggests that jet-mode and wind-mode feedback may be expected to occur in different environments.


Modern hydrodynamic and semi-analytic galaxy evolution models routinely implement feedback from AGNs and match a wide array of observables \citep{2014MNRAS.439..264G, 2014MNRAS.444.1518V, 2015MNRAS.446..521S, 2015ARA&A..53...51S, 2016MNRAS.463.3948D, 2016ApJS..222...22C, 2017MNRAS.467.4739K, 2018MNRAS.475..676S, 2019MNRAS.486.2827D, 2023arXiv230915898R}. However, feedback effects typically take place on smaller scales than the resolution elements of simulations, such that models must assume ``sub-grid'' prescriptions. For a complete understanding of AGN feedback, observations are required to constrain which flavor of feedback dominates, and in which environments. Therefore in this work, we study the environments in which luminous low-frequency radio galaxies occur over the past $\sim \lookbacktime$ Gyr.


Radio galaxies at $z<1$ have been found to cluster strongly, residing in halos of masses $\sim 10^{13.2} - 10^{13.5} h^{-1} M_{\odot}$ \citep{2004MNRAS.350.1485M, 2009MNRAS.393..377M, 2009ApJ...696..891H}, implying they typically deposit their energy into massive group environments. It is important to study the environments radio galaxies occupied during ``cosmic noon'' ($z\sim 2$), when the cosmic star formation efficiency began to decline likely in part due to AGN feedback \citep{2014ARA&A..52..415M}. To date, relatively few studies have explored the clustering of radio galaxies at $z > 1$, which have been confined to small fields, and conducted at GHz frequencies \citep{2014MNRAS.440.2322L, 2017MNRAS.464.3271M, 2018MNRAS.474.4133H}. Recent advances in low-frequency radio surveys such as with the LOw-Frequency ARray \citep[LOFAR;][]{2013A&A...556A...2V} now offer a new window on radio galaxy clustering. The depth and area curate an unprecedentedly large radio AGN sample, and the low frequency observations reveal the old electron populations and thus likely better trace the long-term energy input of jets into their surroundings.

In this work, we study the host halo environments of $\sim \samplesize$ luminous radio galaxies ($L_{150 \ \mathrm{MHz}} \gtrsim$ \lcut W/Hz) selected at low frequencies using the LOFAR telescope with tomographic clustering and lensing measurements out to $z < \maxz$. We find that luminous radio galaxies occupy massive halo environments characteristic of galaxy groups ($M_h \gtrsim 10^{13} h^{-1} M_{\odot}$). We estimate the duty cycle of these systems is of order $\sim 10\%$, and has been increasing with cosmic time since $z\sim 2$, interestingly meanwhile the duty cycle of quasars has been declining. We finally estimate the average kinetic heating power released by these systems into halos as a function of mass, finding that jet-mode heating dominates over quasar/wind-mode in group-scale halos at $z < 2$.


Throughout this work, we adopt a ``Planck 2018'' CMB+BAO $\Lambda$-CDM concordance cosmology \citep{2020A&A...641A...6P}, with $h = H_0/100 \ \mathrm{km \ s}^{-1} \mathrm{Mpc}^{-1} = 0.6766$, $\Omega_{m} = 0.3111$, $\Omega_{\Lambda} = 0.6888$, $\sigma_{8} = 0.8102$, and $n_{s} = 0.9665$. We adopt the convention where synchrotron spectra are parameterized as $S_\nu \propto \nu^{\alpha}$, and perform $K$-corrections throughout using an assumed spectral index for radio galaxies of $\alpha=-0.7$. Any magnitudes are presented in the Vega system, and any logarithms are in base ten unless otherwise stated.

\section{Data} \label{sec:data}

In this work, we study the clustering properties of radio galaxies selected at 150 MHz using LOFAR data. We use Wide-field Survey Explorer \citep[\textit{WISE;}][]{2010AJ....140.1868W} infrared and DESI Legacy Imaging Survey \citep[DLIS; ][]{2019AJ....157..168D} optical counterparts for redshift information of the radio sources, allowing study of the time-evolution of their clustering. Finally, we cross-correlate their positions with spectroscopic galaxy samples  from the Sloan Digital Sky Survey \citep[SDSS;][]{2000AJ....120.1579Y} in several redshift bins for additional resolution in the time-evolution.

\subsection{The LOFAR Two-metre Sky Survey}

The LOFAR is a sensitive interferometric telescope currently revolutionizing surveys of the radio sky. In particular, the pioneering LOFAR Two-metre Sky Survey \citep[LoTSS;][]{2017A&A...598A.104S} is the first to produce deep (RMS $\sim 100 \ \mu$Jy beam$^{-1}$) and high-resolution (6$\arcsec$) maps at low frequencies ($\sim 150$ MHz) over a wide area, and represents an order of magnitude sensitivity improvement for typical sources over the Faint Images of the Radio Sky at Twenty-centimeters (FIRST) survey \citep{1995ApJ...450..559B, 2015ApJ...801...26H}, conducted at 1.4 GHz. The second data release of LoTSS-wide \citep[DR2;][]{2022A&A...659A...1S} covers 5634 deg$^2$ and catalogs 4,396,228 sources.

This is a premier dataset to study radio galaxy clustering \citep{2020A&A...643A.100S, 2022ApJ...928...38T, 2024MNRAS.527.6540H, 2024A&A...681A.105N} for a variety of reasons. The detection depth surpasses the switch point of $\sim 1.5$ mJy below which star-forming galaxies and radio-quiet AGNs become the dominant populations \citep{2023MNRAS.523.1729B}, and therefore a deeper survey would not select a larger sample of radio galaxies on the basis of flux alone. The wide area is useful for statistical power in clustering measurements due to the inherent rarity of luminous radio galaxies, and the survey footprint significantly overlaps with that of the SDSS, enabling cross-correlations with spectroscopic samples. The high angular resolution allows accurate identification of optical/infrared counterparts \citep[e.g.,][]{2019A&A...622A...2W, 2019A&A...622A..12H, 2023A&A...678A.151H} for host galaxy property and redshift estimation. Finally, low-frequency observations are well-suited to tracing the environments in which feedback may be taking place, as they are more sensitive to steep-spectrum, lobe-dominated sources which better trace the long-term integrated power input into their surroundings than GHz frequencies, which are more sensitive to core-dominated or beamed emission.

We adopt the LoTSS DR2 source catalog presented in \citet[][hereafter H23]{2023A&A...678A.151H}, which has attempted to associate double-lobed duplicate detections into single sources, and match to optical \citep[DLIS;][]{2019AJ....157..168D} and/or infrared \citep[unWISE;][]{2019ApJS..240...30S} counterparts when possible. The association results from a combination of likelihood-ratio cross-match methods, machine learning methods, and visual-inspections by astronomers and citizen scientists. Sources with optical counterparts in the DLIS are enhanced by photometric redshift estimates \citep{2022MNRAS.512.3662D}, which we leverage in our study. The samples we construct using optical counterpart photometric redshifts at $z < 1$ in $\S$\ref{sec:sample} are composed entirely of sources brighter than the $>8$ mJy "bright" threshold referred to in H23, which received the highest priority of visual inspections. We thus are confident in the robustness and completeness of our photo-$z$-selected samples, but we refer the reader to H23 for a detailed discussion of the association techniques. 

We choose to use only the LoTSS data in the northern Galactic hemisphere, as this region makes up the majority of the DR2 footprint ($74\%$), has more uniform imaging depth, and better overlap with DLIS imaging and SDSS spectroscopic surveys.


\subsection{eBOSS Quasar Sample}

One of the goals of this work is to study radio galaxy clustering and its evolution at $z > 1$. We therefore utilize quasars with spectroscopic redshifts from SDSS surveys as biased matter tracers for cross-correlations with LoTSS-selected radio galaxies at $1 < z < \maxz$.

The Extended Baryon Oscillation Spectroscopic Survey \citep[eBOSS;][]{2016AJ....151...44D} was a spectroscopic survey designed to measure baryon acoustic oscillations in the distribution of three tracers, star-forming emission-line galaxies, luminous red galaxies, and quasars. The quasar sample tracing high-redshift structure consists of 343,708 uniformly-targeted \citep{2015ApJS..221...27M} systems at $0.8 < z < 2.2$, representing the largest statistical sample of spectroscopic quasars to date. The eBOSS collaboration has produced large-scale structure catalogs including randoms and weights for measuring unbiased correlation functions \citep{2020MNRAS.498.2354R}. We utilize the versions presented in \citet{2021MNRAS.506.3439R}, incorporating updated systematic weights. 



\subsection{LoTSS-deep Data}
\label{sec:deep}

We make use of derived data products from the LoTSS-deep survey DR1 \citep{2021A&A...648A...1T, 2021A&A...648A...2S} to verify our sample selections and redshift distributions. This lies amongst the deepest (RMS $\sim20 \ \mu$Jy bm$^{-1}$) radio surveys to date, but covers more than an order of magnitude wider area than the comparable depth VLA-COSMOS survey \citep{2017A&A...602A...1S}, totaling 25 deg$^2$ with overlapping deep multiwavelength photometry in the Bo{\"o}tes, Lockman-hole, and ELAIS-N1 fields. This photometry allows for host identification \citep{2021A&A...648A...3K}, photometric redshift estimation \citep{2021A&A...648A...4D}, source classification, and host galaxy property constraint through spectral energy decomposition \citep[][hereafter B23]{2023MNRAS.523.1729B}, and the measurement of the radio AGN and star formation luminosity functions \citep{2022MNRAS.513.3742K}. The redshift information is crucial to our clustering and lensing analysis, and the luminosity functions will allow estimation of the occupation statistics when combined with clustering measurements. We update the catalogs with the new Dark Energy Spectroscopic Instrument (DESI) early release data \citep{2024AJ....168...58D} for additional spectroscopic redshifts where available. We prioritize the data in the Bo{\"o}tes field unless otherwise stated, as this region has the best spectroscopic coverage primarily courtesy of the AGN and Galaxy Evolution Survey \citep[AGES;][]{2012ApJS..200....8K}.

\subsection{Planck Cosmic Microwave Background Lensing Map}
\label{sec:cmbdata}

Cosmic microwave background photons emitted during the recombination epoch have been gravitationally-lensed by the intervening structure, and provide a complimentary probe of high-redshift galaxies' host halo properties along with correlation functions. A number of high-resolution CMB experiments have now produced wide-area maps of the lensing convergence $\kappa$, a projected surface mass density tracing structure from $0.5 \lesssim z \lesssim 5$ \citep[e.g.,][]{2020A&A...641A...6P, 2023PhRvD.107b3529O, 
2024ApJ...962..113M}. Only \textit{Planck}'s all-sky survey overlaps with LoTSS DR2, and thus we use the \textit{Planck} final release (PR4) lensing map \citep{2022JCAP...09..039C} as an independent constraint on the host-halo properties of the LoTSS radio galaxies. We also make use of the provided simulated maps to estimate uncertainties. We produce maps from the combined temperature and polarization data (minimum-variance reconstruction) at NSIDE=1024 resolution using $\ell$-modes < 2048 for cross-correlation with radio galaxy overdensity maps.

\subsection{CatWISE Infrared Photometry} \label{sec:wise}

Many luminous radio galaxies at higher redshifts ($z \gtrsim 1)$ are hosted by galaxies too faint to be readily detected in the optical waveband at the DLIS depth. However, their hosts can be easier recovered at near-infrared wavelengths, as aged stellar populations exhibit a negative $K$-correction out to $z \sim 2$ at observed-frame 3-5 $\mu$m. Thus, we update the H23 catalog with deeper \textit{WISE} data than was used in H23 to push the host galaxy detection to higher redshift. \textit{WISE} is a space telescope which has mapped the entire sky in four infrared bands, centered at 3.4, 4.6, 12 and 22 $\mu$m (named W1, W2, W3, and W4 respectively). \textit{WISE} continues to operate in the two shorter-wavelength bands in its post-cryogenic phase \citep{2011ApJ...731...53M}, and the deeper resulting imaging continues to detect fainter and higher-redshift galaxies. The H23 LoTSS catalog has associated radio sources with unWISE counterparts when possible, and 69$\%$ of $S_{150 \ \mathrm{MHz}} > 2$ mJy sources have a W1/W2 measurement. However, the CatWISE2020 \citep{CatWISE, 2021ApJS..253....8M} catalog probes deeper than unWISE, which facilitates detection of host galaxy counterparts of radio galaxies at higher-redshift. We match the H23 catalog with CatWISE2020 using a conservative match radius of $2.5\arcsec$ (from the H23 counterpart if available, then the radio position), and let the CatWISE measurements supersede unWISE where applicable. This raises the fraction of $S_{150 \ \mathrm{MHz}} > 2$ mJy systems with infrared counterparts to $79\%$. As expected, systems detected in CatWISE but not unWISE are found to have a redshift distribution peaking at $z\sim2$ in the LoTSS-deep fields. We will leverage the feature that higher-redshift galaxies will appear both redder and fainter in W1/W2 space \citep{2019ApJS..240...30S, 2020JCAP...05..047K} to select $z>1$ radio galaxies in $\S$\ref{sec:sample}. 


\section{Radio Galaxy Samples} \label{sec:sample}
We aim to study the environments of luminous radio galaxies across cosmic time, including at the relatively unexplored epoch of $z > 1$. However, the subset of LoTSS sources dominated by radio galaxies ($S_{150 \ \mathrm{MHz}} \gtrsim 2$ mJy) exhibits a broad redshift distribution peaking at $z \sim 0.5$ with a long tail towards higher redshifts \citep[][B23]{2021A&A...648A...4D, 2021MNRAS.502..876A}. Thus, in this work we use optical/infrared counterparts to separate the radio galaxies into three broad redshift regimes. We then also utilize a \textit{tomographic} technique whereby we cross-correlate radio sources with tracer populations of galaxies at known redshift for additional time-resolution in the study of their clustering.

The host galaxies of luminous radio galaxies are massive at $z \lesssim 1$ ($M_{\star} \sim 10^{11} M_{\odot}$; B23), and thus readily detectable in the optical waveband at the depth of the DLIS out to $z \lesssim 1$, where their photo-$z$s are reliable \citep{2022MNRAS.512.3662D}. At higher redshifts, host galaxies undetectable in the DLIS can be recovered at infrared wavebands probed by \textit{WISE} ($\S$\ref{sec:wise}). We thus use optical counterparts to LoTSS sources augmented by photo-$z$s to select samples of radio galaxies at $z < 1$, and \textit{WISE} counterparts to select systems at $z > 1$.

We match a flux-limited ($S_{150 \ \mathrm{MHz}} > \hzrgfluxcut$ mJy) subset of the H23 catalog dominated by radio galaxies ($\sim 88\%$) with \textit{WISE} counterparts to the LoTSS-deep catalog in the Bo{\"o}tes field for deep photometric or spectroscopic redshift information \citep{2021A&A...648A...4D}, and display the resulting \textit{WISE} color-magnitude diagram in Figure \ref{fig:wisediagram}. Sources are colored by their broad redshift regime and galaxies belonging to redshift epochs cluster together in the diagram, demonstrating that \textit{WISE} magnitudes and colors can be used to isolate radio galaxies at high redshift.

\subsection{High-z Radio Galaxies (HzRGs)}

\begin{figure}
    \centering
    \includegraphics[width=0.45\textwidth]{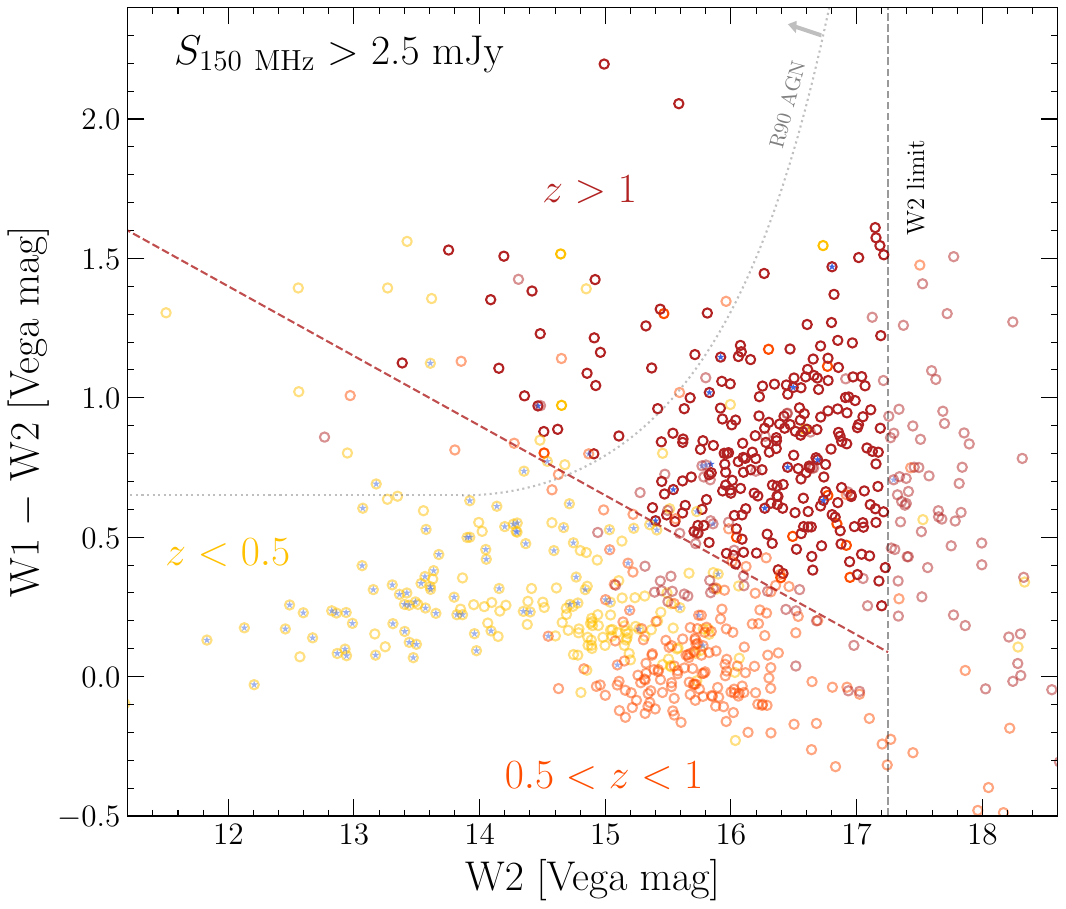}
    \caption{An infrared \textit{WISE} diagram of LoTSS-bright ($S_{150 \ \mathrm{MHz}} > \hzrgfluxcut$ mJy) galaxies in the Bo{\"o}tes field, colored by their broad redshift regime \citep[][B23]{2021A&A...648A...4D}, with yellow, orange, and red corresponding to the redshift ranges $z \in [0, 0.5], [0.5, 1], [1, 3]$, respectively. Galaxies at different redshift epochs broadly cluster together in the diagram. We display our sliding color cut for selecting high-redshift $z>1$ radio galaxies (Eq.\ \ref{eq:hzrg}) with a red dashed line, highlighting the sources satisfying this cut in bold. The \citet{2018ApJS..234...23A} R90 criterion for selecting radiative-mode quasars is drawn with a gray dotted line; most of these radio galaxies would not be selected as \textit{WISE} quasars. Sources with radio emission consistent with their far-infrared star formation rates (B23) are highlighted with an additional internal blue star marker, and are mostly confined to low redshifts. This figure demonstrates that the infrared counterpart properties of LoTSS sources can be used to select a complete and reliable sample of radio galaxies at $1 \lesssim z \lesssim \maxz$.}
    \label{fig:wisediagram}
\end{figure}

To curate the $z>1$ sample, we adopt the sliding color cut form of \citet{2019ApJS..240...30S} designed to select high redshift galaxies shown in Figure \ref{fig:wisediagram} with a red dashed line, along with a $90\%$ W2 completeness cut \citep{catwise2020}. We also incorporate a photometric redshift cut of $z_{\mathrm{phot}} > \hzrgminzphot$ for sources with an optical counterpart to cull low-redshift interlopers. 

\begin{equation}
    \left\{ \begin{aligned}
    &  S_{\mathrm{150 \ MHz}} > \hzrgfluxcut \ \mathrm{mJy} \\
    &\mathrm{W}1 - \mathrm{W}2 > (17-\mathrm{W}2)/4 + 0.15 \\
    &\mathrm{W}2 < \wtwofaint \\
    & z_{\mathrm{phot}} > \hzrgminzphot \ | \ z_{\mathrm{phot}} == \mathrm{Null} 
    \end{aligned} \right.
    \label{eq:hzrg}
\end{equation}

This flux cut corresponds to a luminosity limit of $\sim 10^{25.25}$ W/Hz at $z\sim1.5$. We refer to this sample of bright LoTSS sources with faint/red \textit{WISE} counterparts as our ``high-redshift radio galaxy'' (HzRG;\ $z > 1$) sample throughout the rest of the text. Ninety-five percent of these sources are classified as radio galaxies in B23, verifying the reliability of our $z > 1$ radio galaxy selection.

\subsection{Intermediate-z Radio Galaxies (IzRGs)}

At $z \lesssim$ 1, reliable photo-$z$s enable construction of approximately radio luminosity-limited samples, as opposed to the flux-limited HzRG sample. For broad correspondence with the luminosity distribution probed by the HzRG sample, we select systems with $L_{150 \ \mathrm{MHz}}(z_{\mathrm{phot}}) > 10^{\loglcut}\ \mathrm{W/Hz}$. As $S_{150\mathrm{MHz}} > 8$ mJy sources received the highest-priority of visual associations in H23, the above luminosity limit requires that we limit to $z_{\mathrm{phot}} \lesssim \izrgmaxz$ for uniform selection. Additionally, we use the converse of the \textit{WISE} color cut for HzRGs to remove likely $z>1$ interlopers. Together, these cuts form a selection for intermediate-z radio galaxies (IzRGs, $\izrgminz <z< \izrgmaxz$): 


\begin{equation}
    \left\{ \begin{aligned} 
    & \izrgminz < z_{\mathrm{phot}} < \izrgmaxz \\
    & L_{150 \ \mathrm{MHz}}(z_{\mathrm{phot}}) > 10^{\loglcut}\ \mathrm{W/Hz} \\
    &\mathrm{W}1 - \mathrm{W}2 < (17-\mathrm{W}2)/4 + 0.15
    \end{aligned} \right.
    \label{eq:izrg}
\end{equation}

\subsection{Low-z Radio Galaxies (LzRGs)}



Finally, we extend our study once more to lower redshift with the same strategy presented for IzRGs, forming a selection for low-$z$ radio galaxies (LzRGs; $\lzrgminz < z < \lzrgmaxz$):

\begin{equation}
    \left\{ \begin{aligned} 
    & \lzrgminz < z_{\mathrm{phot}} < \lzrgmaxz \\
    & L_{150 \ \mathrm{MHz}}(z_{\mathrm{phot}}) > 10^{\loglcut}\ \mathrm{W/Hz} \\
    &\mathrm{W}1 - \mathrm{W}2 < (17-\mathrm{W}2)/4 + 0.15
    \end{aligned} \right.
    \label{eq:lzrg}
\end{equation}


\subsection{Luminosity Distributions}

\begin{figure}
    \centering
    \includegraphics[width=0.45\textwidth]{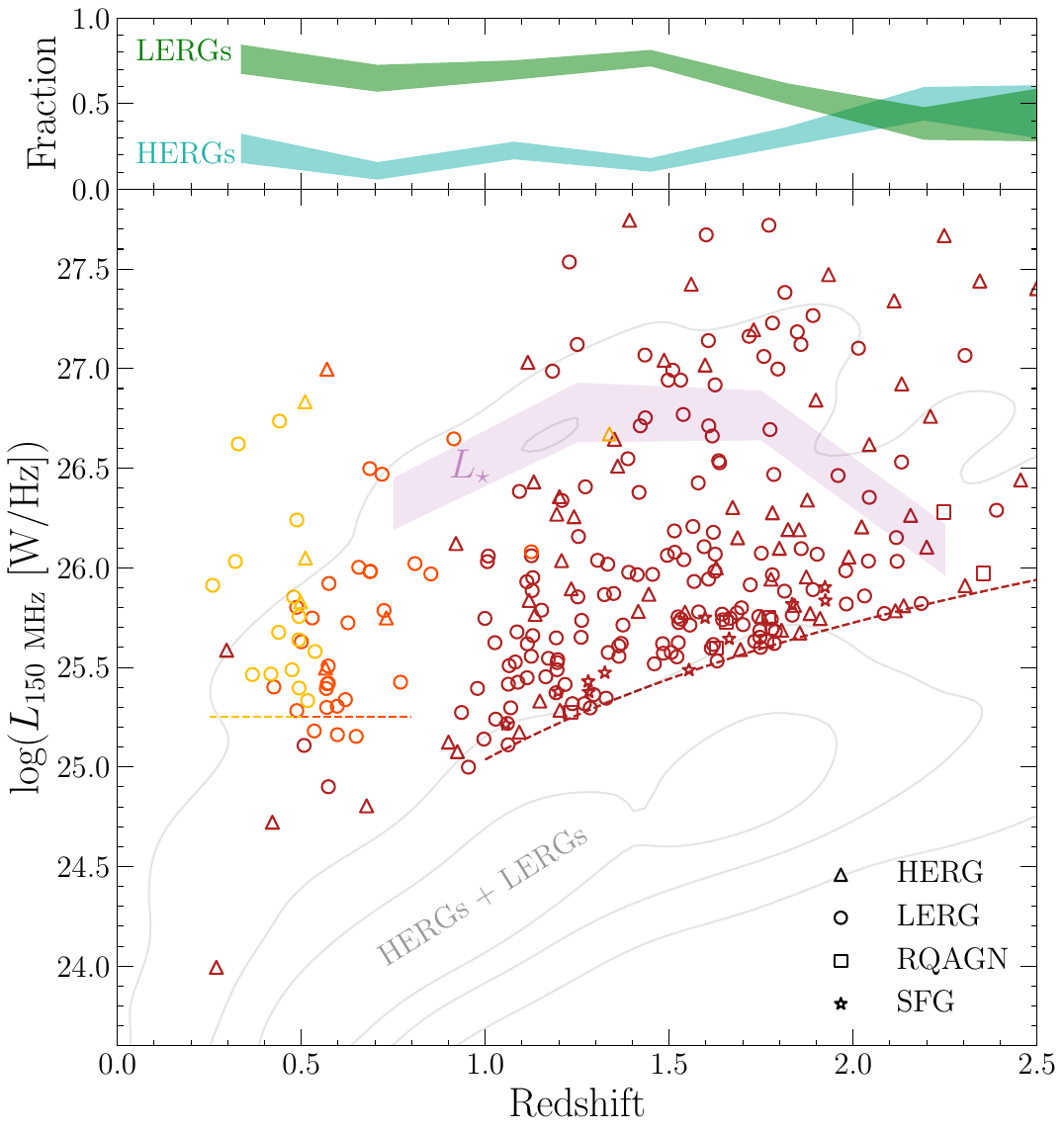}
    \caption{The luminosity-redshift distributions of our samples using redshift estimates from the Bo{\"o}tes LoTSS-deep field. The dashed yellow, orange, and red lines show the luminosity thresholds corresponding to the selections for LzRGs, IzRGs, and HzRGs, respectively. This demonstrates that we are able to curate samples with similar luminosity thresholds at three characteristic redshift epochs. Sources outside the selection boundaries result from discrepancies between the \citep{2022MNRAS.512.3662D} photo-$z$ and the best deep field redshift estimate. Gray contours show the distribution of the full radio galaxy populations recovered in LoTSS-deep. Circular, triangular, square, and star markers denote classifications as LERGs, HERGs, RQAGN, and SFGs, respectively. The top panel shows the LERG and HERG fractions of our combined samples binned in redshift, demonstrating that our sample contains a mix of the two populations, with LERGs dominating at $z < 2$. The characteristic break luminosity of the radio galaxy luminosity function $L_{\star}$ \citep{2023MNRAS.523.5292K} is shown with a purple band, demonstrating that our selections curate representative samples of radio galaxies.}
    \label{fig:lum}
\end{figure}

We show the luminosity-redshift distributions of our samples in Figure \ref{fig:lum}. This demonstrates that our selections define samples with similar luminosity thresholds ($L_{150 \ \mathrm{MHz}} \gtrsim$ \lcut W/Hz) from $\lzrgminz \lesssim z \lesssim \maxz$. These systems are complete to luminosities $\sim 1$ dex below the break of the radio AGN luminosity function \citep{2023MNRAS.523.5292K}, implying that they are representative of the global heating output of jetted AGN. The top panel shows that the samples are dominated by LERGs at $z < 2$, with negligible contamination rates of radio-quiet AGN and SFGs.

\subsection{Redshift Distributions}

\begin{figure}
    \centering
    \includegraphics[width=0.45\textwidth]{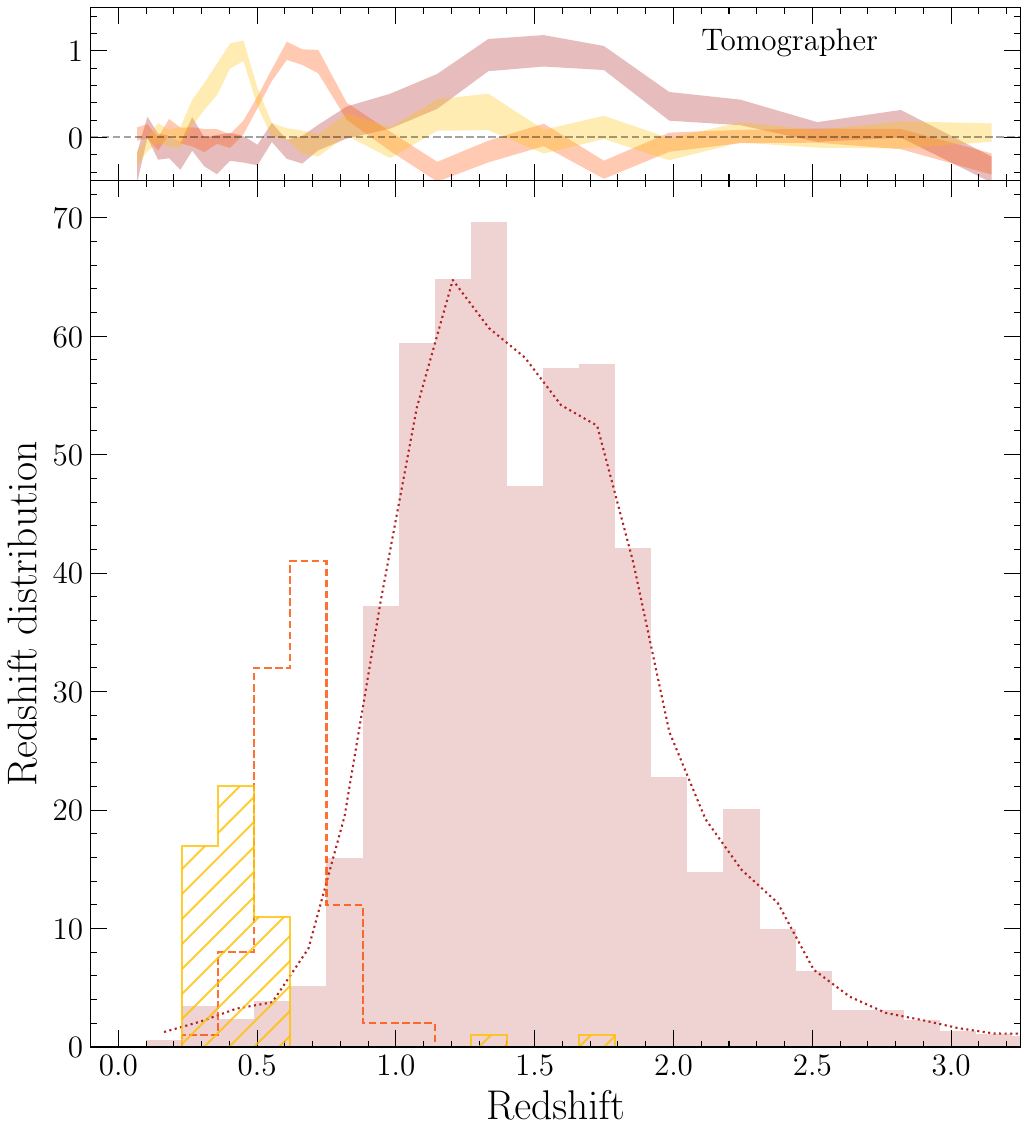}
    \caption{We show the redshift distributions of LzRGs, IzRGs, and HzRGs with yellow hatched, orange dashed, and filled red histograms, respectively. We also show a Savitsky-Golay filter fit to the HzRG distribution with a dotted red line to smooth over the artificial peaks in the distribution, which we use in our analysis. The top panel displays corresponding redshift distributions implied from a clustering-based method using the {\tt Tomographer} assuming a fiducial redshift-bias dependence (see text). This qualitatively verifies the efficacy of the selection techniques and the validity of the cross-match redshift distributions. This figure demonstrates that our selections curate radio galaxies in three largely distinct redshift epochs.}
    \label{fig:dndz}
\end{figure}
Crucial to interpretation of a clustering or lensing measurement is knowledge of a sample's redshift distribution. After curating the HzRG, IzRG, and LzRG samples using optical/infrared information, we estimate their redshift distributions using the LoTSS source-associated redshift catalogs of \citep[][B23]{2021A&A...648A...4D}. We display the resulting redshift distributions in Figure \ref{fig:dndz}. The proposed selections clearly curate samples in distinct redshift regimes, allowing the study of radio galaxy clustering evolution. 

To generate the redshift distribution of HzRGs which relies heavily on photometric rather than spectroscopic estimates, we have incorporated the photo-$z$ probability density functions. Even after this step, we observe that the redshift distribution features ``peaks'', notably at $z \sim 1.1$ and $z\sim1.7$, as noted in \citet{2022MNRAS.513.3742K}. These result from photometric redshift aliasing due to gaps in filter coverage, and are thus unphysical. We thus filter the HzRG redshift distribution with a linear Savitsky-Golay filter, and use this smoothed redshift distribution in all subsequent analyses, though this does not materially affect our results. 

To confirm the validity of these cross-match redshift distributions, we derive independent estimates using the clustering redshift technique \citep{2008ApJ...684...88N, 2013arXiv1303.4722M, 2019ApJ...870..120C}, as implemented in the {\tt Tomographer} web tool.\footnote{\url{http://tomographer.org}} This yields a bias-scaled redshift distribution $ b(z) dN/dz$, but untangling any bias evolution from the redshift distribution is degenerate with the goal of this work. Instead, we use this information as a qualitative check on our redshift distributions and assume that the bias evolves inversely with the cosmological growth factor \citep[$b \propto 1/D(z)$; e.g.,][]{2021MNRAS.502..876A}. We show the results of the clustering redshift technique in the top panel of Figure \ref{fig:dndz}, confirming that our selections generate radio galaxy samples at three largely distinct redshift epochs. We do not use this clustering-based redshift distribution in any further analysis, rather this serves as a qualitative check that our sample selections are performing as intended.

\subsection{Sample Properties}

We thus have constructed samples of luminous ($L_{150 \ \mathrm{MHz}} \gtrsim$ \lcut W/Hz) radio galaxies at low ($z=\lozeff$), intermediate ($z=\izeff$), and high redshift ($z=\hizeff$) using optical and infrared counterparts to LoTSS sources to probe the host halo environments of low-frequency radio galaxies over cosmic time. We note that this sample is comprised of the most luminous radio galaxies, with limits $\sim 1$ dex below the break luminosity $L_{\star}$. As the total kinetic power output of all radio galaxies is dominated by relatively-luminous sources \citep{2023MNRAS.523.5292K}, systems of the luminosities considered here contribute more than half of all the kinetic luminosity density of all radio AGNs. If radio jets are indeed important in shaping galaxy evolution, this sample should thus be significantly representative of the environments in which feedback is taking place. We note that recovering the remainder of fainter radio galaxies over wide swaths of sky will be challenging without corresponding far-infrared data to disentangle star formation processes. The sample has a low contamination rate ($< 5\%$) by pure SFGs and radio quiet AGN, and is dominated by low-excitation radio galaxies (LERGs). As noted in \citet{2022MNRAS.513.3742K}, this sample is also hosted by predominantly star-forming galaxies at high redshift, transitioning to being hosted almost entirely by luminous passive galaxies at $z < 1$.

\subsection{Masking}
\label{sec:mask}
Clustering measurements are sensitive to systematics in the selection function of the given sample of galaxies. It is therefore paramount to characterize the angular selection function of the survey. We characterize the footprint of the LoTSS-wide DR2 using a multi-order coverage (MOC) map.\footnote{\url{https://hips.astron.nl/ASTRON/P/lotss_dr2_high/Moc.fits}} We then stitch together an estimate of the RMS imaging noise from the provided RMS map of each facet,\footnote{\url{https://lofar-surveys.org/dr2_release.html}} by projecting to a {\tt NSIDE=4096} {\tt HEALPix} map. Next, we downgrade the map to {\tt NSIDE=128} using the median of the child pixels. Finally, we mask regions of the sky with local median RMS $> 0.2$ mJy beam$^{-1}$, totalling $1.5\%$ of the LoTSS Galactic north footprint. Our samples of radio sources ($S_{150 \ \mathrm{MHz}} > \hzrgfluxcut$ mJy) are thus expected to be highly complete. 

Requiring optical/infrared counterparts to the radio sources from \textit{WISE} and/or DLIS introduces additional considerations in the selection functions of our samples. We thus leverage a series of {\tt HEALPix} maps\footnote{\url{https://data.desi.lbl.gov/public/ets/target/catalogs/dr9/1.1.1/pixweight/main/resolve/dark/pixweight-1-dark.fits}} characterizing angular systematics generated by the DESI collaboration \citep{2023AJ....165...50M}. First, we limit the footprint to that of the DLIS DR8 using the provided MOC. We mask regions of abnormally shallow \textit{WISE} imaging depth ($5\sigma$ W2 PSF depth < 17.25).  Next, we limit the footprint to where DLIS made at least 3 observations in each of the $g$, $r$, and $z$ bands, and the $5\sigma$ PSF $z$-band depth was at least $z [\mathrm{AB}]>23$. We also construct a veto {\tt NSIDE=4096} mask of regions with DLIS bitmask flags touching a bright star or large galaxy in the Siena Galaxy Atlas \citep{2023ApJS..269....3M}. Finally, we measure trends of each sample's density with Galactic/ecliptic coordinates, stellar density, and Galactic reddening. We observe significant deviations at reddening values $E(B-V) > 0.1$ and at stellar densities $> 2000 \ \mathrm{deg}^{-2}$, and thus mask these regions. In total, 12.5$\%$ of the LoTSS-DR2 north Galactic footprint is masked by our procedure, leaving 3654 deg$^2$ of survey area. In the case of cross-correlations with SDSS spectroscopic quasars, we mask the HzRG sample with the eBOSS masks,\footnote{\url{https://data.sdss.org/sas/dr16/eboss/lss/catalogs/DR16/}} and correspondingly we mask the spectroscopic sample with the LoTSS mask described above. 


\section{Measurements}
\label{sec:measurements}

We perform a variety of measurements to study the halo environments of luminous low-frequency radio galaxies over cosmic time. All correlation function measurements are performed with the {\tt Corrfunc} \citep{2020MNRAS.491.3022S} package. 

\subsection{Radio Galaxy Autocorrelations}

First, we measure the angular autocorrelation functions of the luminous radio galaxy samples. The angular autocorrelation function is the excess probability above that of a Poisson distribution for detecting a pair of galaxies separated by an angle $\theta$ \citep{1980lssu.book.....P}. We mask the data and random catalogs identically ($\S$\ref{sec:mask}), and then measure the angular clustering using the \citep{1993ApJ...412...64L} estimator:

\begin{equation}
    w(\theta) = \frac{DD - 2 DR + RR}{DD}, 
\end{equation}
where $DD$, $DR$, and $RR$ are normalized counts for data-data, data-random, and random-random pairs, respectively. We interpret the autocorrelations both with linear and full halo occupation distribution models in ($\S$\ref{sec:model}). Therefore, we measure the autocorrelations at projected angular scales for the median redshift of the sample corresponding to $0.5 - 25 \ h^{-1}$Mpc for the full halo occupation distribution constraint, but fit only over $5 - 25 \ h^{-1}$Mpc when assuming linear models.

To estimate uncertainties on all correlation functions, we split the survey footprint into 30 equal area patches using $k$-means clustering, then perform a bootstrap resampling of the patches \citep[e.g.,][]{1982jbor.book.....E, 2009MNRAS.396...19N}, calculating a correlation function with the data and randoms from each draw. This process is repeated 500 times and the variance across realizations is taken to be the variance of our measurement. 

\begin{figure}
    \centering
    \includegraphics[width=0.45\textwidth]{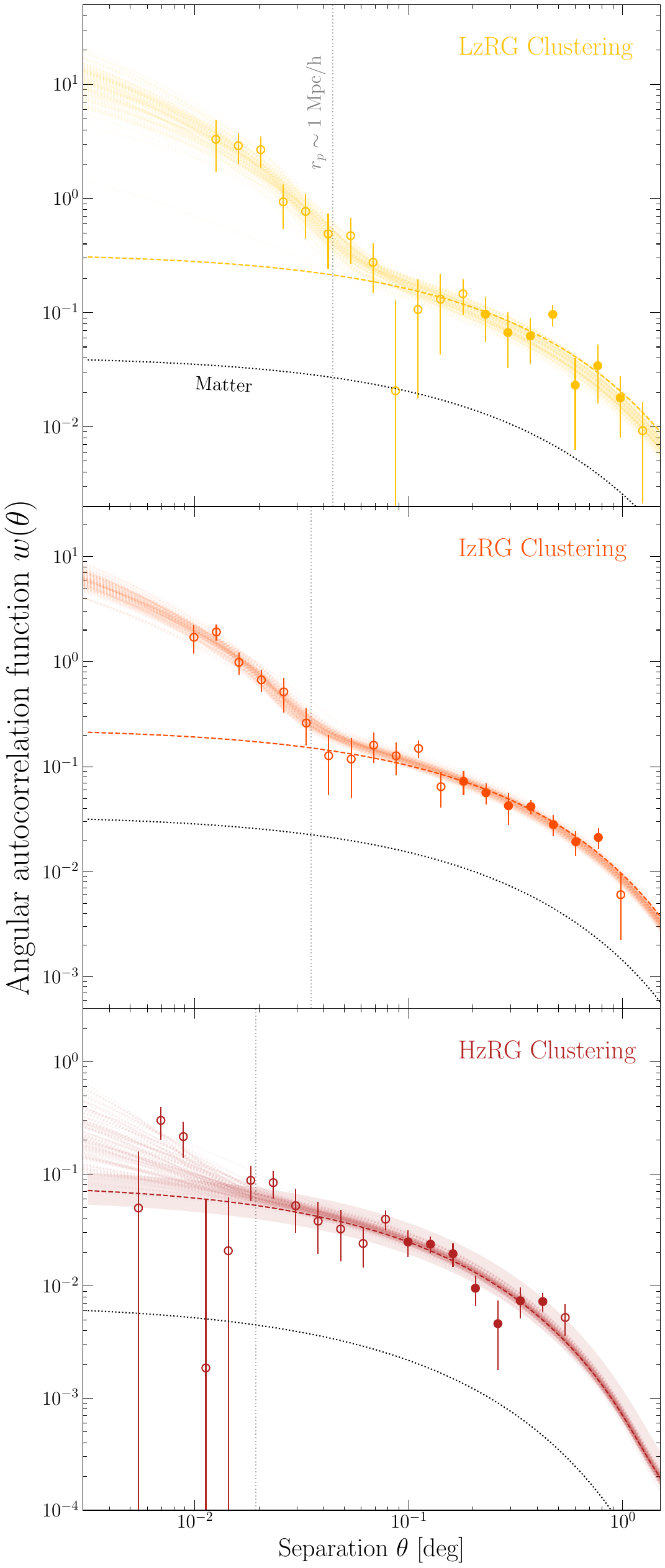}
    \caption{The autocorrelation function measurements of LzRGs (top panel in yellow), IzRGs (middle panel in orange) and HzRGs (bottom panel in red). Filled markers show the measurements on linear projected scales ($5 - 25 \ h^{-1}$ Mpc) at the median redshift of the sample, while open markers represent other scales. The projected scale of $\sim 1$ Mpc/h delineating the boundary of the one and two-halo terms is shown with vertical gray lines. The best fit of a linear model to the linear scales is shown with a dashed colored line. We also show the best fit HOD models to the full range of scales using transparent lines drawn from the posterior distributions. We thus are able to constrain the host halo properties of luminous radio galaxies at three redshift epochs at $z < 2$.}
    \label{fig:autocfs}
\end{figure}

We show the autocorrelation function measurements of our radio galaxy samples in Figure \ref{fig:autocfs}. Each sample's clustering is very well fit by a halo occupation distribution model ($\S$\ref{sec:hodmodel}), allowing the inference of the host halo properties of radio galaxies.

\subsection{Tomographic Cross-Correlations}

\begin{figure}
    \centering
    \includegraphics[width=0.45\textwidth]{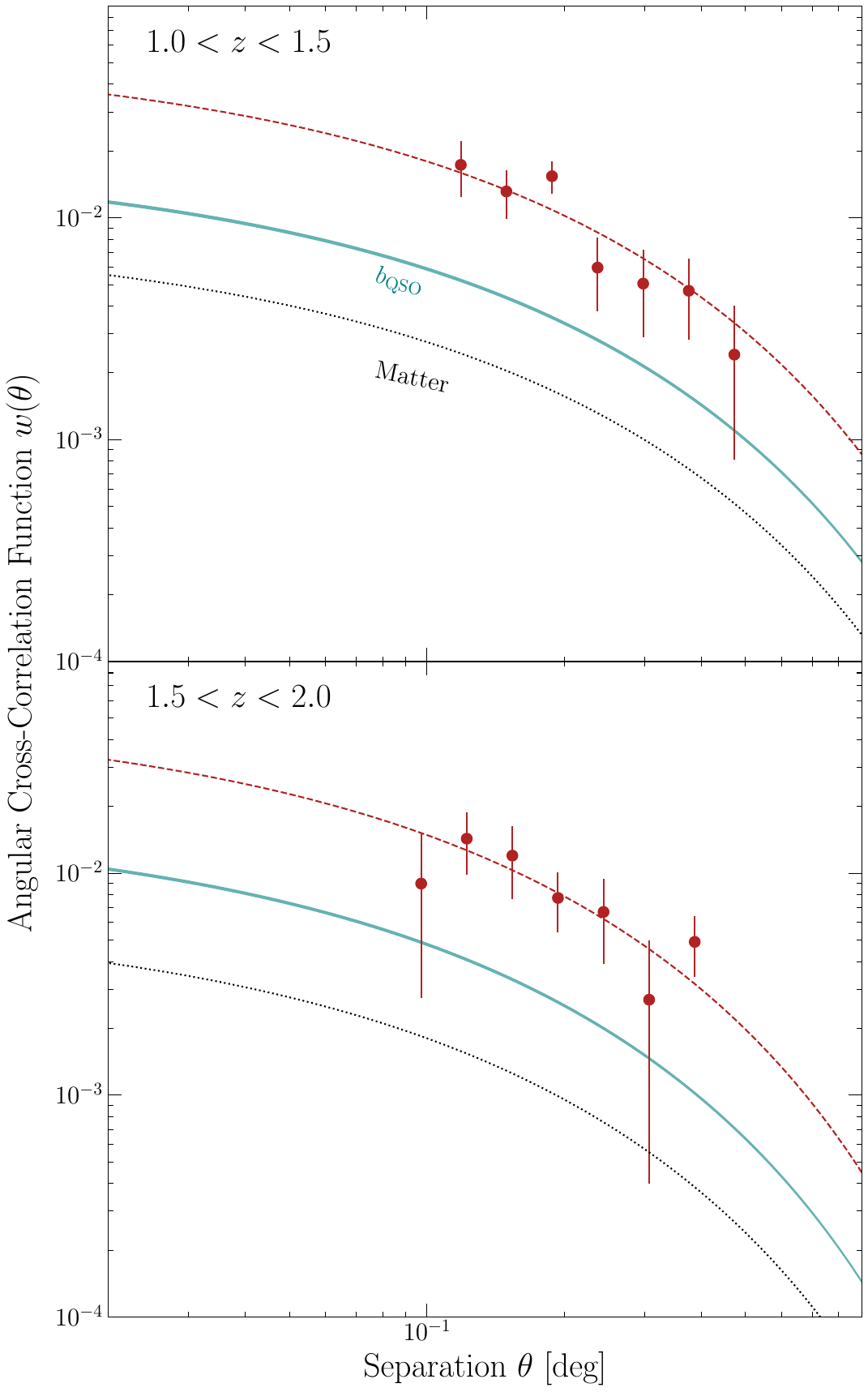}
    \caption{The tomographic angular cross correlation function measurements between HzRGs and eBOSS quasars in two redshift slices, displayed in the upper left corners. The linear matter model for the samples' redshift overlap is shown with a black dotted line, while the 1$\sigma$ confidence interval of the quasar tracer bias is shown with teal bands. HzRGs are best fit with higher bias parameters than quasars, indicating they occupy more massive halos from $1 < z < \maxz$. }
    \label{fig:xcfs}
\end{figure}

To probe the redshift evolution of radio galaxy clustering at $z > 1$, we also cross-correlate the radio galaxies with samples of quasars at known redshift, using a tomographic technique. We thus cross correlate HzRGs with eBOSS quasars in redshift slices. After cleaning each sample with the mask of the other ($\S$\ref{sec:mask}), we bin the eBOSS sample into two slices, $1 < z < 1.5$, and $1.5 < z < 2$. We then measure angular cross-correlations of each slice with the HzRGs, using the \citet{1983ApJ...267..465D} estimator:

\begin{equation}
    w(\theta) = \frac{D_1 D_2}{D_1 R_2} - 1, 
\end{equation}
where HzRGs and quasars make up the first and second samples, respectively. This estimator depends only on the quasar selection function. We use the provided quasar weights as specified by \citet{2021MNRAS.506.3439R}, while the radio galaxies are weighted to unity. We show the results of these cross-correlations in Figure \ref{fig:xcfs}. The measurements are well-fit by linearly biased models, and their bias factors are larger than the bias of quasars being correlated against, implying they occupy more massive halos. 

\subsection{Matter Tracer Autocorrelations}
\label{sec:tracers}

In order to interpret the above cross-correlations, we must estimate the bias parameters of the tracer populations being correlated against. Therefore, we measure the projected spatial clustering of the eBOSS quasars and BOSS galaxies in each redshift slice. We first measure the two-dimensional correlation function $\xi(r_p, \pi)$ using the \citep{1993ApJ...412...64L} estimator, before integrating over the line-of sight to the projected correlation function for mitigation of redshift space distortions \citep{1983ApJ...267..465D, 1987MNRAS.227....1K}, up to a maximum line-of-sight separation $\pi_{\mathrm{max}} = 40$ h$^{-1}$Mpc:

\begin{equation}
    w_p(r_p) = 2 \int^{\pi_{\mathrm{max}}}_{0} d \pi \  \xi(r_p, \pi).
    \label{projcf}
\end{equation}

\subsection{Radio Galaxy CMB Lensing}

Complementing the autocorrelation measurement of HzRGs, we acquire an independent constraint on the halos hosting HzRGs using a cross correlation of their positions with the \textit{Planck} CMB lensing convergence ($\kappa$) map ($\S$\ref{sec:cmbdata}). First, we produce a fractional overdensity ($\delta$) map of radio galaxies at the same NSIDE=1024 resolution as the lensing map:

\begin{equation}
    \delta_{RG} = \frac{\rho - \langle \rho \rangle}{\langle \rho \rangle},
\end{equation}
where $\rho$ represents the galaxy counts-in-cells. We then must estimate the cross-power spectrum of this overdensity map with the lensing map. Harmonic analysis of maps which do not cover the entire sphere induces mode-coupling, such as for surveys with a limited footprint. However, a fast and nearly-optimal estimator to measure an unbiased psuedo-spectrum is implemented in the {\tt NaMASTER} package \citep{2002ApJ...567....2H, 2019MNRAS.484.4127A}. We mask the RG density map ($\S$\ref{sec:mask}), and mask the lensing map with the provided mask after apodizing the edges with a 1$^\circ$ FWHM Gaussian \citep{2020JCAP...05..047K}. With {\tt NaMASTER}, we measure the cross-spectrum in 10 logarithmically-spaced bins from $200 < \ell < 2000$. For error estimation, we repeat the above process with 100 provided realizations of simulated lensing maps. 

\begin{figure}
    \centering
    \includegraphics[width=0.45\textwidth]{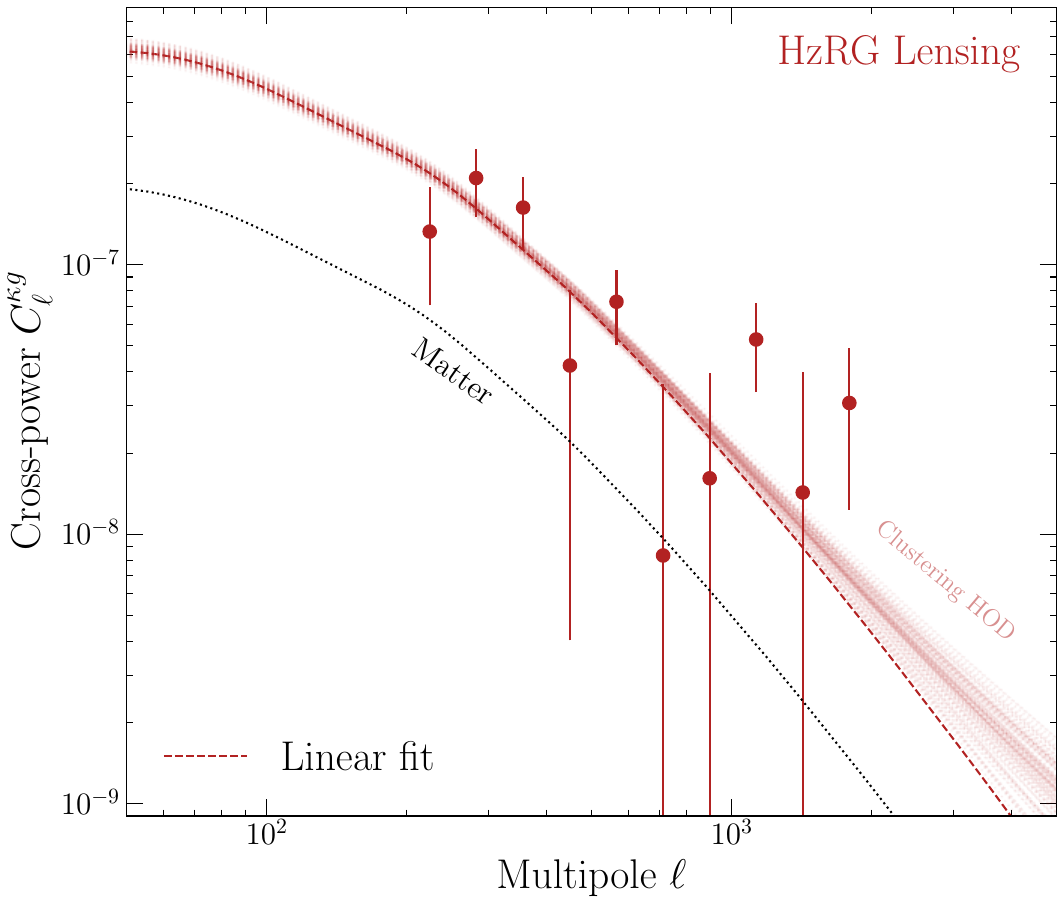}
    \caption{The cross correlation measurement of HzRG densities with \textit{Planck} CMB lensing convergence $\kappa$. The matter spectrum for the redshift distribution is shown with a black dashed line, while the best linearly-biased fit is shown with a red dashed line. We also show the predictions of the lensing spectrum drawn from the HOD posterior fit to the autocorrelation clustering measurement (Fig.\ \ref{fig:autocfs}), demonstrating excellent agreement between the lensing and clustering analyses, and providing a systematic test of the HzRG redshift distribution.}
    \label{fig:lensing}
\end{figure}

We show the resulting cross-spectrum of HzRG density with CMB lensing convergence in Figure \ref{fig:lensing}. This spectrum is well-fit by a linearly-biased halo model ($\S$\ref{sec:model}), allowing inference of the host halo properties in an independent manner from the clustering.

We perform a lensing analysis only of HzRGs for a number of reasons. Firstly, CMB lensing is most efficient at $z\sim 1-2$. Interpretation of the HzRG measurements also relies on constraint of the redshift distribution through photometric redshifts moreso than LzRGs and IzRGs, allowing a comparison of the HzRG lensing and clustering measurements to test for systematics in the redshift distribution \citep[e.g.,][]{2023ApJ...946...27P}. The LzRG sample is too sparse and at too low redshift to detect a significant signal. Finally, we observe that a cross-correlation between IzRG positions and CMB lensing is not well-fit by a linear model. This implies contamination in the CMB lensing map from the sources, which is consistent with the fact that this sample is significantly brighter in the radio band than HzRGs. 

\section{Modeling}
\label{sec:model}

To interpret the clustering measurements and constrain the environments in which radio galaxies release their energy, we model the measurements within a halo model framework \citep{2000MNRAS.318..203S, 2002PhR...372....1C}. Throughout this section, $\chi(z)$ is the comoving distance, $k$ is the comoving wavenumber, and $c$ is the speed of light. We assume that dark matter halos follow the ``NFW'' \citep{1997ApJ...490..493N} density profile, and the mass-concentration relation of \citet{2008MNRAS.390L..64D}, where the mass is defined within a radius containing 200 times the universal critical density. We adopt the halo mass function $dn/dM (M, z)$ of \citet{2008ApJ...688..709T}, and the halo mass-bias relation $b(M, z)$ of \citet{2010ApJ...724..878T}. Finally, we utilize {\tt CAMB} \citep{2000ApJ...538..473L} to compute all linear matter power spectra $P_{\mathrm{mm}}(k, z)$.

An angular correlation function between two matter tracers can be modeled using the \citet{1953ApJ...117..134L} approximation \citep{1980lssu.book.....P, 1991MNRAS.253P...1P}:

\begin{equation}
    w(\theta) = \int dz \frac{dz}{d\chi} \left(\frac{dN}{dz}_{i}  \frac{dN}{dz}_{j}\right) \int \frac{dk \ k}{2 \pi} P_{ij} \  J_{0}(k \theta \chi),
    \label{eq:ang_cf}
\end{equation}
where $J_0$ is the zeroth-order Bessel function of the first kind, and $dN/dz_{i,j}$ are the normalized redshift distributions of the two tracers, identical in the case of an autocorrelation. For a linearly-biased model, the power spectrum is then given by bias factors multiplied by the matter power spectrum:
\begin{equation}
    P_{ij} (k, z) = b_i b_j P_{\mathrm{mm}} (k, z).
\end{equation}

We also model the spatial autocorrelations of eBOSS quasars to interpret the tomographic cross-correlation measurements. This involves deducing the tracer bias by modeling the projected correlation functions ($\S$\ref{sec:tracers}). The spatial correlation function $\xi(r)$ is the Fourier transform of the galaxy/quasar auto spectrum $P_{gg}$. The projected correlation function is then an Abel transform of $\xi(r)$:

\begin{equation}
    w_p(r_p) = 2 \int_{r_p}^{\infty} dr \  \frac{r \xi(r)}{\sqrt{r^2 - r_p^2}}.
\end{equation}

\subsection{Modeling CMB Lensing}

We also model the cross-spectrum between radio galaxy overdensity and CMB lensing convergence. This is again given by a \citet{1953ApJ...117..134L} integral over the lensing and galaxy projection kernels:

\begin{equation}
    C^{\kappa g}_{\ell} =  \int dz \frac{d\chi}{dz} \frac{W^{\kappa} W^{g}}{\chi^2}  P_{\mathrm{mm}}\left(k = \frac{\ell + 1/2}{\chi}, \ z\right),
\end{equation}
where the CMB lensing kernel is given by \citep[e.g.,][]{2000ApJ...534..533C}:

\begin{equation}
    W^{\kappa}(z) = \frac{3}{2} \Omega_{m, 0} \left(\frac{H_0}{c}\right)^2 (1+z) \chi \frac{\chi_{\mathrm{CMB}} - \chi}{\chi_{\mathrm{CMB}}},
    \label{eq:lenskern}
\end{equation}
and the linearly-biased galaxy overdensity kernel is:
\begin{equation}
    W^{g}(z) = b \frac{dz}{d\chi} \frac{dN}{dz}.
    \label{eq:gal_kern}
\end{equation}

The effective redshift at which we report the bias measurement is weighted by the lensing kernel \citep{2017JCAP...08..009M, 2024JCAP...03..021K}:
\begin{equation}
    z_{\mathrm{eff}} = \int dz \frac{d\chi}{dz} \frac{dN}{dz} \frac{z}{\chi^2} W^{\kappa}.
\end{equation}

We do not consider magnification bias, as we measure the response of number counts of our only flux-limited sample (HzRGs) against limiting flux/magnitude ($s_\mu \equiv d \ \mathrm{log}_{10}N / dm = 0.412 \pm 0.001$), which is sufficiently near the value of $s_\mu=0.4$ at which magnification bias is zero.

\subsection{Halo Occupation Distribution Modeling}
\label{sec:hodmodel}

We also model the radio galaxy autocorrelation functions in a halo occupation distribution (HOD) framework \citep{2001ApJ...546...20S, 2002ApJ...575..587B} for a more complete interpretation of how these systems occupy their host halos than the linear bias model. This entails replacing the power $P_{ij}$ in Eq.\ \ref{eq:ang_cf} with an HOD power spectrum.

An HOD power spectrum is the sum of the ``1-halo'' power arising from pairs of galaxies in common halos, and the ``2-halo'' term from pairs between different halos. The HOD $\langle N (M)\rangle$ is the mean number of galaxies belonging to halos of mass $M$, decomposed into contributions from galaxies at the centers of halos, $\langle N_c(M) \rangle$, and secondary or `satellite'' galaxies belonging to the same halos $\langle N_s(M) \rangle$:

We adopt the HOD model of \citet{2007ApJ...667..760Z} and \citet{2011ApJ...736...59Z}, which implements the central occupation as a softened step function:

\begin{equation}
    \langle N_c\rangle = \frac{1}{2} \left[1 + \mathrm{erf}\left(\frac{\mathrm{log}_{10}(M / M_{\mathrm{min}})}{\sigma_{\mathrm{log}_{10}M}}\right)\right]
\end{equation}
where \textbf{erf} is the Gauss error function, $M_{\mathrm{min}}$ is the minimum halo mass required to host a radio galaxy, and $\sigma_{\mathrm{log}_{10}M}$ is the softening parameter. The satellite HOD is given then as:

\begin{equation}
\label{eqn:satellite}
    \langle N_s \rangle = \Theta(M-M_0) \left(\frac{M - M_0}{M_1}\right)^{\alpha}
\end{equation}
where $\Theta$ is the Heaviside step function, $M_0$ is the minimum mass to host a satellite quasar, and $M_1$ is the mass at which the term transitions to the power-law form. We enforce $M_0 = M_{\mathrm{min}}$ to reduce the degrees of freedom, such that the minimum mass to host a central or satellite is identical. Our HOD models thus contain four free parameters, $M_{\mathrm{min}}$, $\sigma_{\mathrm{log}_{10}M}$, $M_1$, and $\alpha$.

We compute the HOD power spectra given a set of parameters using the {\tt Core Cosmology Library} \citep{2019ApJS..242....2C} package. Finally, unphysical one-halo power at large scales is suppressed using the \citet{2021MNRAS.502.1401M} prescription, and a smoothing is applied to intermediate scales between the one and two-halo regimes \citep{2015MNRAS.454.1958M}. We refer the reader to \citet{2023ApJ...946...27P} for additional details in the HOD modeling procedure. The observed autocorrelation functions are fit using the {\tt emcee} \citep{2013PASP..125..306F} sampler, using broad Gaussian (1$\sigma$) priors of log$_{10}M_{\mathrm{min}} = 13 \pm 2$, $\sigma_{\mathrm{log} M} = 0.5 \pm 0.5$, log$_{10}M_1 = 14 \pm 2$, and $\alpha = 1 \pm 0.5$.
\section{Results}

We can now interpret all of the observed autocorrelation functions, cross-correlations with CMB lensing, and cross-correlation functions with spectroscopic galaxy tracers in the halo model framework to constrain the environments in which radio galaxies release their energy.

\subsection{Effective halo masses}

\begin{figure}
    \centering
    \includegraphics[width=0.45\textwidth]{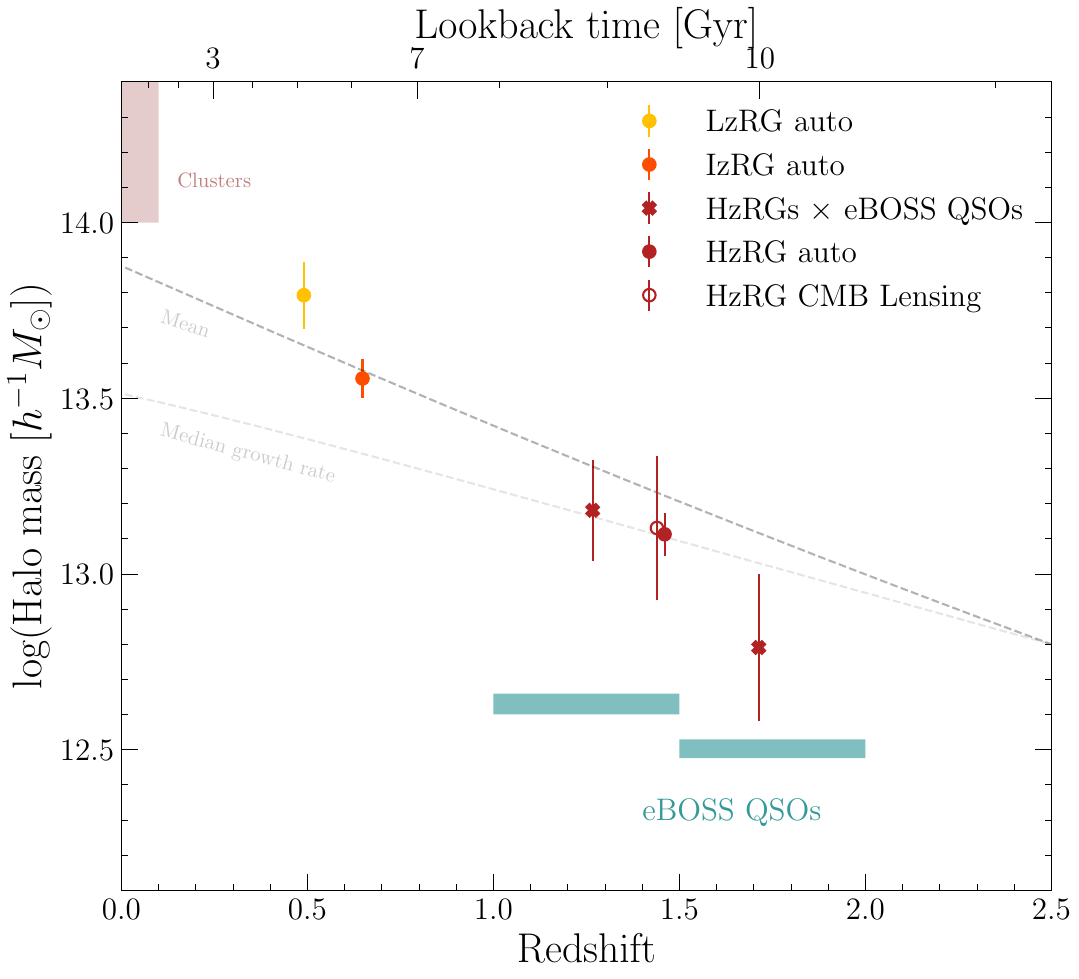}
    \caption{The effective halo masses of luminous low-frequency radio galaxies from $z < \maxz$. The red cross markers show the tomographic cross-correlations between eBOSS QSOs (halo masses shown with blue bands) and HzRGs. The solid circles in yellow, orange, and red show the linear fits to the autocorrelations of LzRGs, IzRGs and HzRGs, and the open circle shows the result from HzRG CMB lensing. Radio galaxies are strongly clustered across cosmic time, residing in group-scale halos. The median and mean halo growth rates from \citet{2010MNRAS.406.2267F} are shown with dashed and dotted lines respectively, for a halo whose progenitor mass was $10^{12.8} h^{-1} M_{\odot}$  at z=2.5. This shows that the effective host halo mass of radio galaxies evolves at a rate consistent with the mean growth rate of halos. The halos hosting luminous low-frequency radio galaxies at cosmic noon are expected to evolve on average into massive groups or clusters by the present.}
    \label{fig:halomass}
\end{figure}

We display the effective host halo mass results from all clustering measurements of luminous low-frequency radio galaxies at $z < \maxz$ in Fig.\ \ref{fig:halomass}. These systems occupy massive group halos of $ \sim 10^{13} - 10^{14} \ h^{-1} M_{\odot}$ for the past $\sim$ \lookbacktime Gyr, with an increasing effective halo mass with cosmic time consistent with the mean growth rate of halos \citep{2010MNRAS.406.2267F}. We conclude that luminous radio galaxies are typically hosted by group-scale halos across time, and that the occurrence of these systems is intimately connected to host halo mass \citep[e.g.;][]{2014MNRAS.445..280H}.  These observations support the paradigm where radiatively-inefficient accretion resulting in jets is driven by gas condensation from the hot halo, and may also reflect the fact that more massive halos contain more gas for jets to collide into and emit at low frequencies via shocks.

Though there are relatively few studies of radio galaxy environments at $z>1$, our results at lower redshifts appear consistent with previous studies at $z < 1$. Clustering studies of radio galaxies at $z < 1$ reveal they exist in group environments \citep{2009MNRAS.393..377M, 2009ApJ...696..891H}. This is consistent with studies of the galaxy overdensities surrounding $z < 1$ radio galaxies, which also show they are on average found in galaxy groups \citep{2004MNRAS.351...70B, 2015MNRAS.453.2682I, 2017MNRAS.469.4584C, 2019A&A...622A..10C}. 

At higher redshift ($z>1$), we find similar agreement with previous results including clustering \citep{2014MNRAS.440.2322L, 2017MNRAS.464.3271M, 2018MNRAS.474.4133H} and environmental \citep{2014MNRAS.445..280H} studies, conducted at GHz frequencies. These results support the paradigm where radio galaxy activity must be intimately connected with the large-scale environment \citep{2022A&ARv..30....6M}. 

\subsection{Halo occupation distributions}

\begin{figure}
    \centering
    \includegraphics[width=0.45\textwidth]{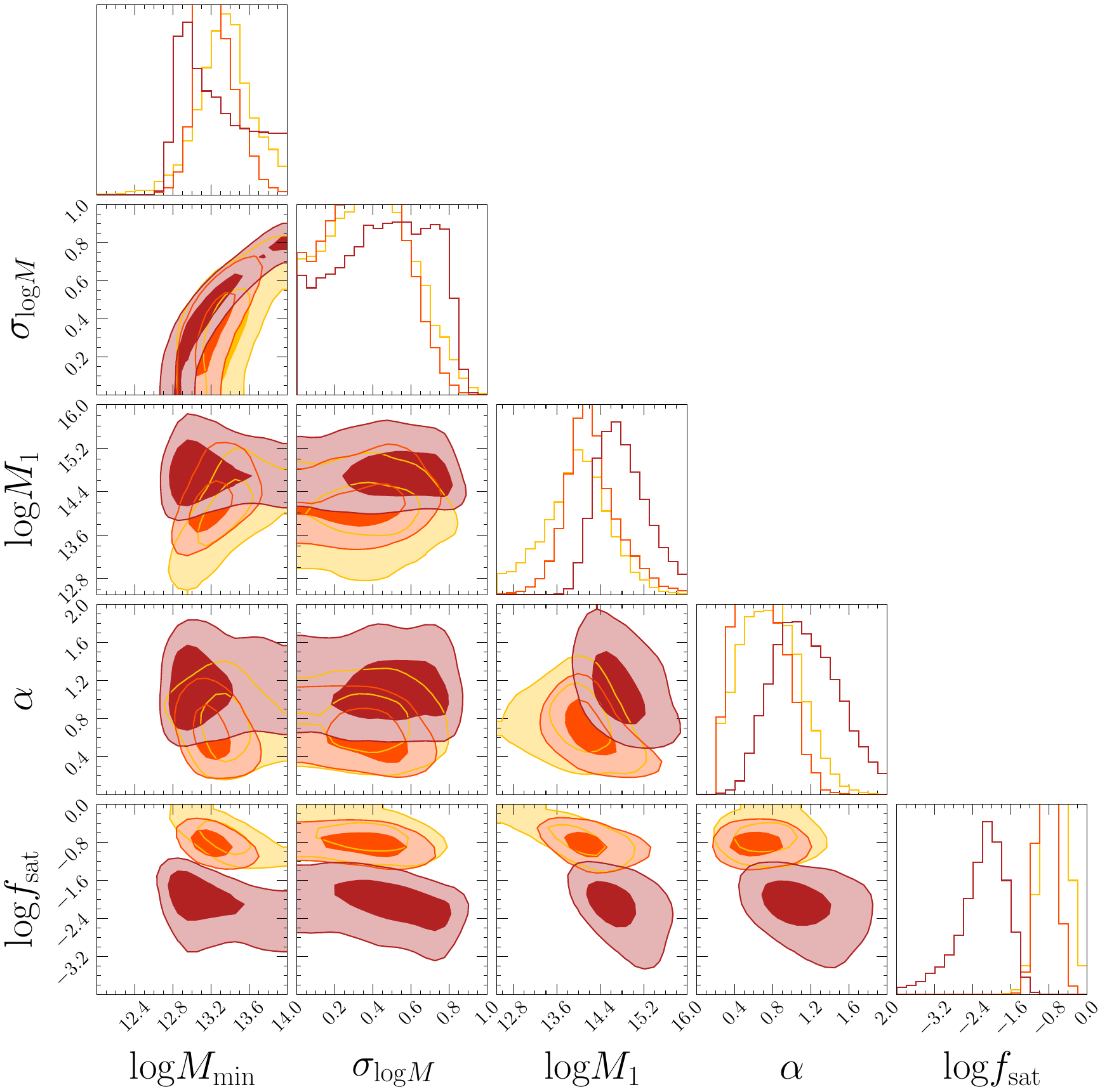}
    \caption{A corner plot of the HOD parameter posterior distributions (68 and 95$\%$ intervals) from fits to the autocorrelation functions of radio galaxies at $z \sim \lozeff$ (yellow), $\sim \izeff$ (orange) and $\sim \hizeff$ (red). Radio galaxies at $z < \maxz$ appear to have central occupations similar across time, hosted by halos more massive than $\gtrsim 10^{13} h^{-1} M_{\odot}$. However, the evolving halo mass function implies that radio galaxies appear to be satellites in their halos more often at $z<1$ than at $z \sim \hizeff$, at rates of $\sim 10\%$ and $\sim 1\%$, respectively.}
    \label{fig:hodparams}
\end{figure}

For a more sophisticated interpretation of the manner in which radio galaxies occupy their host halos than a single effective host mass, we display the posterior distributions of HOD model fits to the autocorrelation functions of LzRGs, IzRGs and HzRGs in Figure \ref{fig:hodparams}. We observe that luminous radio galaxies broadly exhibit similar HODs at $z\lesssim 2$, hosted by halos $M_h \gtrsim 10^{13} h^{-1} M_{\odot}$. Interestingly, the similar effective satellite halo mass of $M_1 \sim 10^{14.25} h^{-1} M_{\odot}$ across redshift may imply a difference in the satellite fraction at $z>1$ compared to $z <1$, because of the fact that these extremely massive halos are increasingly rare at higher redshift. Indeed, we estimate the satellite fraction of HzRGs to be $f_{\mathrm{sat}} = \hzrgfsat \%$, while the satellite fractions of IzRGs are $f_{\mathrm{sat}} = \izrgfsat \%$, and LzRGs are $f_{\mathrm{sat}} = \lzrgfsat \%$. We caution that this is a suggestive lead for future study, and tentatively highlight that this may be connected to the evolution of star-forming properties, FRI/FRII status, and accretion mode from high to low redshift.

\begin{figure}
    \centering
    \includegraphics[width=0.45\textwidth]{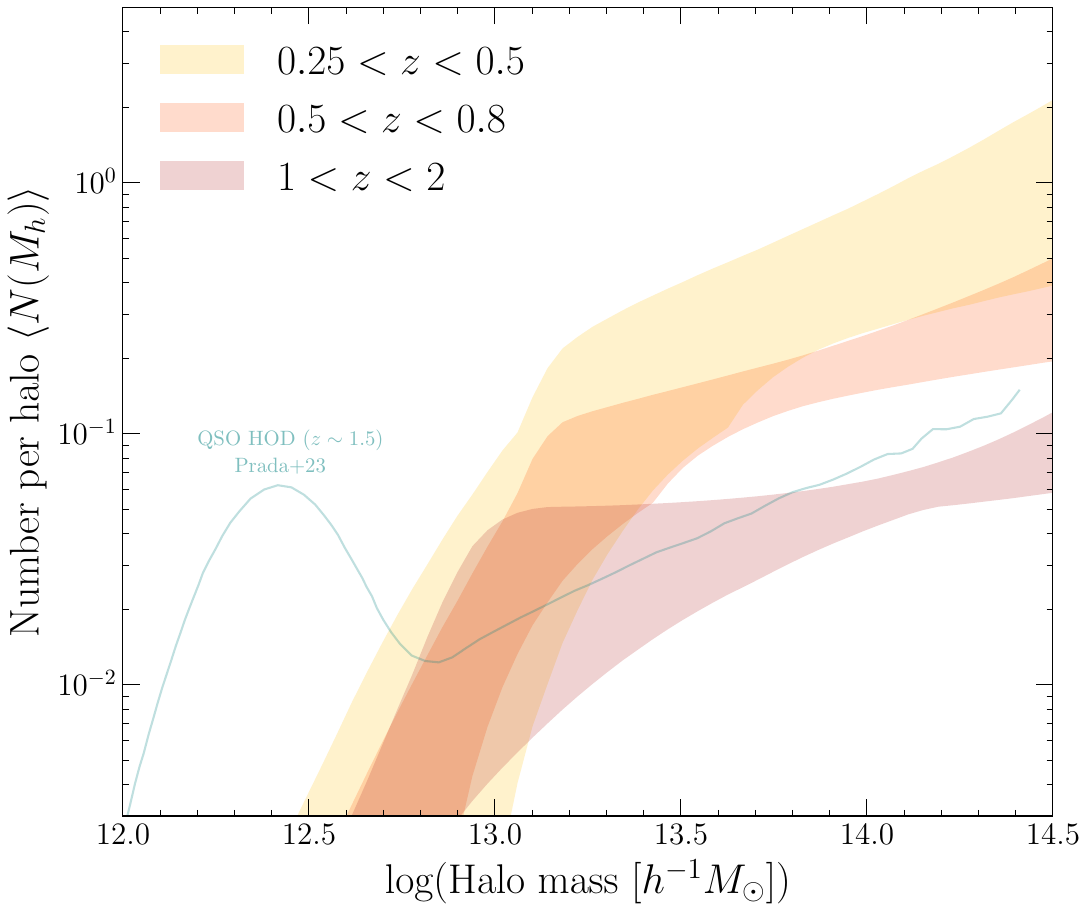}
    \caption{The halo occupation distributions (68$\%$ confidence intervals) inferred from fits to angular autocorrelations of LzRGs, IzRGs, and HzRGs. The populations are strongly clustered, hosted by halos $ \gtrsim 10^{13} h^{-1} M_{\odot}$ across cosmic time. We also show the HOD of Type-1 quasars at $z\sim 1.5$ derived from early DESI data \citep{2023arXiv230606315P} in blue, demonstrating that radio galaxies occupy significantly more massive halos than quasars at $z \sim 1.5$.}
    \label{fig:hods}
\end{figure}

We also display the confidence intervals of the HODs in Figure \ref{fig:hods}. This more clearly demonstrates the similarity of radio galaxies' HODs since $z \lesssim \maxz$. The amplitudes of the HODs are modulated by their respective duty cycles ($\S$\ref{sec:duty}). We also display the best fit HOD for optically-luminous quasars from early DESI data \citep{2023arXiv230606315P}. This shows that radio galaxies occupy systematically more massive halos than quasars, which has implications about which environments jets and quasar winds provide their feedback energy into. We will incorporate this information in weighing the relative contribution of jets and winds as a function of environment in $\S$\ref{sec:heating}.

\subsection{Duty Cycle}
\label{sec:duty}

A constraint on the properties of halos hosting radio galaxies can be combined with the radio AGN luminosity function (LF) to constrain the occupation fraction or duty cycle of these systems. A comparison of the number density of radio galaxies in the sample to the number density of halos massive enough to host them yields the fraction of said halos at a given time which host an observable radio AGN. Assuming that every sufficiently massive halo hosts a central galaxy with a SMBH, this occupation fraction is interpreted as a duty cycle $f_{\mathrm{duty}}$, or fraction of time that a SMBH hosted by these halos is observable as a radio galaxy in the survey \citep[e.g.,][]{2001ApJ...547...27H, 2001ApJ...547...12M}. 

To estimate the radio galaxy duty cycle, we refit all clustering measurements ($\S$\ref{sec:measurements}) with a minimum halo mass parameter $M_{\mathrm{min}}$ required to host a radio galaxy rather than the effective mass used before, using the relation between minimum mass and bias:
\begin{equation}
    b(M > M_{\mathrm{min}}) =  \frac{ \int_{M_{\mathrm{min}}}^{\infty} dM \frac{dn}{dM} b(M)}{ \int_{M_{\mathrm{min}}}^{\infty} dM \frac{dn}{dM}}.
\end{equation}

We then estimate the space density of halos more massive than $M_{\mathrm{min}}$ by integrating the \citet{2008ApJ...688..709T} halo mass function over mass and the considered redshift distribution. To estimate the space densities of the radio galaxies, we integrate the LoTSS-deep radio AGN LFs of \citet{2022MNRAS.513.3742K}. We consider the best-fit LFs including all radio-excess AGNs, allowing for luminosity and density evolution, as presented in \citet{2023MNRAS.523.5292K}. We account for the fact that our HzRG sample is limited by flux rather than luminosity by integrating above an evolving luminosity threshold with interpolation between LFs measured in different redshift epochs. At $z < 0.5$ where a best-fit LF is not provided, we interpolate between the LF measured at $0.5 < z < 1$, and the local ($z < 0.3$) LF from \citet{2019A&A...622A..17S} and \citet{2022MNRAS.513.3742K}. We adopt the local LF of \citet{2022MNRAS.513.3742K} at $L_{150 \ \mathrm{MHz}} < 10^{26}$ W/Hz, and the LF of \citet{2019A&A...622A..17S} above this limit, for reasons discussed in \citet{2022MNRAS.513.3742K}, though choosing one or the other does not change the interpretation of our results.

\begin{figure}
    \centering
    \includegraphics[width=0.45\textwidth]{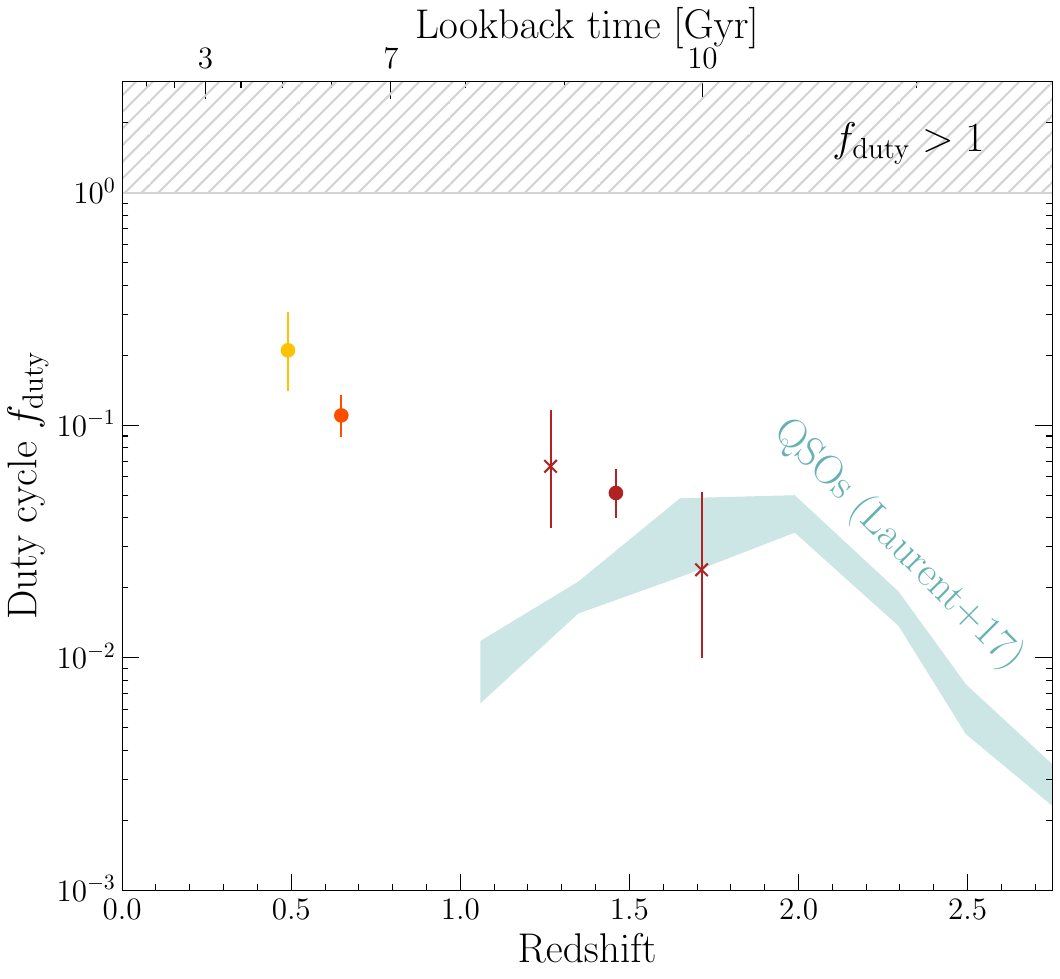}
    \caption{The duty cycle of luminous low-frequency radio galaxies over $z <\maxz$ implied by our clustering measurements and the luminosity functions of \citet{2022MNRAS.513.3742K} are shown in yellow, orange and red markers. These systems appear to be active for $\sim 10\%$ of the Hubble time, and rise from $\sim 5\%$ at $z \sim \hizeff$ to $\sim 20\%$ at $z \sim \lozeff$. We also show the 1$\sigma$ confidence interval for the duty cycle of eBOSS quasars determined by \citet{2017JCAP...07..017L} with a teal band. Interestingly, the luminous radio galaxy and quasar duty cycles appear to diverge after $z \sim 1.5$.}
    \label{fig:fduty}
\end{figure}

We display the resulting duty cycle estimates in Figure \ref{fig:fduty}. Luminous low-frequency radio galaxies exhibit duty cycles of order $\sim 10\%$ since $z < \maxz$, mildly increasing over the past \lookbacktime Gyr. The characteristic observable lifetime is thus $\sim$1 Gyr, consistent with the estimate of \citet{2017MNRAS.464.3271M}. We also display the implied duty cycle of eBOSS quasars from \citep{2017JCAP...07..017L}, which rises and falls before and after cosmic noon. Thus, $z\sim 2$ appears to mark the beginning of the end of the ``quasar epoch'', and the emergence of more frequent radio galaxy activity. 

\subsection{Heating Power per Halo}
\label{sec:heating}

Having constrained the host halo properties of luminous radio galaxies, we can now consider the energetics of the heating they supply to their environments. Here, we estimate the time-averaged kinetic heating power injected per host halo, and compare this to the same quantity for radiatively-driven quasar winds across cosmic time.

\citet{2023MNRAS.523.5292K} recently estimated the kinetic heating rate of radio galaxies as a function of 150 MHz luminosity - known as a specific heating rate $\Psi(L_{150}, z)$ -  by convolving the luminosity functions of \citet{2022MNRAS.513.3742K} with a conversion between luminosity and kinetic jet power. They use the \citet{2014ARA&A..52..589H} relation calibrated using X-ray cluster cavities to trace the work done by jets observed at 1.4 GHz \citep{2008ApJ...686..859B, 2010ApJ...720.1066C}. This heating rate thus specifies the spatially-averaged kinetic power released into a unit volume as a function of radio luminosity. \citet{2023MNRAS.523.5292K} finds that the total heating is dominated by relatively high-luminosity sources ($L_{150} \gtrsim 10^{25}$ W/Hz), similar to our sample. They find that the total heating is dominated by LERGs, and that the global kinetic heating dominates over radiative quasar-mode feedback since $z \lesssim 2$. However, as quasars and radio galaxies occupy halos differently, a more meaningful comparison is between the energy injected by jets versus winds within halos in a certain mass range. 

We integrate the \citet{2023MNRAS.523.5292K} specific heating rates (for all radio-excess AGNs) over luminosity using the flux or luminosity limit of our sample at each redshift, yielding a total kinetic heating power sourced from these jets per unit volume, a kinetic luminosity density. The volume and time-averaged kinetic power injected by jets per halo into halos within some mass interval is given by the division of the kinetic luminosity density with the number density of halos in the mass interval, then multiplied by the fraction of luminous radio galaxies which reside in halos within that mass interval, which depends on the HOD:

\begin{equation}
    P_{\mathrm{kin/halo}}^{\mathrm{Jets}} = \frac{\int_{L_{\mathrm{min}}}^{\infty} dL \Psi(L, z)}{{\int_{M_{\mathrm{min}}}^{M_{\mathrm{max}}}} dM_h \frac{dN_h}{dM_h}}
    \times \frac{{\int_{M_{\mathrm{min}}}^{M_{\mathrm{max}}}} dM_h \langle N(M_h) \rangle \frac{dN_h}{dM_h}}{{\int_{0}^{\infty}} dM_h \langle N(M_h) \rangle \frac{dN_h}{dM_h}}.
    \label{eq:jetpower}
\end{equation}




We can now compare this energy output to that of radiative-mode quasars' kinetic contribution in the form of winds. \citet{2007ApJ...654..731H} estimated the cumulative bolometric power density $U_{\mathrm{bol}}$ released by radiative-mode AGN over time by integrating their AGN luminosity functions under the \citet{1982MNRAS.200..115S} argument, assuming a $10\%$ radiative efficiency. Only a fraction of this energy is ultimately transferred to kinetic energy in gaseous winds, found to vary considerably between studies \citep[0.0001$\% - 1\%$;][]{2017A&A...601A.143F, 2019MNRAS.483L..22L, 2020A&A...633A.134L, 2021MNRAS.504.3890D, 2022MNRAS.511.2105K, 2024ApJ...969...56M}. Like in K23, here we adopt a characteristic range of $f_{\mathrm{wind}} = 0.1 - 0.5 \%$. To compute the average quasar-driven wind power density, we multiply this factor by the bolometric energy density released per unit of cosmological time within the same redshift intervals probed by our radio galaxy samples. Analogous with Eq. \ref{eq:jetpower}, the power released by quasars into halos in a mass interval is:

\begin{equation}
    \begin{split}
    P_{\mathrm{kin/halo}}^{\mathrm{winds}} = & \frac{f_{\mathrm{wind}} U_{\mathrm{bol}}(z_1, z_2)}{{\int_{M_{\mathrm{min}}}^{M_{\mathrm{max}}}} dM_h \frac{dN_h}{dM_h}}
    \times 
    \frac{\int_{45}^{\infty} dL \rho L }{\int_{0}^{\infty} dL \rho L} \\
    & \times \frac{{\int_{M_{\mathrm{min}}}^{M_{\mathrm{max}}}} dM_h \langle N(M_h) \rangle \frac{dN_h}{dM_h}}{{\int_{0}^{\infty}} dM_h \langle N(M_h) \rangle \frac{dN_h}{dM_h}}.
    \end{split}
    \label{eq:windpower}
\end{equation}

The second term modulates the power by the fraction of the total AGN luminosity density contributed by quasars relative to all AGN. For this we adopt the best-fit AGN luminosity function of \citet{2020MNRAS.495.3252S}.

The computation in Eq. \ref{eq:windpower} depends on the HOD of quasars, which we must assume for the purposes of this work. 
The effective masses of halos hosting Type-1 quasars appear remarkably consistent since $z\sim 6$ and across luminosity \citep{2004MNRAS.355.1010P, 2005MNRAS.356..415C, 2006MNRAS.371.1824P, 2007AJ....133.2222S, 2007ApJ...658...85M, 2008MNRAS.383..565D, 2009MNRAS.397.1862P, 2015MNRAS.453.2779E, 2017JCAP...07..017L, 2019ApJ...874...85G, 2018ApJ...859...20T, 2018PASJ...70S..33H, 2022ApJ...927...16P, 2023ApJ...954..210A, 2023arXiv230606315P, 2024MNRAS.530..947Y, 2024arXiv240307986E}, at a few times $10^{12} h^{-1} M_{\odot}$. However, the full quasar HOD \citep{2012ApJ...755...30R, 2019MNRAS.486..274E} and its possible evolution or luminosity dependence has remained difficult to study. Here, we adopt the quasar HOD of \citet{2023arXiv230606315P}, assuming that it is valid across cosmic time and at all luminosities above $L_{\mathrm{bol}} \gtrsim 10^{45}$ erg/s, though we acknowledge our following interpretation may depend on the validity of this assumption.

\begin{figure}
    \centering
    \includegraphics[width=0.45\textwidth]{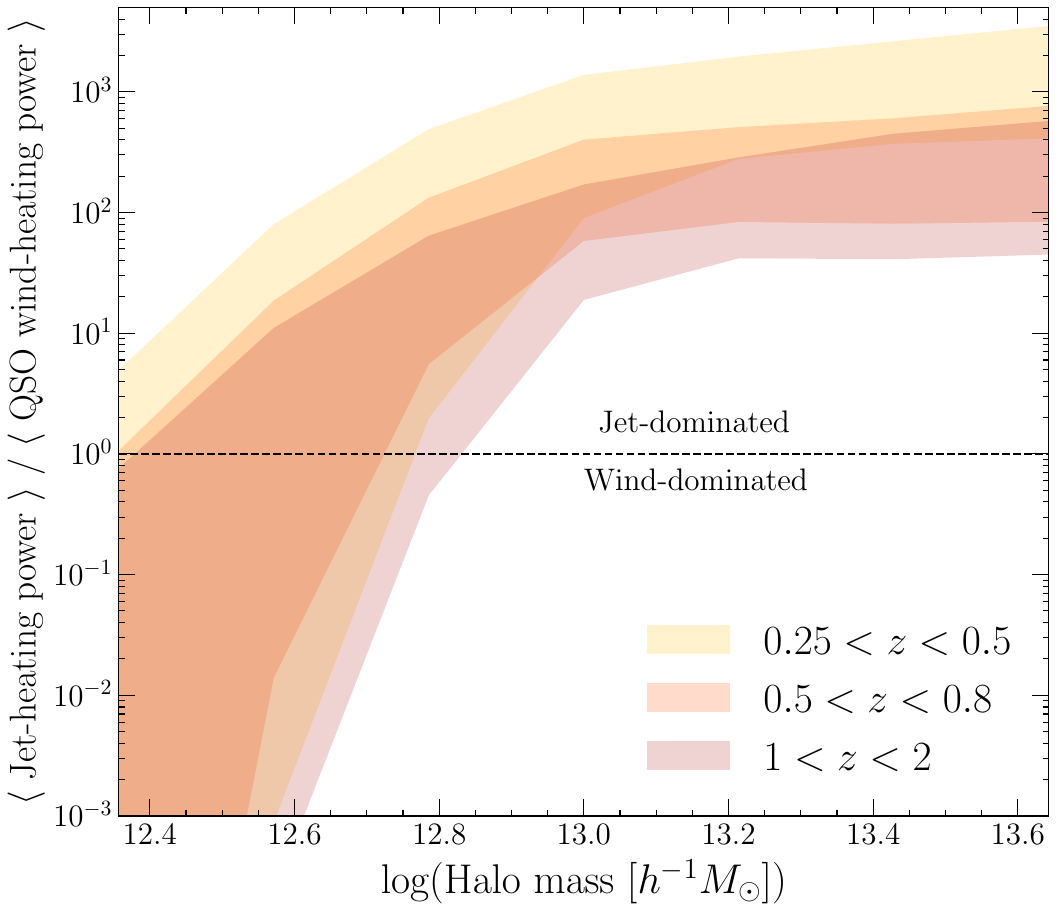}
    \caption{The ratio of the kinetic heating power injected into halos within a mass interval of 0.25 dex by luminous radio galaxies compared with quasar winds. This comparison results from combining the radio galaxy HODs presented in this work, the jet heating rates of K23, the bolometric quasar power densities of \citet{2007ApJ...654..731H}, the quasar HOD of \citet{2023arXiv230606315P} and assuming a wind-coupling range $f_{\mathrm{wind}} = 0.1 - 0.5 \%$ according to equations \ref{eq:jetpower} and \ref{eq:windpower}. Luminous jets appear to dominate the global heating in environments $\gtrsim 10^{\jetdommass} h^{-1} M_{\odot}$ at $z < \maxz$. We note that this computation assumes that the HODs of radio galaxies and quasars are luminosity independent, and the quasar HOD is also redshift-independent. In this picture, jet-mode feedback appears to dominate the global kinetic input into group and cluster-scale halos, while quasar winds may dominate at lower masses. }
    \label{fig:power}
\end{figure}

Dividing equations \ref{eq:jetpower} and \ref{eq:windpower} gives the ratio of kinetic power injected by luminous jets versus luminous quasar winds into halos within a given mass range. We estimate this ratio in ten mass intervals from $10^{12} - 10^{13.75} h^{-1} M_{\odot}$, and display the result in Figure \ref{fig:power}. This demonstrates that the feedback energy injected globally by jets dominates in halos $M_h \gtrsim 10^{13} h^{-1} M_{\odot}$, while quasar winds may dominate in lower-mass halos. This can be understood as a consequence of the result of \citet{2023MNRAS.523.5292K} that jet-heating dominates the total heating budget at $z \lesssim 2$ across all environments, unified with our clustering constraints showing that radio galaxies are typically found in halos $\gtrsim 10^{13} h^{-1} M_\odot$, while quasars are found in lower-mass halos (Fig.\ \ref{fig:hods}).

We raise the caveat that considering only Type-1 / unobscured quasars may provide an incomplete picture of radiatively-driven winds. Obscured quasars are canonically obscured by orientation with the line of sight, but a growing body of evidence may instead favor an evolutionary model of obscuration \citep{1988ApJ...328L..35S, 2012NewAR..56...93A, 2018ARA&A..56..625H}, where some quasars are obscured by galactic or circumnuclear gas during an evolutionary phase. Obscured / reddened quasars are sometimes observed to be more strongly clustered than unobscured systems \citep{2011ApJ...731..117H, 2014ApJ...789...44D, 2017MNRAS.469.4630D, 2023ApJ...946...27P}, occupy galaxies forming stars more vigorously \citep{2015ApJ...802...50C, 2022MNRAS.517.2577A}, and show enhanced radio emission \citep{2019MNRAS.488.3109K}, possibly tracing winds \citep{2021MNRAS.505.5283R}. The high clustering amplitude and duty cycle of $\sim 10\%$ \citep{2023ApJ...946...27P} are similar to the radio galaxies studied in this work, suggesting a possible stronger influence on the halos considered here compared to unobscured quasars. However, further constraints on obscured quasars' clustering evolution, their luminosity functions, and their radiation-wind coupling efficiency will be required to make meaningful constraints on their contribution to kinetic feedback.

We also note that by necessity, we have selected a heterogeneous radio galaxy population, containing a mix of LERGs and HERGs, FRI and FRII systems, which are hosted by a mix of star-forming and quiescent galaxies. 
\citet{2019A&A...622A..10C} finds that FRI systems at $z < 0.4$ occupy systematically richer environments than FRII analogues. Binning samples by FR class may become possible with upcoming sub-arcsecond LOFAR imaging \citep[][]{2022NatAs...6..350S}.

We note that the conversion rate between observed radio luminosity and mechanical power is subject to significant systematic uncertainties \citep{1999MNRAS.309.1017W, 2013MNRAS.430..174H, 2018MNRAS.475.2768H}. Therefore, further calibration of such a relation at 150 MHz and at high redshift, along with more precise clustering measurements are required to uncover which form of feedback dominates in massive halos at high redshift.

\section{Conclusions}

In this work, we select luminous ($L_{150 \ \mathrm{MHz}} \gtrsim$ \lcut W/Hz) low-frequency radio galaxies, bin them into redshift epochs using optical and infrared counterparts, and study the evolution of their clustering properties from $z < \maxz$. We summarize our results as follows:
\begin{itemize}
  \item These systems are strongly clustered at $z < \maxz$, hosted group-scale halos ($  10^{13} - 10^{14} h^{-1} M_{\odot}$). The effective halo mass thus increases with cosmic time at a rate consistent with the mean growth rate of halos. The minimum halo mass of $ \sim 10^{13} h^{-1} M_{\odot}$ derived from HOD modeling is consistent with a model where the triggering of low-frequency radio galaxy activity is strongly linked to the condensation of gas from a hot halo.
  
  \item The coevolution of the radio galaxy clustering and luminosity function implies that the duty cycle of these systems is of order $\sim \duty \%$ for the past \lookbacktime Gyr, mildly increasing during this epoch where the quasar duty cycle has been declining. The characteristic observable lifetime of these systems is thus order $\sim$1 Gyr. 
  
  
  \item The kinetic power injected into group halos ($M_h \gtrsim 10^{13} h^{-1} M_{\odot}$) appears to be dominated by luminous radio galaxies rather than radiatively-efficient quasar wind power since $z < \maxz$. Radio galaxies are thus expected to be more efficient drivers of feedback in groups for at least the last $\sim \lookbacktime$ Gyr. Improved constraints on the relation from radio luminosity to kinetic power, the fraction of quasar luminosity coupled to winds, the HODs of radio galaxies and quasars, and any possible evolution with luminosity to confirm this result.
\end{itemize}

The upcoming WEAVE-LOFAR spectroscopic follow up survey of LoTSS-detected sources  \citep{2016sf2a.conf..271S} will improve precision on the clustering properties of radio galaxies, mitigate against any systematics in the radio galaxy redshift distribution, and possibly support separately studying clustering for LERGs and HERGs. Furthermore, upcoming sub-arcsecond imaging \citep[][]{2022NatAs...6..350S} of the LoTSS-wide area using the International LOFAR baselines should allow purer selection of radio galaxies to fainter fluxes using brightness temperature criteria \citep{2022MNRAS.515.5758M}. Higher-resolution imaging should also allow for more precise cross-matching at alternate wavelengths, and enable clustering studies as a function of FRI/II status. Finally, future X-ray missions such as \textit{Athena} \citep{2013arXiv1306.2307N} will detect the gas in groups and clusters out to $z \sim 2$ heated in part by the systems studied in this work.

This work shows that jet-mode feedback appears to be the dominant source of heating in massive environments at $z < \maxz$. However, further constraints on the evolution of radio galaxy clustering with redshift and luminosity, and the conversions between radio luminosity and mechanical power are required to perform a census of jet-mode feedback, and uncover the complete picture of how supermassive black hole accretion shapes galaxy growth across time.


\acknowledgements

GCP acknowledges support from the Dartmouth Fellowship. RCH acknowledges support from NASA through ADAP grant number 80NSSC23K0485. LKM is grateful for support from the Medical Research Council [MR/T042842/1]. DMA thanks the Science Technology Facilities Council (STFC) for support from the Durham consolidated grant (ST/T000244/1).

\facilities{LOFAR, SDSS, eBOSS, \textit{WISE}, IRSA, \textit{Planck}}

\software{astropy: \citet{2013A&A...558A..33A}, CAMB: \citet{2000ApJ...538..473L}, CCL: \citet{2019ApJS..242....2C}, colossus: \citet[][]{2018ApJS..239...35D}, Corrfunc: \citet{2020MNRAS.491.3022S}, emcee: \citet{2013PASP..125..306F}, HEALPix: \citet[][]{ 2005ApJ...622..759G}}, MANGLE: \citet[][]{2004MNRAS.349..115H}, TOPCAT: \citet{2005ASPC..347...29T}.

\newpage


\bibliography{radiohalos}{}

\begin{thebibliography}{}
\expandafter\ifx\csname natexlab\endcsname\relax\def\natexlab#1{#1}\fi
\providecommand{\url}[1]{\href{#1}{#1}}
\providecommand{\dodoi}[1]{doi:~\href{http://doi.org/#1}{\nolinkurl{#1}}}
\providecommand{\doeprint}[1]{\href{http://ascl.net/#1}{\nolinkurl{http://ascl.net/#1}}}
\providecommand{\doarXiv}[1]{\href{https://arxiv.org/abs/#1}{\nolinkurl{https://arxiv.org/abs/#1}}}

\bibitem[{{Alexander} \& {Hickox}(2012)}]{2012NewAR..56...93A}
{Alexander}, D.~M., \& {Hickox}, R.~C. 2012, \nar, 56, 93,
  \dodoi{10.1016/j.newar.2011.11.003}

\bibitem[{{Alexander} {et~al.}(2010){Alexander}, {Swinbank}, {Smail},
  {McDermid}, \& {Nesvadba}}]{2010MNRAS.402.2211A}
{Alexander}, D.~M., {Swinbank}, A.~M., {Smail}, I., {McDermid}, R., \&
  {Nesvadba}, N.~P.~H. 2010, \mnras, 402, 2211,
  \dodoi{10.1111/j.1365-2966.2009.16046.x}

\bibitem[{{Alonso} {et~al.}(2021){Alonso}, {Bellini}, {Hale}, {Jarvis}, \&
  {Schwarz}}]{2021MNRAS.502..876A}
{Alonso}, D., {Bellini}, E., {Hale}, C., {Jarvis}, M.~J., \& {Schwarz}, D.~J.
  2021, \mnras, 502, 876, \dodoi{10.1093/mnras/stab046}

\bibitem[{{Alonso} {et~al.}(2019){Alonso}, {Sanchez}, {Slosar}, \& {LSST Dark
  Energy Science Collaboration}}]{2019MNRAS.484.4127A}
{Alonso}, D., {Sanchez}, J., {Slosar}, A., \& {LSST Dark Energy Science
  Collaboration}. 2019, \mnras, 484, 4127, \dodoi{10.1093/mnras/stz093}

\bibitem[{{Andonie} {et~al.}(2022){Andonie}, {Alexander}, {Rosario}, {Laloux},
  {Georgakakis}, {Morabito}, {Villforth}, {Avirett-Mackenzie}, {Calistro
  Rivera}, {Del Moro}, {Fotopoulou}, {Harrison}, {Lapi}, {Petley}, {Petter}, \&
  {Shankar}}]{2022MNRAS.517.2577A}
{Andonie}, C., {Alexander}, D.~M., {Rosario}, D., {et~al.} 2022, \mnras, 517,
  2577, \dodoi{10.1093/mnras/stac2800}

\bibitem[{{Arita} {et~al.}(2023){Arita}, {Kashikawa}, {Matsuoka}, {He}, {Ito},
  {Liang}, {Ishimoto}, {Yoshioka}, {Takeda}, {Iwasawa}, {Onoue}, {Toba}, \&
  {Imanishi}}]{2023ApJ...954..210A}
{Arita}, J., {Kashikawa}, N., {Matsuoka}, Y., {et~al.} 2023, \apj, 954, 210,
  \dodoi{10.3847/1538-4357/ace43a}

\bibitem[{{Assef} {et~al.}(2018){Assef}, {Stern}, {Noirot}, {Jun}, {Cutri}, \&
  {Eisenhardt}}]{2018ApJS..234...23A}
{Assef}, R.~J., {Stern}, D., {Noirot}, G., {et~al.} 2018, \apjs, 234, 23,
  \dodoi{10.3847/1538-4365/aaa00a}

\bibitem[{{Astropy Collaboration} {et~al.}(2013){Astropy Collaboration},
  {Robitaille}, {Tollerud}, {Greenfield}, {Droettboom}, {Bray}, {Aldcroft},
  {Davis}, {Ginsburg}, {Price-Whelan}, {Kerzendorf}, {Conley}, {Crighton},
  {Barbary}, {Muna}, {Ferguson}, {Grollier}, {Parikh}, {Nair}, {Unther},
  {Deil}, {Woillez}, {Conseil}, {Kramer}, {Turner}, {Singer}, {Fox}, {Weaver},
  {Zabalza}, {Edwards}, {Azalee Bostroem}, {Burke}, {Casey}, {Crawford},
  {Dencheva}, {Ely}, {Jenness}, {Labrie}, {Lim}, {Pierfederici}, {Pontzen},
  {Ptak}, {Refsdal}, {Servillat}, \& {Streicher}}]{2013A&A...558A..33A}
{Astropy Collaboration}, {Robitaille}, T.~P., {Tollerud}, E.~J., {et~al.} 2013,
  \aap, 558, A33, \dodoi{10.1051/0004-6361/201322068}

\bibitem[{{Balogh} {et~al.}(2001){Balogh}, {Pearce}, {Bower}, \&
  {Kay}}]{2001MNRAS.326.1228B}
{Balogh}, M.~L., {Pearce}, F.~R., {Bower}, R.~G., \& {Kay}, S.~T. 2001, \mnras,
  326, 1228, \dodoi{10.1111/j.1365-2966.2001.04667.x}

\bibitem[{{Becker} {et~al.}(1995){Becker}, {White}, \&
  {Helfand}}]{1995ApJ...450..559B}
{Becker}, R.~H., {White}, R.~L., \& {Helfand}, D.~J. 1995, \apj, 450, 559,
  \dodoi{10.1086/176166}

\bibitem[{{Benson}(2010)}]{2010PhR...495...33B}
{Benson}, A.~J. 2010, \physrep, 495, 33, \dodoi{10.1016/j.physrep.2010.06.001}

\bibitem[{{Benson} {et~al.}(2003){Benson}, {Bower}, {Frenk}, {Lacey}, {Baugh},
  \& {Cole}}]{2003ApJ...599...38B}
{Benson}, A.~J., {Bower}, R.~G., {Frenk}, C.~S., {et~al.} 2003, \apj, 599, 38,
  \dodoi{10.1086/379160}

\bibitem[{{Berlind} \& {Weinberg}(2002)}]{2002ApJ...575..587B}
{Berlind}, A.~A., \& {Weinberg}, D.~H. 2002, \apj, 575, 587,
  \dodoi{10.1086/341469}

\bibitem[{{Best}(2004)}]{2004MNRAS.351...70B}
{Best}, P.~N. 2004, \mnras, 351, 70, \dodoi{10.1111/j.1365-2966.2004.07752.x}

\bibitem[{{Best} {et~al.}(2006){Best}, {Kaiser}, {Heckman}, \&
  {Kauffmann}}]{2006MNRAS.368L..67B}
{Best}, P.~N., {Kaiser}, C.~R., {Heckman}, T.~M., \& {Kauffmann}, G. 2006,
  \mnras, 368, L67, \dodoi{10.1111/j.1745-3933.2006.00159.x}

\bibitem[{{Best} {et~al.}(2023){Best}, {Kondapally}, {Williams}, {Cochrane},
  {Duncan}, {Hale}, {Haskell}, {Ma{\l}ek}, {McCheyne}, {Smith}, {Wang},
  {Botteon}, {Bonato}, {Bondi}, {Rivera}, {Gao}, {G{\"u}rkan}, {Hardcastle},
  {Jarvis}, {Mingo}, {Miraghaei}, {Morabito}, {Nisbet}, {Prandoni},
  {R{\"o}ttgering}, {Sabater}, {Shimwell}, {Tasse}, \& {van
  Weeren}}]{2023MNRAS.523.1729B}
{Best}, P.~N., {Kondapally}, R., {Williams}, W.~L., {et~al.} 2023, \mnras, 523,
  1729, \dodoi{10.1093/mnras/stad1308}

\bibitem[{{Binney} \& {Tabor}(1995)}]{1995MNRAS.276..663B}
{Binney}, J., \& {Tabor}, G. 1995, \mnras, 276, 663,
  \dodoi{10.1093/mnras/276.2.663}

\bibitem[{{B{\^\i}rzan} {et~al.}(2008){B{\^\i}rzan}, {McNamara}, {Nulsen},
  {Carilli}, \& {Wise}}]{2008ApJ...686..859B}
{B{\^\i}rzan}, L., {McNamara}, B.~R., {Nulsen}, P.~E.~J., {Carilli}, C.~L., \&
  {Wise}, M.~W. 2008, \apj, 686, 859, \dodoi{10.1086/591416}

\bibitem[{{B{\^\i}rzan} {et~al.}(2004){B{\^\i}rzan}, {Rafferty}, {McNamara},
  {Wise}, \& {Nulsen}}]{2004ApJ...607..800B}
{B{\^\i}rzan}, L., {Rafferty}, D.~A., {McNamara}, B.~R., {Wise}, M.~W., \&
  {Nulsen}, P.~E.~J. 2004, \apj, 607, 800, \dodoi{10.1086/383519}

\bibitem[{{Boehringer} {et~al.}(1993){Boehringer}, {Voges}, {Fabian}, {Edge},
  \& {Neumann}}]{1993MNRAS.264L..25B}
{Boehringer}, H., {Voges}, W., {Fabian}, A.~C., {Edge}, A.~C., \& {Neumann},
  D.~M. 1993, \mnras, 264, L25, \dodoi{10.1093/mnras/264.1.L25}

\bibitem[{{Borgani} {et~al.}(2004){Borgani}, {Murante}, {Springel}, {Diaferio},
  {Dolag}, {Moscardini}, {Tormen}, {Tornatore}, \&
  {Tozzi}}]{2004MNRAS.348.1078B}
{Borgani}, S., {Murante}, G., {Springel}, V., {et~al.} 2004, \mnras, 348, 1078,
  \dodoi{10.1111/j.1365-2966.2004.07431.x}

\bibitem[{{Bower} {et~al.}(2012){Bower}, {Benson}, \&
  {Crain}}]{2012MNRAS.422.2816B}
{Bower}, R.~G., {Benson}, A.~J., \& {Crain}, R.~A. 2012, \mnras, 422, 2816,
  \dodoi{10.1111/j.1365-2966.2012.20516.x}

\bibitem[{{Bower} {et~al.}(2006){Bower}, {Benson}, {Malbon}, {Helly}, {Frenk},
  {Baugh}, {Cole}, \& {Lacey}}]{2006MNRAS.370..645B}
{Bower}, R.~G., {Benson}, A.~J., {Malbon}, R., {et~al.} 2006, \mnras, 370, 645,
  \dodoi{10.1111/j.1365-2966.2006.10519.x}

\bibitem[{{Brusa} {et~al.}(2015){Brusa}, {Bongiorno}, {Cresci}, {Perna},
  {Marconi}, {Mainieri}, {Maiolino}, {Salvato}, {Lusso}, {Santini}, {Comastri},
  {Fiore}, {Gilli}, {La Franca}, {Lanzuisi}, {Lutz}, {Merloni}, {Mignoli},
  {Onori}, {Piconcelli}, {Rosario}, {Vignali}, \&
  {Zamorani}}]{2015MNRAS.446.2394B}
{Brusa}, M., {Bongiorno}, A., {Cresci}, G., {et~al.} 2015, \mnras, 446, 2394,
  \dodoi{10.1093/mnras/stu2117}

\bibitem[{{Carilli} {et~al.}(1994){Carilli}, {Perley}, \&
  {Harris}}]{1994MNRAS.270..173C}
{Carilli}, C.~L., {Perley}, R.~A., \& {Harris}, D.~E. 1994, \mnras, 270, 173,
  \dodoi{10.1093/mnras/270.1.173}

\bibitem[{{Carron} {et~al.}(2022){Carron}, {Mirmelstein}, \&
  {Lewis}}]{2022JCAP...09..039C}
{Carron}, J., {Mirmelstein}, M., \& {Lewis}, A. 2022, \jcap, 2022, 039,
  \dodoi{10.1088/1475-7516/2022/09/039}

\bibitem[{{Cattaneo} {et~al.}(2009){Cattaneo}, {Faber}, {Binney}, {Dekel},
  {Kormendy}, {Mushotzky}, {Babul}, {Best}, {Br{\"u}ggen}, {Fabian}, {Frenk},
  {Khalatyan}, {Netzer}, {Mahdavi}, {Silk}, {Steinmetz}, \&
  {Wisotzki}}]{2009Natur.460..213C}
{Cattaneo}, A., {Faber}, S.~M., {Binney}, J., {et~al.} 2009, \nat, 460, 213,
  \dodoi{10.1038/nature08135}

\bibitem[{{Cavagnolo} {et~al.}(2008){Cavagnolo}, {Donahue}, {Voit}, \&
  {Sun}}]{2008ApJ...683L.107C}
{Cavagnolo}, K.~W., {Donahue}, M., {Voit}, G.~M., \& {Sun}, M. 2008, \apjl,
  683, L107, \dodoi{10.1086/591665}

\bibitem[{{Cavagnolo} {et~al.}(2010){Cavagnolo}, {McNamara}, {Nulsen},
  {Carilli}, {Jones}, \& {B{\^\i}rzan}}]{2010ApJ...720.1066C}
{Cavagnolo}, K.~W., {McNamara}, B.~R., {Nulsen}, P.~E.~J., {et~al.} 2010, \apj,
  720, 1066, \dodoi{10.1088/0004-637X/720/2/1066}

\bibitem[{{Chen} {et~al.}(2015){Chen}, {Hickox}, {Alberts}, {Harrison},
  {Alexander}, {Assef}, {Brodwin}, {Brown}, {Del Moro}, {Forman}, {Gorjian},
  {Goulding}, {Hainline}, {Jones}, {Kochanek}, {Murray}, {Pope}, {Rovilos}, \&
  {Stern}}]{2015ApJ...802...50C}
{Chen}, C.-T.~J., {Hickox}, R.~C., {Alberts}, S., {et~al.} 2015, \apj, 802, 50,
  \dodoi{10.1088/0004-637X/802/1/50}

\bibitem[{{Chiang} \& {M{\'e}nard}(2019)}]{2019ApJ...870..120C}
{Chiang}, Y.-K., \& {M{\'e}nard}, B. 2019, \apj, 870, 120,
  \dodoi{10.3847/1538-4357/aaf4f6}

\bibitem[{{Ching} {et~al.}(2017){Ching}, {Croom}, {Sadler}, {Robotham},
  {Brough}, {Baldry}, {Bland-Hawthorn}, {Colless}, {Driver}, {Holwerda},
  {Hopkins}, {Jarvis}, {Johnston}, {Kelvin}, {Liske}, {Loveday}, {Norberg},
  {Pracy}, {Steele}, {Thomas}, \& {Wang}}]{2017MNRAS.469.4584C}
{Ching}, J.~H.~Y., {Croom}, S.~M., {Sadler}, E.~M., {et~al.} 2017, \mnras, 469,
  4584, \dodoi{10.1093/mnras/stx1173}

\bibitem[{{Chisari} {et~al.}(2019){Chisari}, {Alonso}, {Krause}, {Leonard},
  {Bull}, {Neveu}, {Villarreal}, {Singh}, {McClintock}, {Ellison}, {Du},
  {Zuntz}, {Mead}, {Joudaki}, {Lorenz}, {Tr{\"o}ster}, {Sanchez}, {Lanusse},
  {Ishak}, {Hlozek}, {Blazek}, {Campagne}, {Almoubayyed}, {Eifler}, {Kirby},
  {Kirkby}, {Plaszczynski}, {Slosar}, {Vrastil}, {Wagoner}, \& {LSST Dark
  Energy Science Collaboration}}]{2019ApJS..242....2C}
{Chisari}, N.~E., {Alonso}, D., {Krause}, E., {et~al.} 2019, \apjs, 242, 2,
  \dodoi{10.3847/1538-4365/ab1658}

\bibitem[{{Churazov} {et~al.}(2000){Churazov}, {Forman}, {Jones}, \&
  {B{\"o}hringer}}]{2000A&A...356..788C}
{Churazov}, E., {Forman}, W., {Jones}, C., \& {B{\"o}hringer}, H. 2000, \aap,
  356, 788, \dodoi{10.48550/arXiv.astro-ph/0002375}

\bibitem[{{Churazov} {et~al.}(2005){Churazov}, {Sazonov}, {Sunyaev}, {Forman},
  {Jones}, \& {B{\"o}hringer}}]{2005MNRAS.363L..91C}
{Churazov}, E., {Sazonov}, S., {Sunyaev}, R., {et~al.} 2005, \mnras, 363, L91,
  \dodoi{10.1111/j.1745-3933.2005.00093.x}

\bibitem[{{Cooray} \& {Hu}(2000)}]{2000ApJ...534..533C}
{Cooray}, A., \& {Hu}, W. 2000, \apj, 534, 533, \dodoi{10.1086/308799}

\bibitem[{{Cooray} \& {Sheth}(2002)}]{2002PhR...372....1C}
{Cooray}, A., \& {Sheth}, R. 2002, \physrep, 372, 1,
  \dodoi{10.1016/S0370-1573(02)00276-4}

\bibitem[{{Croom} {et~al.}(2005){Croom}, {Boyle}, {Shanks}, {Smith}, {Miller},
  {Outram}, {Loaring}, {Hoyle}, \& {da {\^A}ngela}}]{2005MNRAS.356..415C}
{Croom}, S.~M., {Boyle}, B.~J., {Shanks}, T., {et~al.} 2005, \mnras, 356, 415,
  \dodoi{10.1111/j.1365-2966.2004.08379.x}

\bibitem[{{Croston} {et~al.}(2019){Croston}, {Hardcastle}, {Mingo}, {Best},
  {Sabater}, {Shimwell}, {Williams}, {Duncan}, {R{\"o}ttgering}, {Brienza},
  {G{\"u}rkan}, {Ineson}, {Miley}, {Morabito}, {O'Sullivan}, \&
  {Prandoni}}]{2019A&A...622A..10C}
{Croston}, J.~H., {Hardcastle}, M.~J., {Mingo}, B., {et~al.} 2019, \aap, 622,
  A10, \dodoi{10.1051/0004-6361/201834019}

\bibitem[{{Croton} {et~al.}(2006){Croton}, {Springel}, {White}, {De Lucia},
  {Frenk}, {Gao}, {Jenkins}, {Kauffmann}, {Navarro}, \&
  {Yoshida}}]{2006MNRAS.365...11C}
{Croton}, D.~J., {Springel}, V., {White}, S. D.~M., {et~al.} 2006, \mnras, 365,
  11, \dodoi{10.1111/j.1365-2966.2005.09675.x}

\bibitem[{{Croton} {et~al.}(2016){Croton}, {Stevens}, {Tonini}, {Garel},
  {Bernyk}, {Bibiano}, {Hodkinson}, {Mutch}, {Poole}, \&
  {Shattow}}]{2016ApJS..222...22C}
{Croton}, D.~J., {Stevens}, A. R.~H., {Tonini}, C., {et~al.} 2016, \apjs, 222,
  22, \dodoi{10.3847/0067-0049/222/2/22}

\bibitem[{{da {\^A}ngela} {et~al.}(2008){da {\^A}ngela}, {Shanks}, {Croom},
  {Weilbacher}, {Brunner}, {Couch}, {Miller}, {Myers}, {Nichol}, {Pimbblet},
  {de Propris}, {Richards}, {Ross}, {Schneider}, \&
  {Wake}}]{2008MNRAS.383..565D}
{da {\^A}ngela}, J., {Shanks}, T., {Croom}, S.~M., {et~al.} 2008, \mnras, 383,
  565, \dodoi{10.1111/j.1365-2966.2007.12552.x}

\bibitem[{{Dall'Agnol de Oliveira} {et~al.}(2021){Dall'Agnol de Oliveira},
  {Storchi-Bergmann}, {Kraemer}, {Villar Mart{\'\i}n}, {Schnorr-M{\"u}ller},
  {Schmitt}, {Ruschel-Dutra}, {Crenshaw}, \& {Fischer}}]{2021MNRAS.504.3890D}
{Dall'Agnol de Oliveira}, B., {Storchi-Bergmann}, T., {Kraemer}, S.~B.,
  {et~al.} 2021, \mnras, 504, 3890, \dodoi{10.1093/mnras/stab1067}

\bibitem[{{Dav{\'e}} {et~al.}(2019){Dav{\'e}}, {Angl{\'e}s-Alc{\'a}zar},
  {Narayanan}, {Li}, {Rafieferantsoa}, \& {Appleby}}]{2019MNRAS.486.2827D}
{Dav{\'e}}, R., {Angl{\'e}s-Alc{\'a}zar}, D., {Narayanan}, D., {et~al.} 2019,
  \mnras, 486, 2827, \dodoi{10.1093/mnras/stz937}

\bibitem[{{Davis} \& {Peebles}(1983)}]{1983ApJ...267..465D}
{Davis}, M., \& {Peebles}, P.~J.~E. 1983, \apj, 267, 465,
  \dodoi{10.1086/160884}

\bibitem[{{Dawson} {et~al.}(2016){Dawson}, {Kneib}, {Percival}, {Alam},
  {Albareti}, {Anderson}, {Armengaud}, {Aubourg}, {Bailey}, {Bautista},
  {Berlind}, {Bershady}, {Beutler}, {Bizyaev}, {Blanton}, {Blomqvist},
  {Bolton}, {Bovy}, {Brandt}, {Brinkmann}, {Brownstein}, {Burtin}, {Busca},
  {Cai}, {Chuang}, {Clerc}, {Comparat}, {Cope}, {Croft}, {Cruz-Gonzalez}, {da
  Costa}, {Cousinou}, {Darling}, {de la Macorra}, {de la Torre}, {Delubac}, {du
  Mas des Bourboux}, {Dwelly}, {Ealet}, {Eisenstein}, {Eracleous}, {Escoffier},
  {Fan}, {Finoguenov}, {Font-Ribera}, {Frinchaboy}, {Gaulme}, {Georgakakis},
  {Green}, {Guo}, {Guy}, {Ho}, {Holder}, {Huehnerhoff}, {Hutchinson}, {Jing},
  {Jullo}, {Kamble}, {Kinemuchi}, {Kirkby}, {Kitaura}, {Klaene}, {Laher},
  {Lang}, {Laurent}, {Le Goff}, {Li}, {Liang}, {Lima}, {Lin}, {Lin}, {Lin},
  {Long}, {Lundgren}, {MacDonald}, {Geimba Maia}, {Malanushenko},
  {Malanushenko}, {Mariappan}, {McBride}, {McGreer}, {M{\'e}nard}, {Merloni},
  {Meza}, {Montero-Dorta}, {Muna}, {Myers}, {Nandra}, {Naugle}, {Newman},
  {Noterdaeme}, {Nugent}, {Ogando}, {Olmstead}, {Oravetz}, {Oravetz},
  {Padmanabhan}, {Palanque-Delabrouille}, {Pan}, {Parejko}, {P{\^a}ris},
  {Peacock}, {Petitjean}, {Pieri}, {Pisani}, {Prada}, {Prakash}, {Raichoor},
  {Reid}, {Rich}, {Ridl}, {Rodriguez-Torres}, {Carnero Rosell}, {Ross},
  {Rossi}, {Ruan}, {Salvato}, {Sayres}, {Schneider}, {Schlegel}, {Seljak},
  {Seo}, {Sesar}, {Shandera}, {Shu}, {Slosar}, {Sobreira}, {Streblyanska},
  {Suzuki}, {Taylor}, {Tao}, {Tinker}, {Tojeiro}, {Vargas-Maga{\~n}a}, {Wang},
  {Weaver}, {Weinberg}, {White}, {Wood-Vasey}, {Yeche}, {Zhai}, {Zhao}, {Zhao},
  {Zheng}, {Ben Zhu}, \& {Zou}}]{2016AJ....151...44D}
{Dawson}, K.~S., {Kneib}, J.-P., {Percival}, W.~J., {et~al.} 2016, \aj, 151,
  44, \dodoi{10.3847/0004-6256/151/2/44}

\bibitem[{{DESI Collaboration} {et~al.}(2024){DESI Collaboration}, {Adame},
  {Aguilar}, {Ahlen}, {Alam}, {Aldering}, {Alexander}, {Alfarsy}, {Prieto},
  {Alvarez}, {Alves}, {Anand}, {Andrade-Oliveira}, {Armengaud}, {Asorey},
  {Avila}, {Aviles}, {Bailey}, {Balaguera-Antol{\'\i}nez}, {Ballester},
  {Baltay}, {Bault}, {Bautista}, {Behera}, {Beltran}, {BenZvi}, {Beraldo e
  Silva}, {Bermejo-Climent}, {Berti}, {Besuner}, {Beutler}, {Bianchi}, {Blake},
  {Blum}, {Bolton}, {Brieden}, {Brodzeller}, {Brooks}, {Brown}, {Buckley-Geer},
  {Burtin}, {Cabayol-Garcia}, {Cai}, {Canning}, {Cardiel-Sas}, {Carnero
  Rosell}, {Castander}, {Cervantes-Cota}, {Chabanier}, {Chaussidon},
  {Chaves-Montero}, {Chen}, {Chen}, {Chuang}, {Claybaugh}, {Cole}, {Cooper},
  {Cuceu}, {Davis}, {Dawson}, {de Belsunce}, {de la Cruz}, {de la Macorra},
  {Della Costa}, {de Mattia}, {Demina}, {Demirbozan}, {DeRose}, {Dey}, {Dey},
  {Dhungana}, {Ding}, {Ding}, {Doel}, {Doshi}, {Douglass}, {Edge},
  {Eftekharzadeh}, {Eisenstein}, {Elliott}, {Ereza}, {Escoffier}, {Fagrelius},
  {Fan}, {Fanning}, {Fawcett}, {Ferraro}, {Flaugher}, {Font-Ribera},
  {Forero-Romero}, {Forero-S{\'a}nchez}, {Frenk}, {G{\"a}nsicke},
  {Garc{\'\i}a}, {Garc{\'\i}a-Bellido}, {Garcia-Quintero}, {Garrison},
  {Gil-Mar{\'\i}n}, {Golden-Marx}, {Gontcho}, {Gonzalez-Morales},
  {Gonzalez-Perez}, {Gordon}, {Graur}, {Green}, {Gruen}, {Guy}, {Hadzhiyska},
  {Hahn}, {Han}, {Hanif}, {Herrera-Alcantar}, {Honscheid}, {Hou}, {Howlett},
  {Huterer}, {Ir{\v{s}}i{\v{c}}}, {Ishak}, {Jacques}, {Jana}, {Jiang},
  {Jimenez}, {Jing}, {Joudaki}, {Joyce}, {Jullo}, {Juneau},
  {Kara{\c{c}}ayl{\i}}, {Karim}, {Kehoe}, {Kent}, {Khederlarian}, {Kim},
  {Kirkby}, {Kisner}, {Kitaura}, {Kizhuprakkat}, {Kneib}, {Koposov},
  {Kov{\'a}cs}, {Kremin}, {Krolewski}, {L'Huillier}, {Lahav}, {Lambert},
  {Lamman}, {Lan}, {Landriau}, {Lang}, {Lange}, {Lasker}, {Leauthaud}, {Le
  Guillou}, {Levi}, {Li}, {Linder}, {Lyons}, {Magneville}, {Manera}, {Manser},
  {Margala}, {Martini}, {McDonald}, {Medina}, {Medina-Varela}, {Meisner},
  {Mena-Fern{\'a}ndez}, {Meneses-Rizo}, {Mezcua}, {Miquel}, {Montero-Camacho},
  {Moon}, {Moore}, {Moustakas}, {Mueller}, {Mundet}, {Mu{\~n}oz-Guti{\'e}rrez},
  {Myers}, {Nadathur}, {Napolitano}, {Neveux}, {Newman}, {Nie}, {Nikutta},
  {Niz}, {Norberg}, {Noriega}, {Paillas}, {Palanque-Delabrouille}, {Palmese},
  {Pan}, {Parkinson}, {Penmetsa}, {Percival}, {P{\'e}rez-Fern{\'a}ndez},
  {P{\'e}rez-R{\`a}fols}, {Pieri}, {Poppett}, {Porredon}, {Pothier}, {Prada},
  {Pucha}, {Raichoor}, {Ram{\'\i}rez-P{\'e}rez}, {Ramirez-Solano},
  {Rashkovetskyi}, {Ravoux}, {Rocher}, {Rockosi}, {Ross}, {Rossi}, {Ruggeri},
  {Ruhlmann-Kleider}, {Sabiu}, {Said}, {Saintonge}, {Samushia}, {Sanchez},
  {Saulder}, {Schaan}, {Schlafly}, {Schlegel}, {Scholte}, {Schubnell}, {Seo},
  {Shafieloo}, {Sharples}, {Sheu}, {Silber}, {Sinigaglia}, {Siudek}, {Slepian},
  {Smith}, {Soumagnac}, {Sprayberry}, {Stephey}, {Su{\'a}rez-P{\'e}rez}, {Sun},
  {Tan}, {Tarl{\'e}}, {Tojeiro}, {Ure{\~n}a-L{\'o}pez}, {Vaisakh}, {Valcin},
  {Valdes}, {Valluri}, {Vargas-Maga{\~n}a}, {Variu}, {Verde}, {Walther},
  {Wang}, {Wang}, {Weaver}, {Weaverdyck}, {Wechsler}, {White}, {Xie}, {Yang},
  {Y{\`e}che}, {Yu}, {Yuan}, {Zhang}, {Zhang}, {Zhao}, {Zheng}, {Zhou}, {Zhou},
  {Zou}, {Zou}, \& {Zu}}]{2024AJ....168...58D}
{DESI Collaboration}, {Adame}, A.~G., {Aguilar}, J., {et~al.} 2024, \aj, 168,
  58, \dodoi{10.3847/1538-3881/ad3217}

\bibitem[{{Dey} {et~al.}(2019){Dey}, {Schlegel}, {Lang}, {Blum}, {Burleigh},
  {Fan}, {Findlay}, {Finkbeiner}, {Herrera}, {Juneau}, {Landriau}, {Levi},
  {McGreer}, {Meisner}, {Myers}, {Moustakas}, {Nugent}, {Patej}, {Schlafly},
  {Walker}, {Valdes}, {Weaver}, {Y{\`e}che}, {Zou}, {Zhou}, {Abareshi},
  {Abbott}, {Abolfathi}, {Aguilera}, {Alam}, {Allen}, {Alvarez}, {Annis},
  {Ansarinejad}, {Aubert}, {Beechert}, {Bell}, {BenZvi}, {Beutler}, {Bielby},
  {Bolton}, {Brice{\~n}o}, {Buckley-Geer}, {Butler}, {Calamida}, {Carlberg},
  {Carter}, {Casas}, {Castander}, {Choi}, {Comparat}, {Cukanovaite}, {Delubac},
  {DeVries}, {Dey}, {Dhungana}, {Dickinson}, {Ding}, {Donaldson}, {Duan},
  {Duckworth}, {Eftekharzadeh}, {Eisenstein}, {Etourneau}, {Fagrelius},
  {Farihi}, {Fitzpatrick}, {Font-Ribera}, {Fulmer}, {G{\"a}nsicke},
  {Gaztanaga}, {George}, {Gerdes}, {Gontcho}, {Gorgoni}, {Green}, {Guy},
  {Harmer}, {Hernandez}, {Honscheid}, {Huang}, {James}, {Jannuzi}, {Jiang},
  {Joyce}, {Karcher}, {Karkar}, {Kehoe}, {Kneib}, {Kueter-Young}, {Lan},
  {Lauer}, {Le Guillou}, {Le Van Suu}, {Lee}, {Lesser}, {Perreault Levasseur},
  {Li}, {Mann}, {Marshall}, {Mart{\'\i}nez-V{\'a}zquez}, {Martini}, {du Mas des
  Bourboux}, {McManus}, {Meier}, {M{\'e}nard}, {Metcalfe},
  {Mu{\~n}oz-Guti{\'e}rrez}, {Najita}, {Napier}, {Narayan}, {Newman}, {Nie},
  {Nord}, {Norman}, {Olsen}, {Paat}, {Palanque-Delabrouille}, {Peng},
  {Poppett}, {Poremba}, {Prakash}, {Rabinowitz}, {Raichoor}, {Rezaie},
  {Robertson}, {Roe}, {Ross}, {Ross}, {Rudnick}, {Safonova}, {Saha},
  {S{\'a}nchez}, {Savary}, {Schweiker}, {Scott}, {Seo}, {Shan}, {Silva},
  {Slepian}, {Soto}, {Sprayberry}, {Staten}, {Stillman}, {Stupak}, {Summers},
  {Sien Tie}, {Tirado}, {Vargas-Maga{\~n}a}, {Vivas}, {Wechsler}, {Williams},
  {Yang}, {Yang}, {Yapici}, {Zaritsky}, {Zenteno}, {Zhang}, {Zhang}, {Zhou}, \&
  {Zhou}}]{2019AJ....157..168D}
{Dey}, A., {Schlegel}, D.~J., {Lang}, D., {et~al.} 2019, \aj, 157, 168,
  \dodoi{10.3847/1538-3881/ab089d}

\bibitem[{{Di Matteo} {et~al.}(2005){Di Matteo}, {Springel}, \&
  {Hernquist}}]{2005Natur.433..604D}
{Di Matteo}, T., {Springel}, V., \& {Hernquist}, L. 2005, \nat, 433, 604,
  \dodoi{10.1038/nature03335}

\bibitem[{{Diemer}(2018)}]{2018ApJS..239...35D}
{Diemer}, B. 2018, \apjs, 239, 35, \dodoi{10.3847/1538-4365/aaee8c}

\bibitem[{{DiPompeo} {et~al.}(2017){DiPompeo}, {Hickox}, {Eftekharzadeh}, \&
  {Myers}}]{2017MNRAS.469.4630D}
{DiPompeo}, M.~A., {Hickox}, R.~C., {Eftekharzadeh}, S., \& {Myers}, A.~D.
  2017, \mnras, 469, 4630, \dodoi{10.1093/mnras/stx1215}

\bibitem[{{Donahue} \& {Voit}(2022)}]{2022PhR...973....1D}
{Donahue}, M., \& {Voit}, G.~M. 2022, \physrep, 973, 1,
  \dodoi{10.1016/j.physrep.2022.04.005}

\bibitem[{{Donoso} {et~al.}(2014){Donoso}, {Yan}, {Stern}, \&
  {Assef}}]{2014ApJ...789...44D}
{Donoso}, E., {Yan}, L., {Stern}, D., \& {Assef}, R.~J. 2014, \apj, 789, 44,
  \dodoi{10.1088/0004-637X/789/1/44}

\bibitem[{{Dubois} {et~al.}(2016){Dubois}, {Peirani}, {Pichon}, {Devriendt},
  {Gavazzi}, {Welker}, \& {Volonteri}}]{2016MNRAS.463.3948D}
{Dubois}, Y., {Peirani}, S., {Pichon}, C., {et~al.} 2016, \mnras, 463, 3948,
  \dodoi{10.1093/mnras/stw2265}

\bibitem[{{Duffy} {et~al.}(2008){Duffy}, {Schaye}, {Kay}, \& {Dalla
  Vecchia}}]{2008MNRAS.390L..64D}
{Duffy}, A.~R., {Schaye}, J., {Kay}, S.~T., \& {Dalla Vecchia}, C. 2008,
  \mnras, 390, L64, \dodoi{10.1111/j.1745-3933.2008.00537.x}

\bibitem[{{Duncan}(2022)}]{2022MNRAS.512.3662D}
{Duncan}, K.~J. 2022, \mnras, 512, 3662, \dodoi{10.1093/mnras/stac608}

\bibitem[{{Duncan} {et~al.}(2021){Duncan}, {Kondapally}, {Brown}, {Bonato},
  {Best}, {R{\"o}ttgering}, {Bondi}, {Bowler}, {Cochrane}, {G{\"u}rkan},
  {Hardcastle}, {Jarvis}, {Kunert-Bajraszewska}, {Leslie}, {Ma{\l}ek},
  {Morabito}, {O'Sullivan}, {Prandoni}, {Sabater}, {Shimwell}, {Smith}, {Wang},
  {Wo{\l}owska}, \& {Tasse}}]{2021A&A...648A...4D}
{Duncan}, K.~J., {Kondapally}, R., {Brown}, M.~J.~I., {et~al.} 2021, \aap, 648,
  A4, \dodoi{10.1051/0004-6361/202038809}

\bibitem[{{Dunn} \& {Fabian}(2008)}]{2008MNRAS.385..757D}
{Dunn}, R.~J.~H., \& {Fabian}, A.~C. 2008, \mnras, 385, 757,
  \dodoi{10.1111/j.1365-2966.2008.12898.x}

\bibitem[{{Edge} {et~al.}(1992){Edge}, {Stewart}, \&
  {Fabian}}]{1992MNRAS.258..177E}
{Edge}, A.~C., {Stewart}, G.~C., \& {Fabian}, A.~C. 1992, \mnras, 258, 177,
  \dodoi{10.1093/mnras/258.1.177}

\bibitem[{{Efron}(1982)}]{1982jbor.book.....E}
{Efron}, B. 1982, {The Jackknife, the Bootstrap and other resampling plans}

\bibitem[{{Eftekharzadeh} {et~al.}(2019){Eftekharzadeh}, {Myers}, \&
  {Kourkchi}}]{2019MNRAS.486..274E}
{Eftekharzadeh}, S., {Myers}, A.~D., \& {Kourkchi}, E. 2019, \mnras, 486, 274,
  \dodoi{10.1093/mnras/stz770}

\bibitem[{{Eftekharzadeh} {et~al.}(2015){Eftekharzadeh}, {Myers}, {White},
  {Weinberg}, {Schneider}, {Shen}, {Font-Ribera}, {Ross}, {Paris}, \&
  {Streblyanska}}]{2015MNRAS.453.2779E}
{Eftekharzadeh}, S., {Myers}, A.~D., {White}, M., {et~al.} 2015, \mnras, 453,
  2779, \dodoi{10.1093/mnras/stv1763}

\bibitem[{{Eilers} {et~al.}(2024){Eilers}, {Mackenzie}, {Pizzati}, {Matthee},
  {Hennawi}, {Zhang}, {Bordoloi}, {Kashino}, {Lilly}, {Naidu}, {Simcoe}, {Yue},
  {Frenk}, {Helly}, {Schaller}, \& {Schaye}}]{2024arXiv240307986E}
{Eilers}, A.-C., {Mackenzie}, R., {Pizzati}, E., {et~al.} 2024, arXiv e-prints,
  arXiv:2403.07986, \dodoi{10.48550/arXiv.2403.07986}

\bibitem[{{Elvis}(2000)}]{2000ApJ...545...63E}
{Elvis}, M. 2000, \apj, 545, 63, \dodoi{10.1086/317778}

\bibitem[{{Fabian}(2012)}]{2012ARA&A..50..455F}
{Fabian}, A.~C. 2012, \araa, 50, 455,
  \dodoi{10.1146/annurev-astro-081811-125521}

\bibitem[{{Fabian} {et~al.}(2006){Fabian}, {Sanders}, {Taylor}, {Allen},
  {Crawford}, {Johnstone}, \& {Iwasawa}}]{2006MNRAS.366..417F}
{Fabian}, A.~C., {Sanders}, J.~S., {Taylor}, G.~B., {et~al.} 2006, \mnras, 366,
  417, \dodoi{10.1111/j.1365-2966.2005.09896.x}

\bibitem[{{Fabian} {et~al.}(2000){Fabian}, {Sanders}, {Ettori}, {Taylor},
  {Allen}, {Crawford}, {Iwasawa}, {Johnstone}, \& {Ogle}}]{2000MNRAS.318L..65F}
{Fabian}, A.~C., {Sanders}, J.~S., {Ettori}, S., {et~al.} 2000, \mnras, 318,
  L65, \dodoi{10.1046/j.1365-8711.2000.03904.x}

\bibitem[{{Fakhouri} {et~al.}(2010){Fakhouri}, {Ma}, \&
  {Boylan-Kolchin}}]{2010MNRAS.406.2267F}
{Fakhouri}, O., {Ma}, C.-P., \& {Boylan-Kolchin}, M. 2010, \mnras, 406, 2267,
  \dodoi{10.1111/j.1365-2966.2010.16859.x}

\bibitem[{{Fiore} {et~al.}(2017){Fiore}, {Feruglio}, {Shankar}, {Bischetti},
  {Bongiorno}, {Brusa}, {Carniani}, {Cicone}, {Duras}, {Lamastra}, {Mainieri},
  {Marconi}, {Menci}, {Maiolino}, {Piconcelli}, {Vietri}, \&
  {Zappacosta}}]{2017A&A...601A.143F}
{Fiore}, F., {Feruglio}, C., {Shankar}, F., {et~al.} 2017, \aap, 601, A143,
  \dodoi{10.1051/0004-6361/201629478}

\bibitem[{{Foreman-Mackey} {et~al.}(2013){Foreman-Mackey}, {Hogg}, {Lang}, \&
  {Goodman}}]{2013PASP..125..306F}
{Foreman-Mackey}, D., {Hogg}, D.~W., {Lang}, D., \& {Goodman}, J. 2013, \pasp,
  125, 306, \dodoi{10.1086/670067}

\bibitem[{{Geach} {et~al.}(2019){Geach}, {Peacock}, {Myers}, {Hickox},
  {Burchard}, \& {Jones}}]{2019ApJ...874...85G}
{Geach}, J.~E., {Peacock}, J.~A., {Myers}, A.~D., {et~al.} 2019, \apj, 874, 85,
  \dodoi{10.3847/1538-4357/ab0894}

\bibitem[{{Gonzalez-Perez} {et~al.}(2014){Gonzalez-Perez}, {Lacey}, {Baugh},
  {Lagos}, {Helly}, {Campbell}, \& {Mitchell}}]{2014MNRAS.439..264G}
{Gonzalez-Perez}, V., {Lacey}, C.~G., {Baugh}, C.~M., {et~al.} 2014, \mnras,
  439, 264, \dodoi{10.1093/mnras/stt2410}

\bibitem[{{G{\'o}rski} {et~al.}(2005){G{\'o}rski}, {Hivon}, {Banday},
  {Wandelt}, {Hansen}, {Reinecke}, \& {Bartelmann}}]{2005ApJ...622..759G}
{G{\'o}rski}, K.~M., {Hivon}, E., {Banday}, A.~J., {et~al.} 2005, \apj, 622,
  759, \dodoi{10.1086/427976}

\bibitem[{{Greene} {et~al.}(2014){Greene}, {Pooley}, {Zakamska}, {Comerford},
  \& {Sun}}]{2014ApJ...788...54G}
{Greene}, J.~E., {Pooley}, D., {Zakamska}, N.~L., {Comerford}, J.~M., \& {Sun},
  A.-L. 2014, \apj, 788, 54, \dodoi{10.1088/0004-637X/788/1/54}

\bibitem[{{Haiman} \& {Hui}(2001)}]{2001ApJ...547...27H}
{Haiman}, Z., \& {Hui}, L. 2001, \apj, 547, 27, \dodoi{10.1086/318330}

\bibitem[{{Hale} {et~al.}(2018){Hale}, {Jarvis}, {Delvecchio}, {Hatfield},
  {Novak}, {Smol{\v{c}}i{\'c}}, \& {Zamorani}}]{2018MNRAS.474.4133H}
{Hale}, C.~L., {Jarvis}, M.~J., {Delvecchio}, I., {et~al.} 2018, \mnras, 474,
  4133, \dodoi{10.1093/mnras/stx2954}

\bibitem[{{Hale} {et~al.}(2024){Hale}, {Schwarz}, {Best}, {Nakoneczny},
  {Alonso}, {Bacon}, {B{\"o}hme}, {Bhardwaj}, {Bilicki}, {Camera}, {Heneka},
  {Pashapour-Ahmadabadi}, {Tiwari}, {Zheng}, {Duncan}, {Jarvis}, {Kondapally},
  {Magliocchetti}, {Rottgering}, \& {Shimwell}}]{2024MNRAS.527.6540H}
{Hale}, C.~L., {Schwarz}, D.~J., {Best}, P.~N., {et~al.} 2024, \mnras, 527,
  6540, \dodoi{10.1093/mnras/stad3088}

\bibitem[{{Hamilton} \& {Tegmark}(2004)}]{2004MNRAS.349..115H}
{Hamilton}, A.~J.~S., \& {Tegmark}, M. 2004, \mnras, 349, 115,
  \dodoi{10.1111/j.1365-2966.2004.07490.x}

\bibitem[{{Hardcastle}(2018)}]{2018MNRAS.475.2768H}
{Hardcastle}, M.~J. 2018, \mnras, 475, 2768, \dodoi{10.1093/mnras/stx3358}

\bibitem[{{Hardcastle} \& {Croston}(2020)}]{2020NewAR..8801539H}
{Hardcastle}, M.~J., \& {Croston}, J.~H. 2020, \nar, 88, 101539,
  \dodoi{10.1016/j.newar.2020.101539}

\bibitem[{{Hardcastle} \& {Krause}(2013)}]{2013MNRAS.430..174H}
{Hardcastle}, M.~J., \& {Krause}, M.~G.~H. 2013, \mnras, 430, 174,
  \dodoi{10.1093/mnras/sts564}

\bibitem[{{Hardcastle} {et~al.}(2019){Hardcastle}, {Williams}, {Best},
  {Croston}, {Duncan}, {R{\"o}ttgering}, {Sabater}, {Shimwell}, {Tasse},
  {Callingham}, {Cochrane}, {de Gasperin}, {G{\"u}rkan}, {Jarvis}, {Mahatma},
  {Miley}, {Mingo}, {Mooney}, {Morabito}, {O'Sullivan}, {Prandoni},
  {Shulevski}, \& {Smith}}]{2019A&A...622A..12H}
{Hardcastle}, M.~J., {Williams}, W.~L., {Best}, P.~N., {et~al.} 2019, \aap,
  622, A12, \dodoi{10.1051/0004-6361/201833893}

\bibitem[{{Hardcastle} {et~al.}(2023){Hardcastle}, {Horton}, {Williams},
  {Duncan}, {Alegre}, {Barkus}, {Croston}, {Dickinson}, {Osinga},
  {R{\"o}ttgering}, {Sabater}, {Shimwell}, {Smith}, {Best}, {Botteon},
  {Br{\"u}ggen}, {Drabent}, {de Gasperin}, {G{\"u}rkan}, {Hajduk}, {Hale},
  {Hoeft}, {Jamrozy}, {Kunert-Bajraszewska}, {Kondapally}, {Magliocchetti},
  {Mahatma}, {Mostert}, {O'Sullivan}, {Pajdosz-{\'S}mierciak}, {Petley},
  {Pierce}, {Prandoni}, {Schwarz}, {Shulewski}, {Siewert}, {Stott}, {Tang},
  {Vaccari}, {Zheng}, {Bailey}, {Desbled}, {Goyal}, {Gonano}, {Hanset},
  {Kurtz}, {Lim}, {Mielle}, {Molloy}, {Roth}, {Terentev}, \&
  {Torres}}]{2023A&A...678A.151H}
{Hardcastle}, M.~J., {Horton}, M.~A., {Williams}, W.~L., {et~al.} 2023, \aap,
  678, A151, \dodoi{10.1051/0004-6361/202347333}

\bibitem[{{Harrison} {et~al.}(2014){Harrison}, {Alexander}, {Mullaney}, \&
  {Swinbank}}]{2014MNRAS.441.3306H}
{Harrison}, C.~M., {Alexander}, D.~M., {Mullaney}, J.~R., \& {Swinbank}, A.~M.
  2014, \mnras, 441, 3306, \dodoi{10.1093/mnras/stu515}

\bibitem[{{Harrison} \& {Ramos Almeida}(2024)}]{2024Galax..12...17H}
{Harrison}, C.~M., \& {Ramos Almeida}, C. 2024, Galaxies, 12, 17,
  \dodoi{10.3390/galaxies12020017}

\bibitem[{{Harrison} {et~al.}(2012){Harrison}, {Alexander}, {Swinbank},
  {Smail}, {Alaghband-Zadeh}, {Bauer}, {Chapman}, {Del Moro}, {Hickox},
  {Ivison}, {Men{\'e}ndez-Delmestre}, {Mullaney}, \&
  {Nesvadba}}]{2012MNRAS.426.1073H}
{Harrison}, C.~M., {Alexander}, D.~M., {Swinbank}, A.~M., {et~al.} 2012,
  \mnras, 426, 1073, \dodoi{10.1111/j.1365-2966.2012.21723.x}

\bibitem[{{Hatch} {et~al.}(2014){Hatch}, {Wylezalek}, {Kurk}, {Stern}, {De
  Breuck}, {Jarvis}, {Galametz}, {Gonzalez}, {Hartley}, {Mortlock}, {Seymour},
  \& {Stevens}}]{2014MNRAS.445..280H}
{Hatch}, N.~A., {Wylezalek}, D., {Kurk}, J.~D., {et~al.} 2014, \mnras, 445,
  280, \dodoi{10.1093/mnras/stu1725}

\bibitem[{{He} {et~al.}(2018){He}, {Akiyama}, {Bosch}, {Enoki}, {Harikane},
  {Ikeda}, {Kashikawa}, {Kawaguchi}, {Komiyama}, {Lee}, {Matsuoka}, {Miyazaki},
  {Nagao}, {Nagashima}, {Niida}, {Nishizawa}, {Oguri}, {Onoue}, {Oogi},
  {Ouchi}, {Schulze}, {Shirasaki}, {Silverman}, {Tanaka}, {Tanaka}, {Toba},
  {Uchiyama}, \& {Yamashita}}]{2018PASJ...70S..33H}
{He}, W., {Akiyama}, M., {Bosch}, J., {et~al.} 2018, \pasj, 70, S33,
  \dodoi{10.1093/pasj/psx129}

\bibitem[{{Heckman} \& {Best}(2014)}]{2014ARA&A..52..589H}
{Heckman}, T.~M., \& {Best}, P.~N. 2014, \araa, 52, 589,
  \dodoi{10.1146/annurev-astro-081913-035722}

\bibitem[{{Helfand} {et~al.}(2015){Helfand}, {White}, \&
  {Becker}}]{2015ApJ...801...26H}
{Helfand}, D.~J., {White}, R.~L., \& {Becker}, R.~H. 2015, \apj, 801, 26,
  \dodoi{10.1088/0004-637X/801/1/26}

\bibitem[{{Hickox} \& {Alexander}(2018)}]{2018ARA&A..56..625H}
{Hickox}, R.~C., \& {Alexander}, D.~M. 2018, \araa, 56, 625,
  \dodoi{10.1146/annurev-astro-081817-051803}

\bibitem[{{Hickox} {et~al.}(2009){Hickox}, {Jones}, {Forman}, {Murray},
  {Kochanek}, {Eisenstein}, {Jannuzi}, {Dey}, {Brown}, {Stern}, {Eisenhardt},
  {Gorjian}, {Brodwin}, {Narayan}, {Cool}, {Kenter}, {Caldwell}, \&
  {Anderson}}]{2009ApJ...696..891H}
{Hickox}, R.~C., {Jones}, C., {Forman}, W.~R., {et~al.} 2009, \apj, 696, 891,
  \dodoi{10.1088/0004-637X/696/1/891}

\bibitem[{{Hickox} {et~al.}(2011){Hickox}, {Myers}, {Brodwin}, {Alexander},
  {Forman}, {Jones}, {Murray}, {Brown}, {Cool}, {Kochanek}, {Dey}, {Jannuzi},
  {Eisenstein}, {Assef}, {Eisenhardt}, {Gorjian}, {Stern}, {Le Floc'h},
  {Caldwell}, {Goulding}, \& {Mullaney}}]{2011ApJ...731..117H}
{Hickox}, R.~C., {Myers}, A.~D., {Brodwin}, M., {et~al.} 2011, \apj, 731, 117,
  \dodoi{10.1088/0004-637X/731/2/117}

\bibitem[{{Hivon} {et~al.}(2002){Hivon}, {G{\'o}rski}, {Netterfield}, {Crill},
  {Prunet}, \& {Hansen}}]{2002ApJ...567....2H}
{Hivon}, E., {G{\'o}rski}, K.~M., {Netterfield}, C.~B., {et~al.} 2002, \apj,
  567, 2, \dodoi{10.1086/338126}

\bibitem[{{Hopkins} {et~al.}(2006){Hopkins}, {Hernquist}, {Cox}, {Di Matteo},
  {Robertson}, \& {Springel}}]{2006ApJS..163....1H}
{Hopkins}, P.~F., {Hernquist}, L., {Cox}, T.~J., {et~al.} 2006, \apjs, 163, 1,
  \dodoi{10.1086/499298}

\bibitem[{{Hopkins} {et~al.}(2007){Hopkins}, {Richards}, \&
  {Hernquist}}]{2007ApJ...654..731H}
{Hopkins}, P.~F., {Richards}, G.~T., \& {Hernquist}, L. 2007, \apj, 654, 731,
  \dodoi{10.1086/509629}

\bibitem[{{Ineson} {et~al.}(2015){Ineson}, {Croston}, {Hardcastle}, {Kraft},
  {Evans}, \& {Jarvis}}]{2015MNRAS.453.2682I}
{Ineson}, J., {Croston}, J.~H., {Hardcastle}, M.~J., {et~al.} 2015, \mnras,
  453, 2682, \dodoi{10.1093/mnras/stv1807}

\bibitem[{{Kaiser}(1987)}]{1987MNRAS.227....1K}
{Kaiser}, N. 1987, \mnras, 227, 1, \dodoi{10.1093/mnras/227.1.1}

\bibitem[{{Kakkad} {et~al.}(2022){Kakkad}, {Sani}, {Rojas}, {Mallmann},
  {Veilleux}, {Bauer}, {Ricci}, {Mushotzky}, {Koss}, {Ricci}, {Treister},
  {Privon}, {Nguyen}, {B{\"a}r}, {Harrison}, {Oh}, {Powell}, {Riffel}, {Stern},
  {Trakhtenbrot}, \& {Urry}}]{2022MNRAS.511.2105K}
{Kakkad}, D., {Sani}, E., {Rojas}, A.~F., {et~al.} 2022, \mnras, 511, 2105,
  \dodoi{10.1093/mnras/stac103}

\bibitem[{{Kaviraj} {et~al.}(2017){Kaviraj}, {Laigle}, {Kimm}, {Devriendt},
  {Dubois}, {Pichon}, {Slyz}, {Chisari}, \& {Peirani}}]{2017MNRAS.467.4739K}
{Kaviraj}, S., {Laigle}, C., {Kimm}, T., {et~al.} 2017, \mnras, 467, 4739,
  \dodoi{10.1093/mnras/stx126}

\bibitem[{{Kere{\v{s}}} {et~al.}(2009){Kere{\v{s}}}, {Katz}, {Fardal},
  {Dav{\'e}}, \& {Weinberg}}]{2009MNRAS.395..160K}
{Kere{\v{s}}}, D., {Katz}, N., {Fardal}, M., {Dav{\'e}}, R., \& {Weinberg},
  D.~H. 2009, \mnras, 395, 160, \dodoi{10.1111/j.1365-2966.2009.14541.x}

\bibitem[{{Klindt} {et~al.}(2019){Klindt}, {Alexander}, {Rosario}, {Lusso}, \&
  {Fotopoulou}}]{2019MNRAS.488.3109K}
{Klindt}, L., {Alexander}, D.~M., {Rosario}, D.~J., {Lusso}, E., \&
  {Fotopoulou}, S. 2019, \mnras, 488, 3109, \dodoi{10.1093/mnras/stz1771}

\bibitem[{{Kochanek} {et~al.}(2012){Kochanek}, {Eisenstein}, {Cool},
  {Caldwell}, {Assef}, {Jannuzi}, {Jones}, {Murray}, {Forman}, {Dey}, {Brown},
  {Eisenhardt}, {Gonzalez}, {Green}, \& {Stern}}]{2012ApJS..200....8K}
{Kochanek}, C.~S., {Eisenstein}, D.~J., {Cool}, R.~J., {et~al.} 2012, \apjs,
  200, 8, \dodoi{10.1088/0067-0049/200/1/8}

\bibitem[{{Kondapally} {et~al.}(2021){Kondapally}, {Best}, {Hardcastle},
  {Nisbet}, {Bonato}, {Sabater}, {Duncan}, {McCheyne}, {Cochrane}, {Bowler},
  {Williams}, {Shimwell}, {Tasse}, {Croston}, {Goyal}, {Jamrozy}, {Jarvis},
  {Mahatma}, {R{\"o}ttgering}, {Smith}, {Wo{\l}owska}, {Bondi}, {Brienza},
  {Brown}, {Br{\"u}ggen}, {Chambers}, {Garrett}, {G{\"u}rkan}, {Huber},
  {Kunert-Bajraszewska}, {Magnier}, {Mingo}, {Mostert},
  {Nikiel-Wroczy{\'n}ski}, {O'Sullivan}, {Paladino}, {Ploeckinger}, {Prandoni},
  {Rosenthal}, {Schwarz}, {Shulevski}, {Wagenveld}, \&
  {Wang}}]{2021A&A...648A...3K}
{Kondapally}, R., {Best}, P.~N., {Hardcastle}, M.~J., {et~al.} 2021, \aap, 648,
  A3, \dodoi{10.1051/0004-6361/202038813}

\bibitem[{{Kondapally} {et~al.}(2022){Kondapally}, {Best}, {Cochrane},
  {Sabater}, {Duncan}, {Hardcastle}, {Haskell}, {Mingo}, {R{\"o}ttgering},
  {Smith}, {Williams}, {Bonato}, {Calistro Rivera}, {Gao}, {Hale}, {Ma{\l}ek},
  {Miley}, {Prandoni}, \& {Wang}}]{2022MNRAS.513.3742K}
{Kondapally}, R., {Best}, P.~N., {Cochrane}, R.~K., {et~al.} 2022, \mnras, 513,
  3742, \dodoi{10.1093/mnras/stac1128}

\bibitem[{{Kondapally} {et~al.}(2023){Kondapally}, {Best}, {Raouf}, {Thomas},
  {Dav{\'e}}, {Shabala}, {R{\"o}ttgering}, {Hardcastle}, {Bonato}, {Cochrane},
  {Ma{\l}ek}, {Morabito}, {Prandoni}, \& {Smith}}]{2023MNRAS.523.5292K}
{Kondapally}, R., {Best}, P.~N., {Raouf}, M., {et~al.} 2023, \mnras, 523, 5292,
  \dodoi{10.1093/mnras/stad1813}

\bibitem[{{Kormendy} \& {Ho}(2013)}]{2013ARA&A..51..511K}
{Kormendy}, J., \& {Ho}, L.~C. 2013, \araa, 51, 511,
  \dodoi{10.1146/annurev-astro-082708-101811}

\bibitem[{{Krolewski} {et~al.}(2020){Krolewski}, {Ferraro}, {Schlafly}, \&
  {White}}]{2020JCAP...05..047K}
{Krolewski}, A., {Ferraro}, S., {Schlafly}, E.~F., \& {White}, M. 2020, \jcap,
  2020, 047, \dodoi{10.1088/1475-7516/2020/05/047}

\bibitem[{{Krolewski} {et~al.}(2024){Krolewski}, {Percival}, {Ferraro},
  {Chaussidon}, {Rezaie}, {Aguilar}, {Ahlen}, {Brooks}, {Dawson}, {de la
  Macorra}, {Doel}, {Fanning}, {Font-Ribera}, {a Gontcho}, {Guy}, {Honscheid},
  {Kehoe}, {Kisner}, {Kremin}, {Landriau}, {Levi}, {Martini}, {Meisner},
  {Miquel}, {Nie}, {Poppett}, {Ross}, {Rossi}, {Schubnell}, {Seo}, {Tarl{\'e}},
  {Vargas-Maga{\~n}a}, {Weaver}, {Y{\`e}che}, {Zhou}, \&
  {Zhou}}]{2024JCAP...03..021K}
{Krolewski}, A., {Percival}, W.~J., {Ferraro}, S., {et~al.} 2024, \jcap, 2024,
  021, \dodoi{10.1088/1475-7516/2024/03/021}

\bibitem[{{Lacy} {et~al.}(2019){Lacy}, {Mason}, {Sarazin}, {Chatterjee},
  {Nyland}, {Kimball}, {Rocha}, {Rowe}, \& {Surace}}]{2019MNRAS.483L..22L}
{Lacy}, M., {Mason}, B., {Sarazin}, C., {et~al.} 2019, \mnras, 483, L22,
  \dodoi{10.1093/mnrasl/sly215}

\bibitem[{{Laha} {et~al.}(2021){Laha}, {Reynolds}, {Reeves}, {Kriss},
  {Guainazzi}, {Smith}, {Veilleux}, \& {Proga}}]{2021NatAs...5...13L}
{Laha}, S., {Reynolds}, C.~S., {Reeves}, J., {et~al.} 2021, Nature Astronomy,
  5, 13, \dodoi{10.1038/s41550-020-01255-2}

\bibitem[{{Landy} \& {Szalay}(1993)}]{1993ApJ...412...64L}
{Landy}, S.~D., \& {Szalay}, A.~S. 1993, \apj, 412, 64, \dodoi{10.1086/172900}

\bibitem[{{Laurent} {et~al.}(2017){Laurent}, {Eftekharzadeh}, {Le Goff},
  {Myers}, {Burtin}, {White}, {Ross}, {Tinker}, {Tojeiro}, {Bautista},
  {Brinkmann}, {Comparat}, {Dawson}, {du Mas des Bourboux}, {Kneib}, {McGreer},
  {Palanque-Delabrouille}, {Percival}, {Prada}, {Rossi}, {Schneider},
  {Weinberg}, {Y{\`e}che}, {Zarrouk}, \& {Zhao}}]{2017JCAP...07..017L}
{Laurent}, P., {Eftekharzadeh}, S., {Le Goff}, J.-M., {et~al.} 2017, \jcap,
  2017, 017, \dodoi{10.1088/1475-7516/2017/07/017}

\bibitem[{{Lewis} {et~al.}(2000){Lewis}, {Challinor}, \&
  {Lasenby}}]{2000ApJ...538..473L}
{Lewis}, A., {Challinor}, A., \& {Lasenby}, A. 2000, \apj, 538, 473,
  \dodoi{10.1086/309179}

\bibitem[{{Limber}(1953)}]{1953ApJ...117..134L}
{Limber}, D.~N. 1953, \apj, 117, 134, \dodoi{10.1086/145672}

\bibitem[{{Lindsay} {et~al.}(2014){Lindsay}, {Jarvis}, \&
  {McAlpine}}]{2014MNRAS.440.2322L}
{Lindsay}, S.~N., {Jarvis}, M.~J., \& {McAlpine}, K. 2014, \mnras, 440, 2322,
  \dodoi{10.1093/mnras/stu453}

\bibitem[{{Liu} {et~al.}(2013){Liu}, {Zakamska}, {Greene}, {Nesvadba}, \&
  {Liu}}]{2013MNRAS.436.2576L}
{Liu}, G., {Zakamska}, N.~L., {Greene}, J.~E., {Nesvadba}, N. P.~H., \& {Liu},
  X. 2013, \mnras, 436, 2576, \dodoi{10.1093/mnras/stt1755}

\bibitem[{{Lutz} {et~al.}(2020){Lutz}, {Sturm}, {Janssen}, {Veilleux}, {Aalto},
  {Cicone}, {Contursi}, {Davies}, {Feruglio}, {Fischer}, {Fluetsch},
  {Garcia-Burillo}, {Genzel}, {Gonz{\'a}lez-Alfonso}, {Graci{\'a}-Carpio},
  {Herrera-Camus}, {Maiolino}, {Schruba}, {Shimizu}, {Sternberg}, {Tacconi}, \&
  {Wei{\ss}}}]{2020A&A...633A.134L}
{Lutz}, D., {Sturm}, E., {Janssen}, A., {et~al.} 2020, \aap, 633, A134,
  \dodoi{10.1051/0004-6361/201936803}

\bibitem[{{Madau} \& {Dickinson}(2014)}]{2014ARA&A..52..415M}
{Madau}, P., \& {Dickinson}, M. 2014, \araa, 52, 415,
  \dodoi{10.1146/annurev-astro-081811-125615}

\bibitem[{{Madhavacheril} {et~al.}(2024){Madhavacheril}, {Qu}, {Sherwin},
  {MacCrann}, {Li}, {Abril-Cabezas}, {Ade}, {Aiola}, {Alford}, {Amiri},
  {Amodeo}, {An}, {Atkins}, {Austermann}, {Battaglia}, {Battistelli}, {Beall},
  {Bean}, {Beringue}, {Bhandarkar}, {Biermann}, {Bolliet}, {Bond}, {Cai},
  {Calabrese}, {Calafut}, {Capalbo}, {Carrero}, {Challinor}, {Chesmore}, {Cho},
  {Choi}, {Clark}, {C{\'o}rdova Rosado}, {Cothard}, {Coughlin}, {Coulton},
  {Crowley}, {Dalal}, {Darwish}, {Devlin}, {Dicker}, {Doze}, {Duell}, {Duff},
  {Duivenvoorden}, {Dunkley}, {D{\"u}nner}, {Fanfani}, {Fankhanel}, {Farren},
  {Ferraro}, {Freundt}, {Fuzia}, {Gallardo}, {Garrido}, {Givans}, {Gluscevic},
  {Golec}, {Guan}, {Hall}, {Halpern}, {Han}, {Harrison}, {Hasselfield},
  {Healy}, {Henderson}, {Hensley}, {Herv{\'\i}as-Caimapo}, {Hill}, {Hilton},
  {Hilton}, {Hincks}, {Hlo{\v{z}}ek}, {Ho}, {Huber}, {Hubmayr}, {Huffenberger},
  {Hughes}, {Irwin}, {Isopi}, {Jense}, {Keller}, {Kim}, {Knowles}, {Koopman},
  {Kosowsky}, {Kramer}, {Kusiak}, {La Posta}, {Lague}, {Lakey}, {Lee}, {Li},
  {Limon}, {Lokken}, {Louis}, {Lungu}, {MacInnis}, {Maldonado}, {Maldonado},
  {Mallaby-Kay}, {Marques}, {McMahon}, {Mehta}, {Menanteau}, {Moodley},
  {Morris}, {Mroczkowski}, {Naess}, {Namikawa}, {Nati}, {Newburgh}, {Nicola},
  {Niemack}, {Nolta}, {Orlowski-Scherer}, {Page}, {Pandey}, {Partridge},
  {Prince}, {Puddu}, {Radiconi}, {Robertson}, {Rojas}, {Sakuma}, {Salatino},
  {Schaan}, {Schmitt}, {Sehgal}, {Shaikh}, {Sierra}, {Sievers}, {Sif{\'o}n},
  {Simon}, {Sonka}, {Spergel}, {Staggs}, {Storer}, {Switzer}, {Tampier},
  {Thornton}, {Trac}, {Treu}, {Tucker}, {Ullom}, {Vale}, {Van Engelen}, {Van
  Lanen}, {van Marrewijk}, {Vargas}, {Vavagiakis}, {Wagoner}, {Wang}, {Wenzl},
  {Wollack}, {Xu}, {Zago}, \& {Zheng}}]{2024ApJ...962..113M}
{Madhavacheril}, M.~S., {Qu}, F.~J., {Sherwin}, B.~D., {et~al.} 2024, \apj,
  962, 113, \dodoi{10.3847/1538-4357/acff5f}

\bibitem[{{Magliocchetti}(2022)}]{2022A&ARv..30....6M}
{Magliocchetti}, M. 2022, \aapr, 30, 6, \dodoi{10.1007/s00159-022-00142-1}

\bibitem[{{Magliocchetti} {et~al.}(2017){Magliocchetti}, {Popesso}, {Brusa},
  {Salvato}, {Laigle}, {McCracken}, \& {Ilbert}}]{2017MNRAS.464.3271M}
{Magliocchetti}, M., {Popesso}, P., {Brusa}, M., {et~al.} 2017, \mnras, 464,
  3271, \dodoi{10.1093/mnras/stw2541}

\bibitem[{{Magliocchetti} {et~al.}(2004){Magliocchetti}, {Maddox}, {Hawkins},
  {Peacock}, {Bland-Hawthorn}, {Bridges}, {Cannon}, {Cole}, {Colless},
  {Collins}, {Couch}, {Dalton}, {de Propris}, {Driver}, {Efstathiou}, {Ellis},
  {Frenk}, {Glazebrook}, {Jackson}, {Jones}, {Lahav}, {Lewis}, {Lumsden},
  {Norberg}, {Peterson}, {Sutherland}, {Taylor}, \& {2dFGRS
  Team}}]{2004MNRAS.350.1485M}
{Magliocchetti}, M., {Maddox}, S.~J., {Hawkins}, E., {et~al.} 2004, \mnras,
  350, 1485, \dodoi{10.1111/j.1365-2966.2004.07751.x}

\bibitem[{{Magorrian} {et~al.}(1998){Magorrian}, {Tremaine}, {Richstone},
  {Bender}, {Bower}, {Dressler}, {Faber}, {Gebhardt}, {Green}, {Grillmair},
  {Kormendy}, \& {Lauer}}]{1998AJ....115.2285M}
{Magorrian}, J., {Tremaine}, S., {Richstone}, D., {et~al.} 1998, \aj, 115,
  2285, \dodoi{10.1086/300353}

\bibitem[{{Mainzer} {et~al.}(2011){Mainzer}, {Bauer}, {Grav}, {Masiero},
  {Cutri}, {Dailey}, {Eisenhardt}, {McMillan}, {Wright}, {Walker}, {Jedicke},
  {Spahr}, {Tholen}, {Alles}, {Beck}, {Brandenburg}, {Conrow}, {Evans},
  {Fowler}, {Jarrett}, {Marsh}, {Masci}, {McCallon}, {Wheelock}, {Wittman},
  {Wyatt}, {DeBaun}, {Elliott}, {Elsbury}, {Gautier}, {Gomillion}, {Leisawitz},
  {Maleszewski}, {Micheli}, \& {Wilkins}}]{2011ApJ...731...53M}
{Mainzer}, A., {Bauer}, J., {Grav}, T., {et~al.} 2011, \apj, 731, 53,
  \dodoi{10.1088/0004-637X/731/1/53}

\bibitem[{{Mandelbaum} {et~al.}(2009){Mandelbaum}, {Li}, {Kauffmann}, \&
  {White}}]{2009MNRAS.393..377M}
{Mandelbaum}, R., {Li}, C., {Kauffmann}, G., \& {White}, S. D.~M. 2009, \mnras,
  393, 377, \dodoi{10.1111/j.1365-2966.2008.14235.x}

\bibitem[{{Marocco} {et~al.}(2021{\natexlab{a}}){Marocco}, {Eisenhardt},
  {Fowler}, {Kirkpatrick}, {Meisner}, {Schlafly}, {Stanford}, {Garcia},
  {Caselden}, {Cushing}, {Cutri}, {Faherty}, {Gelino}, {Gonzalez}, {Jarrett},
  {Koontz}, {Mainzer}, {Marchese}, {Mobasher}, {Schlegel}, {Stern}, {Teplitz},
  \& {Wright}}]{2021ApJS..253....8M}
{Marocco}, F., {Eisenhardt}, P. R.~M., {Fowler}, J.~W., {et~al.}
  2021{\natexlab{a}}, \apjs, 253, 8, \dodoi{10.3847/1538-4365/abd805}

\bibitem[{{Marocco} {et~al.}(2021{\natexlab{b}}){Marocco}, {Eisenhardt},
  {Fowler}, {Kirkpatrick}, {Meisner}, {Schlafly}, {Stanford}, {Garcia},
  {Caselden}, {Cushing}, {Cutri}, {Faherty}, {Gelino}, {Gonzalez}, {Jarrett},
  {Koontz}, {Mainzer}, {Marchese}, {Mobasher}, {Schlegel}, {Stern}, {Teplitz},
  \& {Wright}}]{catwise2020}
---. 2021{\natexlab{b}}, \apjs, 253, 8, \dodoi{10.3847/1538-4365/abd805}

\bibitem[{{Marocco et al.}(2020)}]{CatWISE}
{Marocco et al.} 2020, CatWISE2020 Catalog,  IPAC, \dodoi{10.26131/IRSA551}

\bibitem[{{Martini} \& {Weinberg}(2001)}]{2001ApJ...547...12M}
{Martini}, P., \& {Weinberg}, D.~H. 2001, \apj, 547, 12, \dodoi{10.1086/318331}

\bibitem[{{Massingill} {et~al.}(2024){Massingill}, {Mason}, {Lacy}, {Emonts},
  {Yoon}, {Li}, \& {Sarazin}}]{2024ApJ...969...56M}
{Massingill}, K., {Mason}, B., {Lacy}, M., {et~al.} 2024, \apj, 969, 56,
  \dodoi{10.3847/1538-4357/ad3a67}

\bibitem[{{McCarthy} {et~al.}(2010){McCarthy}, {Schaye}, {Ponman}, {Bower},
  {Booth}, {Dalla Vecchia}, {Crain}, {Springel}, {Theuns}, \&
  {Wiersma}}]{2010MNRAS.406..822M}
{McCarthy}, I.~G., {Schaye}, J., {Ponman}, T.~J., {et~al.} 2010, \mnras, 406,
  822, \dodoi{10.1111/j.1365-2966.2010.16750.x}

\bibitem[{{McNamara} \& {Nulsen}(2007)}]{2007ARA&A..45..117M}
{McNamara}, B.~R., \& {Nulsen}, P.~E.~J. 2007, \araa, 45, 117,
  \dodoi{10.1146/annurev.astro.45.051806.110625}

\bibitem[{{McNamara} \& {Nulsen}(2012)}]{2012NJPh...14e5023M}
---. 2012, New Journal of Physics, 14, 055023,
  \dodoi{10.1088/1367-2630/14/5/055023}

\bibitem[{{McNamara} {et~al.}(2000){McNamara}, {Wise}, {Nulsen}, {David},
  {Sarazin}, {Bautz}, {Markevitch}, {Vikhlinin}, {Forman}, {Jones}, \&
  {Harris}}]{2000ApJ...534L.135M}
{McNamara}, B.~R., {Wise}, M., {Nulsen}, P.~E.~J., {et~al.} 2000, \apjl, 534,
  L135, \dodoi{10.1086/312662}

\bibitem[{{Mead} {et~al.}(2021){Mead}, {Brieden}, {Tr{\"o}ster}, \&
  {Heymans}}]{2021MNRAS.502.1401M}
{Mead}, A.~J., {Brieden}, S., {Tr{\"o}ster}, T., \& {Heymans}, C. 2021, \mnras,
  502, 1401, \dodoi{10.1093/mnras/stab082}

\bibitem[{{Mead} {et~al.}(2015){Mead}, {Peacock}, {Heymans}, {Joudaki}, \&
  {Heavens}}]{2015MNRAS.454.1958M}
{Mead}, A.~J., {Peacock}, J.~A., {Heymans}, C., {Joudaki}, S., \& {Heavens},
  A.~F. 2015, \mnras, 454, 1958, \dodoi{10.1093/mnras/stv2036}

\bibitem[{{M{\'e}nard} {et~al.}(2013){M{\'e}nard}, {Scranton}, {Schmidt},
  {Morrison}, {Jeong}, {Budavari}, \& {Rahman}}]{2013arXiv1303.4722M}
{M{\'e}nard}, B., {Scranton}, R., {Schmidt}, S., {et~al.} 2013, arXiv e-prints,
  arXiv:1303.4722, \dodoi{10.48550/arXiv.1303.4722}

\bibitem[{{Modi} {et~al.}(2017){Modi}, {White}, \&
  {Vlah}}]{2017JCAP...08..009M}
{Modi}, C., {White}, M., \& {Vlah}, Z. 2017, \jcap, 2017, 009,
  \dodoi{10.1088/1475-7516/2017/08/009}

\bibitem[{{Morabito} {et~al.}(2022){Morabito}, {Sweijen}, {Radcliffe}, {Best},
  {Kondapally}, {Bondi}, {Bonato}, {Duncan}, {Prandoni}, {Shimwell},
  {Williams}, {van Weeren}, {Conway}, \& {Calistro
  Rivera}}]{2022MNRAS.515.5758M}
{Morabito}, L.~K., {Sweijen}, F., {Radcliffe}, J.~F., {et~al.} 2022, \mnras,
  515, 5758, \dodoi{10.1093/mnras/stac2129}

\bibitem[{{Moustakas} {et~al.}(2023){Moustakas}, {Lang}, {Dey}, {Juneau},
  {Meisner}, {Myers}, {Schlafly}, {Schlegel}, {Valdes}, {Weaver}, \&
  {Zhou}}]{2023ApJS..269....3M}
{Moustakas}, J., {Lang}, D., {Dey}, A., {et~al.} 2023, \apjs, 269, 3,
  \dodoi{10.3847/1538-4365/acfaa2}

\bibitem[{{Myers} {et~al.}(2007){Myers}, {Brunner}, {Nichol}, {Richards},
  {Schneider}, \& {Bahcall}}]{2007ApJ...658...85M}
{Myers}, A.~D., {Brunner}, R.~J., {Nichol}, R.~C., {et~al.} 2007, \apj, 658,
  85, \dodoi{10.1086/511519}

\bibitem[{{Myers} {et~al.}(2015){Myers}, {Palanque-Delabrouille}, {Prakash},
  {P{\^a}ris}, {Yeche}, {Dawson}, {Bovy}, {Lang}, {Schlegel}, {Newman},
  {Petitjean}, {Kneib}, {Laurent}, {Percival}, {Ross}, {Seo}, {Tinker},
  {Armengaud}, {Brownstein}, {Burtin}, {Cai}, {Comparat}, {Kasliwal},
  {Kulkarni}, {Laher}, {Levitan}, {McBride}, {McGreer}, {Miller}, {Nugent},
  {Ofek}, {Rossi}, {Ruan}, {Schneider}, {Sesar}, {Streblyanska}, \&
  {Surace}}]{2015ApJS..221...27M}
{Myers}, A.~D., {Palanque-Delabrouille}, N., {Prakash}, A., {et~al.} 2015,
  \apjs, 221, 27, \dodoi{10.1088/0067-0049/221/2/27}

\bibitem[{{Myers} {et~al.}(2023){Myers}, {Moustakas}, {Bailey}, {Weaver},
  {Cooper}, {Forero-Romero}, {Abolfathi}, {Alexander}, {Brooks}, {Chaussidon},
  {Chuang}, {Dawson}, {Dey}, {Dey}, {Dhungana}, {Doel}, {Fanning},
  {Gazta{\~n}aga}, {Gontcho A Gontcho}, {Gonzalez-Morales}, {Hahn},
  {Herrera-Alcantar}, {Honscheid}, {Ishak}, {Karim}, {Kirkby}, {Kisner},
  {Koposov}, {Kremin}, {Lan}, {Landriau}, {Lang}, {Levi}, {Magneville},
  {Napolitano}, {Martini}, {Meisner}, {Newman}, {Palanque-Delabrouille},
  {Percival}, {Poppett}, {Prada}, {Raichoor}, {Ross}, {Schlafly}, {Schlegel},
  {Schubnell}, {Tan}, {Tarle}, {Wilson}, {Y{\`e}che}, {Zhou}, {Zhou}, \&
  {Zou}}]{2023AJ....165...50M}
{Myers}, A.~D., {Moustakas}, J., {Bailey}, S., {et~al.} 2023, \aj, 165, 50,
  \dodoi{10.3847/1538-3881/aca5f9}

\bibitem[{{Nakoneczny} {et~al.}(2024){Nakoneczny}, {Alonso}, {Bilicki},
  {Schwarz}, {Hale}, {Pollo}, {Heneka}, {Tiwari}, {Zheng}, {Br{\"u}ggen},
  {Jarvis}, \& {Shimwell}}]{2024A&A...681A.105N}
{Nakoneczny}, S.~J., {Alonso}, D., {Bilicki}, M., {et~al.} 2024, \aap, 681,
  A105, \dodoi{10.1051/0004-6361/202347728}

\bibitem[{{Nandra} {et~al.}(2013){Nandra}, {Barret}, {Barcons}, {Fabian}, {den
  Herder}, {Piro}, {Watson}, {Adami}, {Aird}, {Afonso}, {Alexander},
  {Argiroffi}, {Amati}, {Arnaud}, {Atteia}, {Audard}, {Badenes}, {Ballet},
  {Ballo}, {Bamba}, {Bhardwaj}, {Stefano Battistelli}, {Becker}, {De Becker},
  {Behar}, {Bianchi}, {Biffi}, {B{\^\i}rzan}, {Bocchino}, {Bogdanov}, {Boirin},
  {Boller}, {Borgani}, {Borm}, {Bouch{\'e}}, {Bourdin}, {Bower}, {Braito},
  {Branchini}, {Branduardi-Raymont}, {Bregman}, {Brenneman}, {Brightman},
  {Br{\"u}ggen}, {Buchner}, {Bulbul}, {Brusa}, {Bursa}, {Caccianiga},
  {Cackett}, {Campana}, {Cappelluti}, {Cappi}, {Carrera}, {Ceballos},
  {Christensen}, {Chu}, {Churazov}, {Clerc}, {Corbel}, {Corral}, {Comastri},
  {Costantini}, {Croston}, {Dadina}, {D'Ai}, {Decourchelle}, {Della Ceca},
  {Dennerl}, {Dolag}, {Done}, {Dovciak}, {Drake}, {Eckert}, {Edge}, {Ettori},
  {Ezoe}, {Feigelson}, {Fender}, {Feruglio}, {Finoguenov}, {Fiore}, {Galeazzi},
  {Gallagher}, {Gandhi}, {Gaspari}, {Gastaldello}, {Georgakakis},
  {Georgantopoulos}, {Gilfanov}, {Gitti}, {Gladstone}, {Goosmann}, {Gosset},
  {Grosso}, {Guedel}, {Guerrero}, {Haberl}, {Hardcastle}, {Heinz}, {Alonso
  Herrero}, {Herv{\'e}}, {Holmstrom}, {Iwasawa}, {Jonker}, {Kaastra}, {Kara},
  {Karas}, {Kastner}, {King}, {Kosenko}, {Koutroumpa}, {Kraft}, {Kreykenbohm},
  {Lallement}, {Lanzuisi}, {Lee}, {Lemoine-Goumard}, {Lobban}, {Lodato},
  {Lovisari}, {Lotti}, {McCharthy}, {McNamara}, {Maggio}, {Maiolino}, {De
  Marco}, {de Martino}, {Mateos}, {Matt}, {Maughan}, {Mazzotta}, {Mendez},
  {Merloni}, {Micela}, {Miceli}, {Mignani}, {Miller}, {Miniutti}, {Molendi},
  {Montez}, {Moretti}, {Motch}, {Naz{\'e}}, {Nevalainen}, {Nicastro}, {Nulsen},
  {Ohashi}, {O'Brien}, {Osborne}, {Oskinova}, {Pacaud}, {Paerels}, {Page},
  {Papadakis}, {Pareschi}, {Petre}, {Petrucci}, {Piconcelli}, {Pillitteri},
  {Pinto}, {de Plaa}, {Pointecouteau}, {Ponman}, {Ponti}, {Porquet}, {Pounds},
  {Pratt}, {Predehl}, {Proga}, {Psaltis}, {Rafferty}, {Ramos-Ceja}, {Ranalli},
  {Rasia}, {Rau}, {Rauw}, {Rea}, {Read}, {Reeves}, {Reiprich}, {Renaud},
  {Reynolds}, {Risaliti}, {Rodriguez}, {Rodriguez Hidalgo}, {Roncarelli},
  {Rosario}, {Rossetti}, {Rozanska}, {Rovilos}, {Salvaterra}, {Salvato}, {Di
  Salvo}, {Sanders}, {Sanz-Forcada}, {Schawinski}, {Schaye}, {Schwope},
  {Sciortino}, {Severgnini}, {Shankar}, {Sijacki}, {Sim}, {Schmid}, {Smith},
  {Steiner}, {Stelzer}, {Stewart}, {Strohmayer}, {Str{\"u}der}, {Sun}, {Takei},
  {Tatischeff}, {Tiengo}, {Tombesi}, {Trinchieri}, {Tsuru}, {Ud-Doula},
  {Ursino}, {Valencic}, {Vanzella}, {Vaughan}, {Vignali}, {Vink}, {Vito},
  {Volonteri}, {Wang}, {Webb}, {Willingale}, {Wilms}, {Wise}, {Worrall},
  {Young}, {Zampieri}, {In't Zand}, {Zane}, {Zezas}, {Zhang}, \&
  {Zhuravleva}}]{2013arXiv1306.2307N}
{Nandra}, K., {Barret}, D., {Barcons}, X., {et~al.} 2013, arXiv e-prints,
  arXiv:1306.2307, \dodoi{10.48550/arXiv.1306.2307}

\bibitem[{{Narayan} \& {Yi}(1994)}]{1994ApJ...428L..13N}
{Narayan}, R., \& {Yi}, I. 1994, \apjl, 428, L13, \dodoi{10.1086/187381}

\bibitem[{{Navarro} {et~al.}(1997){Navarro}, {Frenk}, \&
  {White}}]{1997ApJ...490..493N}
{Navarro}, J.~F., {Frenk}, C.~S., \& {White}, S. D.~M. 1997, \apj, 490, 493,
  \dodoi{10.1086/304888}

\bibitem[{{Newman}(2008)}]{2008ApJ...684...88N}
{Newman}, J.~A. 2008, \apj, 684, 88, \dodoi{10.1086/589982}

\bibitem[{{Norberg} {et~al.}(2009){Norberg}, {Baugh}, {Gazta{\~n}aga}, \&
  {Croton}}]{2009MNRAS.396...19N}
{Norberg}, P., {Baugh}, C.~M., {Gazta{\~n}aga}, E., \& {Croton}, D.~J. 2009,
  \mnras, 396, 19, \dodoi{10.1111/j.1365-2966.2009.14389.x}

\bibitem[{{Omori} {et~al.}(2023){Omori}, {Baxter}, {Chang}, {Friedrich},
  {Alarcon}, {Alves}, {Amon}, {Andrade-Oliveira}, {Bechtol}, {Becker},
  {Bernstein}, {Blazek}, {Bleem}, {Camacho}, {Campos}, {Carnero Rosell},
  {Carrasco Kind}, {Cawthon}, {Chen}, {Choi}, {Cordero}, {Crawford}, {Crocce},
  {Davis}, {DeRose}, {Dodelson}, {Doux}, {Drlica-Wagner}, {Eckert}, {Eifler},
  {Elsner}, {Elvin-Poole}, {Everett}, {Fang}, {Fert{\'e}}, {Fosalba}, {Gatti},
  {Giannini}, {Gruen}, {Gruendl}, {Harrison}, {Herner}, {Huang}, {Huff},
  {Huterer}, {Jarvis}, {Krause}, {Kuropatkin}, {Leget}, {Lemos}, {Liddle},
  {MacCrann}, {McCullough}, {Muir}, {Myles}, {Navarro-Alsina}, {Pandey},
  {Park}, {Porredon}, {Prat}, {Raveri}, {Rollins}, {Roodman}, {Rosenfeld},
  {Ross}, {Rykoff}, {S{\'a}nchez}, {Sanchez}, {Secco}, {Sevilla-Noarbe},
  {Sheldon}, {Shin}, {Troxel}, {Tutusaus}, {Varga}, {Weaverdyck}, {Wechsler},
  {Wu}, {Yanny}, {Yin}, {Zhang}, {Zuntz}, {Abbott}, {Aguena}, {Allam}, {Annis},
  {Bacon}, {Benson}, {Bertin}, {Bocquet}, {Brooks}, {Burke}, {Carlstrom},
  {Carretero}, {Chang}, {Chown}, {Costanzi}, {da Costa}, {Crites}, {Pereira},
  {de Haan}, {De Vicente}, {Desai}, {Diehl}, {Dobbs}, {Doel}, {Everett},
  {Ferrero}, {Flaugher}, {Friedel}, {Frieman}, {Garc{\'\i}a-Bellido},
  {Gaztanaga}, {George}, {Giannantonio}, {Halverson}, {Hinton}, {Holder},
  {Hollowood}, {Holzapfel}, {Honscheid}, {Hrubes}, {James}, {Knox}, {Kuehn},
  {Lahav}, {Lee}, {Lima}, {Luong-Van}, {March}, {McMahon}, {Melchior},
  {Menanteau}, {Meyer}, {Miquel}, {Mocanu}, {Mohr}, {Morgan}, {Natoli},
  {Padin}, {Palmese}, {Paz-Chinch{\'o}n}, {Pieres}, {Plazas Malag{\'o}n},
  {Pryke}, {Reichardt}, {Romer}, {Ruhl}, {Sanchez}, {Schaffer}, {Schubnell},
  {Serrano}, {Shirokoff}, {Smith}, {Staniszewski}, {Stark}, {Suchyta}, {Tarle},
  {Thomas}, {To}, {Vieira}, {Weller}, {Williamson}, {DES}, \& {SPT
  Collaborations}}]{2023PhRvD.107b3529O}
{Omori}, Y., {Baxter}, E.~J., {Chang}, C., {et~al.} 2023, \prd, 107, 023529,
  \dodoi{10.1103/PhysRevD.107.023529}

\bibitem[{{Padmanabhan} {et~al.}(2009){Padmanabhan}, {White}, {Norberg}, \&
  {Porciani}}]{2009MNRAS.397.1862P}
{Padmanabhan}, N., {White}, M., {Norberg}, P., \& {Porciani}, C. 2009, \mnras,
  397, 1862, \dodoi{10.1111/j.1365-2966.2008.14071.x}

\bibitem[{{Peacock}(1991)}]{1991MNRAS.253P...1P}
{Peacock}, J.~A. 1991, \mnras, 253, 1P, \dodoi{10.1093/mnras/253.1.1P}

\bibitem[{{Peebles}(1980)}]{1980lssu.book.....P}
{Peebles}, P.~J.~E. 1980, {The large-scale structure of the universe}

\bibitem[{{Perrotta} {et~al.}(2019){Perrotta}, {Hamann}, {Zakamska},
  {Alexandroff}, {Rupke}, \& {Wylezalek}}]{2019MNRAS.488.4126P}
{Perrotta}, S., {Hamann}, F., {Zakamska}, N.~L., {et~al.} 2019, \mnras, 488,
  4126, \dodoi{10.1093/mnras/stz1993}

\bibitem[{{Petter} {et~al.}(2023){Petter}, {Hickox}, {Alexander}, {Myers},
  {Geach}, {Whalen}, \& {Andonie}}]{2023ApJ...946...27P}
{Petter}, G.~C., {Hickox}, R.~C., {Alexander}, D.~M., {et~al.} 2023, \apj, 946,
  27, \dodoi{10.3847/1538-4357/acb7ef}

\bibitem[{{Petter} {et~al.}(2022){Petter}, {Hickox}, {Alexander}, {Geach},
  {Myers}, {Rosario}, {Fawcett}, {Klindt}, \& {Whalen}}]{2022ApJ...927...16P}
---. 2022, \apj, 927, 16, \dodoi{10.3847/1538-4357/ac4d31}

\bibitem[{{Planck Collaboration} {et~al.}(2020){Planck Collaboration},
  {Aghanim}, {Akrami}, {Ashdown}, {Aumont}, {Baccigalupi}, {Ballardini},
  {Banday}, {Barreiro}, {Bartolo}, {Basak}, {Battye}, {Benabed}, {Bernard},
  {Bersanelli}, {Bielewicz}, {Bock}, {Bond}, {Borrill}, {Bouchet}, {Boulanger},
  {Bucher}, {Burigana}, {Butler}, {Calabrese}, {Cardoso}, {Carron},
  {Challinor}, {Chiang}, {Chluba}, {Colombo}, {Combet}, {Contreras}, {Crill},
  {Cuttaia}, {de Bernardis}, {de Zotti}, {Delabrouille}, {Delouis}, {Di
  Valentino}, {Diego}, {Dor{\'e}}, {Douspis}, {Ducout}, {Dupac}, {Dusini},
  {Efstathiou}, {Elsner}, {En{\ss}lin}, {Eriksen}, {Fantaye}, {Farhang},
  {Fergusson}, {Fernandez-Cobos}, {Finelli}, {Forastieri}, {Frailis},
  {Fraisse}, {Franceschi}, {Frolov}, {Galeotta}, {Galli}, {Ganga},
  {G{\'e}nova-Santos}, {Gerbino}, {Ghosh}, {Gonz{\'a}lez-Nuevo}, {G{\'o}rski},
  {Gratton}, {Gruppuso}, {Gudmundsson}, {Hamann}, {Handley}, {Hansen},
  {Herranz}, {Hildebrandt}, {Hivon}, {Huang}, {Jaffe}, {Jones}, {Karakci},
  {Keih{\"a}nen}, {Keskitalo}, {Kiiveri}, {Kim}, {Kisner}, {Knox},
  {Krachmalnicoff}, {Kunz}, {Kurki-Suonio}, {Lagache}, {Lamarre}, {Lasenby},
  {Lattanzi}, {Lawrence}, {Le Jeune}, {Lemos}, {Lesgourgues}, {Levrier},
  {Lewis}, {Liguori}, {Lilje}, {Lilley}, {Lindholm}, {L{\'o}pez-Caniego},
  {Lubin}, {Ma}, {Mac{\'\i}as-P{\'e}rez}, {Maggio}, {Maino}, {Mandolesi},
  {Mangilli}, {Marcos-Caballero}, {Maris}, {Martin}, {Martinelli},
  {Mart{\'\i}nez-Gonz{\'a}lez}, {Matarrese}, {Mauri}, {McEwen}, {Meinhold},
  {Melchiorri}, {Mennella}, {Migliaccio}, {Millea}, {Mitra},
  {Miville-Desch{\^e}nes}, {Molinari}, {Montier}, {Morgante}, {Moss}, {Natoli},
  {N{\o}rgaard-Nielsen}, {Pagano}, {Paoletti}, {Partridge}, {Patanchon},
  {Peiris}, {Perrotta}, {Pettorino}, {Piacentini}, {Polastri}, {Polenta},
  {Puget}, {Rachen}, {Reinecke}, {Remazeilles}, {Renzi}, {Rocha}, {Rosset},
  {Roudier}, {Rubi{\~n}o-Mart{\'\i}n}, {Ruiz-Granados}, {Salvati}, {Sandri},
  {Savelainen}, {Scott}, {Shellard}, {Sirignano}, {Sirri}, {Spencer},
  {Sunyaev}, {Suur-Uski}, {Tauber}, {Tavagnacco}, {Tenti}, {Toffolatti},
  {Tomasi}, {Trombetti}, {Valenziano}, {Valiviita}, {Van Tent}, {Vibert},
  {Vielva}, {Villa}, {Vittorio}, {Wandelt}, {Wehus}, {White}, {White},
  {Zacchei}, \& {Zonca}}]{2020A&A...641A...6P}
{Planck Collaboration}, {Aghanim}, N., {Akrami}, Y., {et~al.} 2020, \aap, 641,
  A6, \dodoi{10.1051/0004-6361/201833910}

\bibitem[{{Porciani} {et~al.}(2004){Porciani}, {Magliocchetti}, \&
  {Norberg}}]{2004MNRAS.355.1010P}
{Porciani}, C., {Magliocchetti}, M., \& {Norberg}, P. 2004, \mnras, 355, 1010,
  \dodoi{10.1111/j.1365-2966.2004.08408.x}

\bibitem[{{Porciani} \& {Norberg}(2006)}]{2006MNRAS.371.1824P}
{Porciani}, C., \& {Norberg}, P. 2006, \mnras, 371, 1824,
  \dodoi{10.1111/j.1365-2966.2006.10813.x}

\bibitem[{{Prada} {et~al.}(2023){Prada}, {Ereza}, {Smith}, {Lasker}, {Vaisakh},
  {Kehoe}, {Dong-P{\'a}ez}, {Siudek}, {Wang}, {Alam}, {Beutler}, {Bianchi},
  {Cole}, {Dey}, {Kirkby}, {Norberg}, {Aguilar}, {Ahlen}, {Brooks},
  {Claybaugh}, {Dawson}, {de la Macorra}, {Fanning}, {Forero-Romero},
  {Gontcho}, {Hahn}, {Honscheid}, {Ishak}, {Kisner}, {Landriau}, {Manera},
  {Meisner}, {Miquel}, {Moustakas}, {Mueller}, {Nie}, {Percival}, {Poppett},
  {Rezaie}, {Rossi}, {Sanchez}, {Schubnell}, {Tarl{\'e}}, {Vargas-Maga{\~n}a},
  {Weaver}, {Yuan}, \& {Zhou}}]{2023arXiv230606315P}
{Prada}, F., {Ereza}, J., {Smith}, A., {et~al.} 2023, arXiv e-prints,
  arXiv:2306.06315, \dodoi{10.48550/arXiv.2306.06315}

\bibitem[{{Rennehan} {et~al.}(2023){Rennehan}, {Babul}, {Moa}, \&
  {Dav{\'e}}}]{2023arXiv230915898R}
{Rennehan}, D., {Babul}, A., {Moa}, B., \& {Dav{\'e}}, R. 2023, arXiv e-prints,
  arXiv:2309.15898.
\newblock \doarXiv{2309.15898}

\bibitem[{{Rezaie} {et~al.}(2021){Rezaie}, {Ross}, {Seo}, {Mueller},
  {Percival}, {Merz}, {Katebi}, {Bunescu}, {Bautista}, {Brownstein}, {Burtin},
  {Dawson}, {Gil-Mar{\'\i}n}, {Hou}, {Lyke}, {de la Macorra}, {Rossi},
  {Schneider}, {Zarrouk}, \& {Zhao}}]{2021MNRAS.506.3439R}
{Rezaie}, M., {Ross}, A.~J., {Seo}, H.-J., {et~al.} 2021, \mnras, 506, 3439,
  \dodoi{10.1093/mnras/stab1730}

\bibitem[{{Richardson} {et~al.}(2012){Richardson}, {Zheng}, {Chatterjee},
  {Nagai}, \& {Shen}}]{2012ApJ...755...30R}
{Richardson}, J., {Zheng}, Z., {Chatterjee}, S., {Nagai}, D., \& {Shen}, Y.
  2012, \apj, 755, 30, \dodoi{10.1088/0004-637X/755/1/30}

\bibitem[{{Rosario} {et~al.}(2021){Rosario}, {Alexander}, {Moldon}, {Klindt},
  {Thomson}, {Morabito}, {Fawcett}, \& {Harrison}}]{2021MNRAS.505.5283R}
{Rosario}, D.~J., {Alexander}, D.~M., {Moldon}, J., {et~al.} 2021, \mnras, 505,
  5283, \dodoi{10.1093/mnras/stab1653}

\bibitem[{{Ross} {et~al.}(2020){Ross}, {Bautista}, {Tojeiro}, {Alam}, {Bailey},
  {Burtin}, {Comparat}, {Dawson}, {de Mattia}, {du Mas des Bourboux},
  {Gil-Mar{\'\i}n}, {Hou}, {Kong}, {Lyke}, {Mohammad}, {Moustakas}, {Mueller},
  {Myers}, {Percival}, {Raichoor}, {Rezaie}, {Seo}, {Smith}, {Tinker},
  {Zarrouk}, {Zhao}, {Zhao}, {Bizyaev}, {Brinkmann}, {Brownstein}, {Rosell},
  {Chabanier}, {Choi}, {Chuang}, {Cruz-Gonzalez}, {de la Macorra}, {de la
  Torre}, {Escoffier}, {Fromenteau}, {Higley}, {Jullo}, {Kneib}, {McLane},
  {Mu{\~n}oz-Guti{\'e}rrez}, {Neveux}, {Newman}, {Nitschelm},
  {Palanque-Delabrouille}, {Paviot}, {Pullen}, {Rossi}, {Ruhlmann-Kleider},
  {Schneider}, {Maga{\~n}a}, {Vivek}, \& {Zhang}}]{2020MNRAS.498.2354R}
{Ross}, A.~J., {Bautista}, J., {Tojeiro}, R., {et~al.} 2020, \mnras, 498, 2354,
  \dodoi{10.1093/mnras/staa2416}

\bibitem[{{Rupke} \& {Veilleux}(2011)}]{2011ApJ...729L..27R}
{Rupke}, D. S.~N., \& {Veilleux}, S. 2011, \apjl, 729, L27,
  \dodoi{10.1088/2041-8205/729/2/L27}

\bibitem[{{Sabater} {et~al.}(2019){Sabater}, {Best}, {Hardcastle}, {Shimwell},
  {Tasse}, {Williams}, {Br{\"u}ggen}, {Cochrane}, {Croston}, {de Gasperin},
  {Duncan}, {G{\"u}rkan}, {Mechev}, {Morabito}, {Prandoni}, {R{\"o}ttgering},
  {Smith}, {Harwood}, {Mingo}, {Mooney}, \& {Saxena}}]{2019A&A...622A..17S}
{Sabater}, J., {Best}, P.~N., {Hardcastle}, M.~J., {et~al.} 2019, \aap, 622,
  A17, \dodoi{10.1051/0004-6361/201833883}

\bibitem[{{Sabater} {et~al.}(2021){Sabater}, {Best}, {Tasse}, {Hardcastle},
  {Shimwell}, {Nisbet}, {Jelic}, {Callingham}, {R{\"o}ttgering}, {Bonato},
  {Bondi}, {Ciardi}, {Cochrane}, {Jarvis}, {Kondapally}, {Koopmans},
  {O'Sullivan}, {Prandoni}, {Schwarz}, {Smith}, {Wang}, {Williams}, \&
  {Zaroubi}}]{2021A&A...648A...2S}
{Sabater}, J., {Best}, P.~N., {Tasse}, C., {et~al.} 2021, \aap, 648, A2,
  \dodoi{10.1051/0004-6361/202038828}

\bibitem[{{Sanders} {et~al.}(1988){Sanders}, {Soifer}, {Elias}, {Neugebauer},
  \& {Matthews}}]{1988ApJ...328L..35S}
{Sanders}, D.~B., {Soifer}, B.~T., {Elias}, J.~H., {Neugebauer}, G., \&
  {Matthews}, K. 1988, \apjl, 328, L35, \dodoi{10.1086/185155}

\bibitem[{{Schaye} {et~al.}(2015){Schaye}, {Crain}, {Bower}, {Furlong},
  {Schaller}, {Theuns}, {Dalla Vecchia}, {Frenk}, {McCarthy}, {Helly},
  {Jenkins}, {Rosas-Guevara}, {White}, {Baes}, {Booth}, {Camps}, {Navarro},
  {Qu}, {Rahmati}, {Sawala}, {Thomas}, \& {Trayford}}]{2015MNRAS.446..521S}
{Schaye}, J., {Crain}, R.~A., {Bower}, R.~G., {et~al.} 2015, \mnras, 446, 521,
  \dodoi{10.1093/mnras/stu2058}

\bibitem[{{Schlafly} {et~al.}(2019){Schlafly}, {Meisner}, \&
  {Green}}]{2019ApJS..240...30S}
{Schlafly}, E.~F., {Meisner}, A.~M., \& {Green}, G.~M. 2019, \apjs, 240, 30,
  \dodoi{10.3847/1538-4365/aafbea}

\bibitem[{{Scoccimarro} {et~al.}(2001){Scoccimarro}, {Sheth}, {Hui}, \&
  {Jain}}]{2001ApJ...546...20S}
{Scoccimarro}, R., {Sheth}, R.~K., {Hui}, L., \& {Jain}, B. 2001, \apj, 546,
  20, \dodoi{10.1086/318261}

\bibitem[{{Seljak}(2000)}]{2000MNRAS.318..203S}
{Seljak}, U. 2000, \mnras, 318, 203, \dodoi{10.1046/j.1365-8711.2000.03715.x}

\bibitem[{{Shakura} \& {Sunyaev}(1973)}]{1973A&A....24..337S}
{Shakura}, N.~I., \& {Sunyaev}, R.~A. 1973, \aap, 24, 337

\bibitem[{{Shen} {et~al.}(2020){Shen}, {Hopkins}, {Faucher-Gigu{\`e}re},
  {Alexander}, {Richards}, {Ross}, \& {Hickox}}]{2020MNRAS.495.3252S}
{Shen}, X., {Hopkins}, P.~F., {Faucher-Gigu{\`e}re}, C.-A., {et~al.} 2020,
  \mnras, 495, 3252, \dodoi{10.1093/mnras/staa1381}

\bibitem[{{Shen} {et~al.}(2007){Shen}, {Strauss}, {Oguri}, {Hennawi}, {Fan},
  {Richards}, {Hall}, {Gunn}, {Schneider}, {Szalay}, {Thakar}, {Vanden Berk},
  {Anderson}, {Bahcall}, {Connolly}, \& {Knapp}}]{2007AJ....133.2222S}
{Shen}, Y., {Strauss}, M.~A., {Oguri}, M., {et~al.} 2007, \aj, 133, 2222,
  \dodoi{10.1086/513517}

\bibitem[{{Shimwell} {et~al.}(2017){Shimwell}, {R{\"o}ttgering}, {Best},
  {Williams}, {Dijkema}, {de Gasperin}, {Hardcastle}, {Heald}, {Hoang},
  {Horneffer}, {Intema}, {Mahony}, {Mandal}, {Mechev}, {Morabito}, {Oonk},
  {Rafferty}, {Retana-Montenegro}, {Sabater}, {Tasse}, {van Weeren},
  {Br{\"u}ggen}, {Brunetti}, {Chy{\.z}y}, {Conway}, {Haverkorn}, {Jackson},
  {Jarvis}, {McKean}, {Miley}, {Morganti}, {White}, {Wise}, {van Bemmel},
  {Beck}, {Brienza}, {Bonafede}, {Calistro Rivera}, {Cassano}, {Clarke},
  {Cseh}, {Deller}, {Drabent}, {van Driel}, {Engels}, {Falcke}, {Ferrari},
  {Fr{\"o}hlich}, {Garrett}, {Harwood}, {Heesen}, {Hoeft}, {Horellou},
  {Israel}, {Kapi{\'n}ska}, {Kunert-Bajraszewska}, {McKay}, {Mohan},
  {Orr{\'u}}, {Pizzo}, {Prandoni}, {Schwarz}, {Shulevski}, {Sipior}, {Smith},
  {Sridhar}, {Steinmetz}, {Stroe}, {Varenius}, {van der Werf}, {Zensus}, \&
  {Zwart}}]{2017A&A...598A.104S}
{Shimwell}, T.~W., {R{\"o}ttgering}, H.~J.~A., {Best}, P.~N., {et~al.} 2017,
  \aap, 598, A104, \dodoi{10.1051/0004-6361/201629313}

\bibitem[{{Shimwell} {et~al.}(2022){Shimwell}, {Hardcastle}, {Tasse}, {Best},
  {R{\"o}ttgering}, {Williams}, {Botteon}, {Drabent}, {Mechev}, {Shulevski},
  {van Weeren}, {Bester}, {Br{\"u}ggen}, {Brunetti}, {Callingham}, {Chy{\.z}y},
  {Conway}, {Dijkema}, {Duncan}, {de Gasperin}, {Hale}, {Haverkorn}, {Hugo},
  {Jackson}, {Mevius}, {Miley}, {Morabito}, {Morganti}, {Offringa}, {Oonk},
  {Rafferty}, {Sabater}, {Smith}, {Schwarz}, {Smirnov}, {O'Sullivan},
  {Vedantham}, {White}, {Albert}, {Alegre}, {Asabere}, {Bacon}, {Bonafede},
  {Bonnassieux}, {Brienza}, {Bilicki}, {Bonato}, {Calistro Rivera}, {Cassano},
  {Cochrane}, {Croston}, {Cuciti}, {Dallacasa}, {Danezi}, {Dettmar}, {Di
  Gennaro}, {Edler}, {En{\ss}lin}, {Emig}, {Franzen}, {Garc{\'\i}a-Vergara},
  {Grange}, {G{\"u}rkan}, {Hajduk}, {Heald}, {Heesen}, {Hoang}, {Hoeft},
  {Horellou}, {Iacobelli}, {Jamrozy}, {Jeli{\'c}}, {Kondapally}, {Kukreti},
  {Kunert-Bajraszewska}, {Magliocchetti}, {Mahatma}, {Ma{\l}ek}, {Mandal},
  {Massaro}, {Meyer-Zhao}, {Mingo}, {Mostert}, {Nair}, {Nakoneczny},
  {Nikiel-Wroczy{\'n}ski}, {Orr{\'u}}, {Pajdosz-{\'S}mierciak}, {Pasini},
  {Prandoni}, {van Piggelen}, {Rajpurohit}, {Retana-Montenegro}, {Riseley},
  {Rowlinson}, {Saxena}, {Schrijvers}, {Sweijen}, {Siewert}, {Timmerman},
  {Vaccari}, {Vink}, {West}, {Wo{\l}owska}, {Zhang}, \&
  {Zheng}}]{2022A&A...659A...1S}
{Shimwell}, T.~W., {Hardcastle}, M.~J., {Tasse}, C., {et~al.} 2022, \aap, 659,
  A1, \dodoi{10.1051/0004-6361/202142484}

\bibitem[{{Siewert} {et~al.}(2020){Siewert}, {Hale}, {Bhardwaj}, {Biermann},
  {Bacon}, {Jarvis}, {R{\"o}ttgering}, {Schwarz}, {Shimwell}, {Best}, {Duncan},
  {Hardcastle}, {Sabater}, {Tasse}, {White}, \&
  {Williams}}]{2020A&A...643A.100S}
{Siewert}, T.~M., {Hale}, C., {Bhardwaj}, N., {et~al.} 2020, \aap, 643, A100,
  \dodoi{10.1051/0004-6361/201936592}

\bibitem[{{Sijacki} {et~al.}(2007){Sijacki}, {Springel}, {Di Matteo}, \&
  {Hernquist}}]{2007MNRAS.380..877S}
{Sijacki}, D., {Springel}, V., {Di Matteo}, T., \& {Hernquist}, L. 2007,
  \mnras, 380, 877, \dodoi{10.1111/j.1365-2966.2007.12153.x}

\bibitem[{{Silk} \& {Rees}(1998)}]{1998A&A...331L...1S}
{Silk}, J., \& {Rees}, M.~J. 1998, \aap, 331, L1,
  \dodoi{10.48550/arXiv.astro-ph/9801013}

\bibitem[{{Sinha} \& {Garrison}(2020)}]{2020MNRAS.491.3022S}
{Sinha}, M., \& {Garrison}, L.~H. 2020, \mnras, 491, 3022,
  \dodoi{10.1093/mnras/stz3157}

\bibitem[{{Smith} {et~al.}(2016){Smith}, {Best}, {Duncan}, {Hatch}, {Jarvis},
  {R{\"o}ttgering}, {Simpson}, {Stott}, {Cochrane}, {Coppin}, {Dannerbauer},
  {Davis}, {Geach}, {Hale}, {Hardcastle}, {Hatfield}, {Houghton}, {Maddox},
  {McGee}, {Morabito}, {Nisbet}, {Pandey-Pommier}, {Prandoni}, {Saxena},
  {Shimwell}, {Tarr}, {van Bemmel}, {Verma}, {White}, \&
  {Williams}}]{2016sf2a.conf..271S}
{Smith}, D.~J.~B., {Best}, P.~N., {Duncan}, K.~J., {et~al.} 2016, in SF2A-2016:
  Proceedings of the Annual meeting of the French Society of Astronomy and
  Astrophysics, ed. C.~{Reyl{\'e}}, J.~{Richard}, L.~{Cambr{\'e}sy},
  M.~{Deleuil}, E.~{P{\'e}contal}, L.~{Tresse}, \& I.~{Vauglin}, 271--280

\bibitem[{{Smol{\v{c}}i{\'c}} {et~al.}(2017){Smol{\v{c}}i{\'c}}, {Novak},
  {Bondi}, {Ciliegi}, {Mooley}, {Schinnerer}, {Zamorani}, {Navarrete},
  {Bourke}, {Karim}, {Vardoulaki}, {Leslie}, {Delhaize}, {Carilli}, {Myers},
  {Baran}, {Delvecchio}, {Miettinen}, {Banfield}, {Balokovi{\'c}}, {Bertoldi},
  {Capak}, {Frail}, {Hallinan}, {Hao}, {Herrera Ruiz}, {Horesh}, {Ilbert},
  {Intema}, {Jeli{\'c}}, {Kl{\"o}ckner}, {Krpan}, {Kulkarni}, {McCracken},
  {Laigle}, {Middleberg}, {Murphy}, {Sargent}, {Scoville}, \&
  {Sheth}}]{2017A&A...602A...1S}
{Smol{\v{c}}i{\'c}}, V., {Novak}, M., {Bondi}, M., {et~al.} 2017, \aap, 602,
  A1, \dodoi{10.1051/0004-6361/201628704}

\bibitem[{{Soltan}(1982)}]{1982MNRAS.200..115S}
{Soltan}, A. 1982, \mnras, 200, 115, \dodoi{10.1093/mnras/200.1.115}

\bibitem[{{Somerville} \& {Dav{\'e}}(2015)}]{2015ARA&A..53...51S}
{Somerville}, R.~S., \& {Dav{\'e}}, R. 2015, \araa, 53, 51,
  \dodoi{10.1146/annurev-astro-082812-140951}

\bibitem[{{Somerville} {et~al.}(2008){Somerville}, {Hopkins}, {Cox},
  {Robertson}, \& {Hernquist}}]{2008MNRAS.391..481S}
{Somerville}, R.~S., {Hopkins}, P.~F., {Cox}, T.~J., {Robertson}, B.~E., \&
  {Hernquist}, L. 2008, \mnras, 391, 481,
  \dodoi{10.1111/j.1365-2966.2008.13805.x}

\bibitem[{{Springel} {et~al.}(2005){Springel}, {Di Matteo}, \&
  {Hernquist}}]{2005MNRAS.361..776S}
{Springel}, V., {Di Matteo}, T., \& {Hernquist}, L. 2005, \mnras, 361, 776,
  \dodoi{10.1111/j.1365-2966.2005.09238.x}

\bibitem[{{Springel} {et~al.}(2018){Springel}, {Pakmor}, {Pillepich},
  {Weinberger}, {Nelson}, {Hernquist}, {Vogelsberger}, {Genel}, {Torrey},
  {Marinacci}, \& {Naiman}}]{2018MNRAS.475..676S}
{Springel}, V., {Pakmor}, R., {Pillepich}, A., {et~al.} 2018, \mnras, 475, 676,
  \dodoi{10.1093/mnras/stx3304}

\bibitem[{{Sweijen} {et~al.}(2022){Sweijen}, {van Weeren}, {R{\"o}ttgering},
  {Morabito}, {Jackson}, {Offringa}, {van der Tol}, {Veenboer}, {Oonk}, {Best},
  {Bondi}, {Shimwell}, {Tasse}, \& {Thomson}}]{2022NatAs...6..350S}
{Sweijen}, F., {van Weeren}, R.~J., {R{\"o}ttgering}, H.~J.~A., {et~al.} 2022,
  Nature Astronomy, 6, 350, \dodoi{10.1038/s41550-021-01573-z}

\bibitem[{{Tabor} \& {Binney}(1993)}]{1993MNRAS.263..323T}
{Tabor}, G., \& {Binney}, J. 1993, \mnras, 263, 323,
  \dodoi{10.1093/mnras/263.2.323}

\bibitem[{{Tasse} {et~al.}(2021){Tasse}, {Shimwell}, {Hardcastle},
  {O'Sullivan}, {van Weeren}, {Best}, {Bester}, {Hugo}, {Smirnov}, {Sabater},
  {Calistro-Rivera}, {de Gasperin}, {Morabito}, {R{\"o}ttgering}, {Williams},
  {Bonato}, {Bondi}, {Botteon}, {Br{\"u}ggen}, {Brunetti}, {Chy{\.z}y},
  {Garrett}, {G{\"u}rkan}, {Jarvis}, {Kondapally}, {Mandal}, {Prandoni},
  {Repetti}, {Retana-Montenegro}, {Schwarz}, {Shulevski}, \&
  {Wiaux}}]{2021A&A...648A...1T}
{Tasse}, C., {Shimwell}, T., {Hardcastle}, M.~J., {et~al.} 2021, \aap, 648, A1,
  \dodoi{10.1051/0004-6361/202038804}

\bibitem[{{Taylor}(2005)}]{2005ASPC..347...29T}
{Taylor}, M.~B. 2005, in Astronomical Society of the Pacific Conference Series,
  Vol. 347, Astronomical Data Analysis Software and Systems XIV, ed.
  P.~{Shopbell}, M.~{Britton}, \& R.~{Ebert}, 29

\bibitem[{{Thacker} {et~al.}(2006){Thacker}, {Scannapieco}, \&
  {Couchman}}]{2006ApJ...653...86T}
{Thacker}, R.~J., {Scannapieco}, E., \& {Couchman}, H.~M.~P. 2006, \apj, 653,
  86, \dodoi{10.1086/508650}

\bibitem[{{Timlin} {et~al.}(2018){Timlin}, {Ross}, {Richards}, {Myers},
  {Pellegrino}, {Bauer}, {Lacy}, {Schneider}, {Wollack}, \&
  {Zakamska}}]{2018ApJ...859...20T}
{Timlin}, J.~D., {Ross}, N.~P., {Richards}, G.~T., {et~al.} 2018, \apj, 859,
  20, \dodoi{10.3847/1538-4357/aab9ac}

\bibitem[{{Timmerman} {et~al.}(2022){Timmerman}, {van Weeren}, {Botteon},
  {R{\"o}ttgering}, {McNamara}, {Sweijen}, {B{\^\i}rzan}, \&
  {Morabito}}]{2022A&A...668A..65T}
{Timmerman}, R., {van Weeren}, R.~J., {Botteon}, A., {et~al.} 2022, \aap, 668,
  A65, \dodoi{10.1051/0004-6361/202243936}

\bibitem[{{Tinker} {et~al.}(2008){Tinker}, {Kravtsov}, {Klypin}, {Abazajian},
  {Warren}, {Yepes}, {Gottl{\"o}ber}, \& {Holz}}]{2008ApJ...688..709T}
{Tinker}, J., {Kravtsov}, A.~V., {Klypin}, A., {et~al.} 2008, \apj, 688, 709,
  \dodoi{10.1086/591439}

\bibitem[{{Tinker} {et~al.}(2010){Tinker}, {Robertson}, {Kravtsov}, {Klypin},
  {Warren}, {Yepes}, \& {Gottl{\"o}ber}}]{2010ApJ...724..878T}
{Tinker}, J.~L., {Robertson}, B.~E., {Kravtsov}, A.~V., {et~al.} 2010, \apj,
  724, 878, \dodoi{10.1088/0004-637X/724/2/878}

\bibitem[{{Tiwari} {et~al.}(2022){Tiwari}, {Zhao}, {Zheng}, {Zhao}, {Bacon}, \&
  {Schwarz}}]{2022ApJ...928...38T}
{Tiwari}, P., {Zhao}, R., {Zheng}, J., {et~al.} 2022, \apj, 928, 38,
  \dodoi{10.3847/1538-4357/ac5748}

\bibitem[{{van Haarlem} {et~al.}(2013){van Haarlem}, {Wise}, {Gunst}, {Heald},
  {McKean}, {Hessels}, {de Bruyn}, {Nijboer}, {Swinbank}, {Fallows},
  {Brentjens}, {Nelles}, {Beck}, {Falcke}, {Fender}, {H{\"o}randel},
  {Koopmans}, {Mann}, {Miley}, {R{\"o}ttgering}, {Stappers}, {Wijers},
  {Zaroubi}, {van den Akker}, {Alexov}, {Anderson}, {Anderson}, {van Ardenne},
  {Arts}, {Asgekar}, {Avruch}, {Batejat}, {B{\"a}hren}, {Bell}, {Bell}, {van
  Bemmel}, {Bennema}, {Bentum}, {Bernardi}, {Best}, {B{\^\i}rzan}, {Bonafede},
  {Boonstra}, {Braun}, {Bregman}, {Breitling}, {van de Brink}, {Broderick},
  {Broekema}, {Brouw}, {Br{\"u}ggen}, {Butcher}, {van Cappellen}, {Ciardi},
  {Coenen}, {Conway}, {Coolen}, {Corstanje}, {Damstra}, {Davies}, {Deller},
  {Dettmar}, {van Diepen}, {Dijkstra}, {Donker}, {Doorduin}, {Dromer}, {Drost},
  {van Duin}, {Eisl{\"o}ffel}, {van Enst}, {Ferrari}, {Frieswijk}, {Gankema},
  {Garrett}, {de Gasperin}, {Gerbers}, {de Geus}, {Grie{\ss}meier}, {Grit},
  {Gruppen}, {Hamaker}, {Hassall}, {Hoeft}, {Holties}, {Horneffer}, {van der
  Horst}, {van Houwelingen}, {Huijgen}, {Iacobelli}, {Intema}, {Jackson},
  {Jelic}, {de Jong}, {Juette}, {Kant}, {Karastergiou}, {Koers}, {Kollen},
  {Kondratiev}, {Kooistra}, {Koopman}, {Koster}, {Kuniyoshi}, {Kramer},
  {Kuper}, {Lambropoulos}, {Law}, {van Leeuwen}, {Lemaitre}, {Loose}, {Maat},
  {Macario}, {Markoff}, {Masters}, {McFadden}, {McKay-Bukowski}, {Meijering},
  {Meulman}, {Mevius}, {Middelberg}, {Millenaar}, {Miller-Jones}, {Mohan},
  {Mol}, {Morawietz}, {Morganti}, {Mulcahy}, {Mulder}, {Munk}, {Nieuwenhuis},
  {van Nieuwpoort}, {Noordam}, {Norden}, {Noutsos}, {Offringa}, {Olofsson},
  {Omar}, {Orr{\'u}}, {Overeem}, {Paas}, {Pandey-Pommier}, {Pandey}, {Pizzo},
  {Polatidis}, {Rafferty}, {Rawlings}, {Reich}, {de Reijer}, {Reitsma},
  {Renting}, {Riemers}, {Rol}, {Romein}, {Roosjen}, {Ruiter}, {Scaife}, {van
  der Schaaf}, {Scheers}, {Schellart}, {Schoenmakers}, {Schoonderbeek},
  {Serylak}, {Shulevski}, {Sluman}, {Smirnov}, {Sobey}, {Spreeuw}, {Steinmetz},
  {Sterks}, {Stiepel}, {Stuurwold}, {Tagger}, {Tang}, {Tasse}, {Thomas},
  {Thoudam}, {Toribio}, {van der Tol}, {Usov}, {van Veelen}, {van der Veen},
  {ter Veen}, {Verbiest}, {Vermeulen}, {Vermaas}, {Vocks}, {Vogt}, {de Vos},
  {van der Wal}, {van Weeren}, {Weggemans}, {Weltevrede}, {White}, {Wijnholds},
  {Wilhelmsson}, {Wucknitz}, {Yatawatta}, {Zarka}, {Zensus}, \& {van
  Zwieten}}]{2013A&A...556A...2V}
{van Haarlem}, M.~P., {Wise}, M.~W., {Gunst}, A.~W., {et~al.} 2013, \aap, 556,
  A2, \dodoi{10.1051/0004-6361/201220873}

\bibitem[{{Veilleux} {et~al.}(2005){Veilleux}, {Cecil}, \&
  {Bland-Hawthorn}}]{2005ARA&A..43..769V}
{Veilleux}, S., {Cecil}, G., \& {Bland-Hawthorn}, J. 2005, \araa, 43, 769,
  \dodoi{10.1146/annurev.astro.43.072103.150610}

\bibitem[{{Vogelsberger} {et~al.}(2014){Vogelsberger}, {Genel}, {Springel},
  {Torrey}, {Sijacki}, {Xu}, {Snyder}, {Nelson}, \&
  {Hernquist}}]{2014MNRAS.444.1518V}
{Vogelsberger}, M., {Genel}, S., {Springel}, V., {et~al.} 2014, \mnras, 444,
  1518, \dodoi{10.1093/mnras/stu1536}

\bibitem[{{White} \& {Frenk}(1991)}]{1991ApJ...379...52W}
{White}, S. D.~M., \& {Frenk}, C.~S. 1991, \apj, 379, 52,
  \dodoi{10.1086/170483}

\bibitem[{{White} \& {Rees}(1978)}]{1978MNRAS.183..341W}
{White}, S.~D.~M., \& {Rees}, M.~J. 1978, \mnras, 183, 341,
  \dodoi{10.1093/mnras/183.3.341}

\bibitem[{{Whittam} {et~al.}(2022){Whittam}, {Jarvis}, {Hale}, {Prescott},
  {Morabito}, {Heywood}, {Adams}, {Afonso}, {An}, {Ao}, {Bowler}, {Collier},
  {Deane}, {Delhaize}, {Frank}, {Glowacki}, {Hatfield}, {Maddox}, {Marchetti},
  {Matthews}, {Prandoni}, {Randriamampandry}, {Randriamanakoto}, {Smith},
  {Taylor}, {Thomas}, \& {Vaccari}}]{2022MNRAS.516..245W}
{Whittam}, I.~H., {Jarvis}, M.~J., {Hale}, C.~L., {et~al.} 2022, \mnras, 516,
  245, \dodoi{10.1093/mnras/stac2140}

\bibitem[{{Williams} {et~al.}(2019){Williams}, {Hardcastle}, {Best}, {Sabater},
  {Croston}, {Duncan}, {Shimwell}, {R{\"o}ttgering}, {Nisbet}, {G{\"u}rkan},
  {Alegre}, {Cochrane}, {Goyal}, {Hale}, {Jackson}, {Jamrozy}, {Kondapally},
  {Kunert-Bajraszewska}, {Mahatma}, {Mingo}, {Morabito}, {Prandoni},
  {Roskowinski}, {Shulevski}, {Smith}, {Tasse}, {Urquhart}, {Webster}, {White},
  {Beswick}, {Callingham}, {Chy{\.z}y}, {de Gasperin}, {Harwood}, {Hoeft},
  {Iacobelli}, {McKean}, {Mechev}, {Miley}, {Schwarz}, \& {van
  Weeren}}]{2019A&A...622A...2W}
{Williams}, W.~L., {Hardcastle}, M.~J., {Best}, P.~N., {et~al.} 2019, \aap,
  622, A2, \dodoi{10.1051/0004-6361/201833564}

\bibitem[{{Willott} {et~al.}(1999){Willott}, {Rawlings}, {Blundell}, \&
  {Lacy}}]{1999MNRAS.309.1017W}
{Willott}, C.~J., {Rawlings}, S., {Blundell}, K.~M., \& {Lacy}, M. 1999,
  \mnras, 309, 1017, \dodoi{10.1046/j.1365-8711.1999.02907.x}

\bibitem[{{Wright} {et~al.}(2010){Wright}, {Eisenhardt}, {Mainzer}, {Ressler},
  {Cutri}, {Jarrett}, {Kirkpatrick}, {Padgett}, {McMillan}, {Skrutskie},
  {Stanford}, {Cohen}, {Walker}, {Mather}, {Leisawitz}, {Gautier}, {McLean},
  {Benford}, {Lonsdale}, {Blain}, {Mendez}, {Irace}, {Duval}, {Liu}, {Royer},
  {Heinrichsen}, {Howard}, {Shannon}, {Kendall}, {Walsh}, {Larsen}, {Cardon},
  {Schick}, {Schwalm}, {Abid}, {Fabinsky}, {Naes}, \&
  {Tsai}}]{2010AJ....140.1868W}
{Wright}, E.~L., {Eisenhardt}, P. R.~M., {Mainzer}, A.~K., {et~al.} 2010, \aj,
  140, 1868, \dodoi{10.1088/0004-6256/140/6/1868}

\bibitem[{{York} {et~al.}(2000){York}, {Adelman}, {Anderson}, {Anderson},
  {Annis}, {Bahcall}, {Bakken}, {Barkhouser}, {Bastian}, {Berman}, {Boroski},
  {Bracker}, {Briegel}, {Briggs}, {Brinkmann}, {Brunner}, {Burles}, {Carey},
  {Carr}, {Castander}, {Chen}, {Colestock}, {Connolly}, {Crocker}, {Csabai},
  {Czarapata}, {Davis}, {Doi}, {Dombeck}, {Eisenstein}, {Ellman}, {Elms},
  {Evans}, {Fan}, {Federwitz}, {Fiscelli}, {Friedman}, {Frieman}, {Fukugita},
  {Gillespie}, {Gunn}, {Gurbani}, {de Haas}, {Haldeman}, {Harris}, {Hayes},
  {Heckman}, {Hennessy}, {Hindsley}, {Holm}, {Holmgren}, {Huang}, {Hull},
  {Husby}, {Ichikawa}, {Ichikawa}, {Ivezi{\'c}}, {Kent}, {Kim}, {Kinney},
  {Klaene}, {Kleinman}, {Kleinman}, {Knapp}, {Korienek}, {Kron}, {Kunszt},
  {Lamb}, {Lee}, {Leger}, {Limmongkol}, {Lindenmeyer}, {Long}, {Loomis},
  {Loveday}, {Lucinio}, {Lupton}, {MacKinnon}, {Mannery}, {Mantsch}, {Margon},
  {McGehee}, {McKay}, {Meiksin}, {Merelli}, {Monet}, {Munn}, {Narayanan},
  {Nash}, {Neilsen}, {Neswold}, {Newberg}, {Nichol}, {Nicinski}, {Nonino},
  {Okada}, {Okamura}, {Ostriker}, {Owen}, {Pauls}, {Peoples}, {Peterson},
  {Petravick}, {Pier}, {Pope}, {Pordes}, {Prosapio}, {Rechenmacher}, {Quinn},
  {Richards}, {Richmond}, {Rivetta}, {Rockosi}, {Ruthmansdorfer}, {Sandford},
  {Schlegel}, {Schneider}, {Sekiguchi}, {Sergey}, {Shimasaku}, {Siegmund},
  {Smee}, {Smith}, {Snedden}, {Stone}, {Stoughton}, {Strauss}, {Stubbs},
  {SubbaRao}, {Szalay}, {Szapudi}, {Szokoly}, {Thakar}, {Tremonti}, {Tucker},
  {Uomoto}, {Vanden Berk}, {Vogeley}, {Waddell}, {Wang}, {Watanabe},
  {Weinberg}, {Yanny}, {Yasuda}, \& {SDSS Collaboration}}]{2000AJ....120.1579Y}
{York}, D.~G., {Adelman}, J., {Anderson}, John~E., J., {et~al.} 2000, \aj, 120,
  1579, \dodoi{10.1086/301513}

\bibitem[{{Yuan} {et~al.}(2024){Yuan}, {Zhang}, {Ross}, {Donald-McCann},
  {Hadzhiyska}, {Wechsler}, {Zheng}, {Alam}, {Gonzalez-Perez}, {Aguilar},
  {Ahlen}, {Bianchi}, {Brooks}, {de la Macorra}, {Fanning}, {Forero-Romero},
  {Honscheid}, {Ishak}, {Kehoe}, {Lasker}, {Landriau}, {Manera}, {Martini},
  {Meisner}, {Miquel}, {Moustakas}, {Nadathur}, {Newman}, {Nie}, {Percival},
  {Poppett}, {Rocher}, {Rossi}, {Sanchez}, {Samushia}, {Schubnell}, {Seo},
  {Tarl{\'e}}, {Weaver}, {Yu}, {Zhou}, \& {Zou}}]{2024MNRAS.530..947Y}
{Yuan}, S., {Zhang}, H., {Ross}, A.~J., {et~al.} 2024, \mnras, 530, 947,
  \dodoi{10.1093/mnras/stae359}

\bibitem[{{Zakamska} {et~al.}(2016){Zakamska}, {Hamann}, {P{\^a}ris}, {Brandt},
  {Greene}, {Strauss}, {Villforth}, {Wylezalek}, {Alexandroff}, \&
  {Ross}}]{2016MNRAS.459.3144Z}
{Zakamska}, N.~L., {Hamann}, F., {P{\^a}ris}, I., {et~al.} 2016, \mnras, 459,
  3144, \dodoi{10.1093/mnras/stw718}

\bibitem[{{Zehavi} {et~al.}(2011){Zehavi}, {Zheng}, {Weinberg}, {Blanton},
  {Bahcall}, {Berlind}, {Brinkmann}, {Frieman}, {Gunn}, {Lupton}, {Nichol},
  {Percival}, {Schneider}, {Skibba}, {Strauss}, {Tegmark}, \&
  {York}}]{2011ApJ...736...59Z}
{Zehavi}, I., {Zheng}, Z., {Weinberg}, D.~H., {et~al.} 2011, \apj, 736, 59,
  \dodoi{10.1088/0004-637X/736/1/59}

\bibitem[{{Zheng} {et~al.}(2007){Zheng}, {Coil}, \&
  {Zehavi}}]{2007ApJ...667..760Z}
{Zheng}, Z., {Coil}, A.~L., \& {Zehavi}, I. 2007, \apj, 667, 760,
  \dodoi{10.1086/521074}

\end{thebibliography}
\bibliographystyle{aasjournal}



\end{document}